\documentclass{pasj01}
\Received{}
\Accepted{}
\usepackage[switch,mathlines]{lineno}
\usepackage{array} 
\usepackage{geometry}
\usepackage{setspace}
\usepackage{subfigure}
\usepackage{booktabs}
\usepackage{multirow}
\usepackage[round]{natbib}
\usepackage[table,xcdraw,svgnames]{xcolor}
\usepackage{color,xcolor}
\usepackage{CJK}
\usepackage{ulem,xpatch}

\begin{document}
\begin{CJK*}{UTF8}{gkai}

\title{A 3\,mm Molecular Line Survey toward The C-star Envelope CIT\,6}

\author{Kuilu \textsc{Yang}\altaffilmark{1}}%

\author{Yong \textsc{Zhang}\altaffilmark{1,2,3}}%
\email{zhangyong5@mail.sysu.edu.cn}
\author{Jianjie \textsc{Qiu}\altaffilmark{1,3,4}}%
\author{Yina \textsc{Ao}\altaffilmark{1}}%
\author{Xiaohu  \textsc{Li}\altaffilmark{2,5}}%
\altaffiltext{1}{School of Physics and Astronomy, 
                Sun Yat-sen University, 2 Da Xue Road, 
                Tangjia, Zhuhai 519000, 
                Guangdong Province, China}
                \altaffiltext{2}{Xinjiang Astronomical Observatory, Chinese Academy of Sciences, 150 Science 1 − Street, Urumqi, Xinjiang 830011, PR China}
\altaffiltext{3}{CSST Science Center for the Guangdong-Hongkong-Macau Greater Bay Area, Sun Yat-Sen University, Zhuhai, PR China}
\altaffiltext{4}{Key Laboratory of Modern Astronomy and Astrophysics (Nanjing University), Ministry of Education, PR China}
\altaffiltext{5}{Key Laboratory of Radio Astronomy, Chinese Academy of Sciences, Urumqi, Xinjiang 830011, PR China}

\KeyWords{circumstellar matter --- ISM: molecules --- radio lines: stars --- stars: AGB and post-AGB}

\maketitle

\begin{abstract}
We present an unbiased molecular line survey toward the carbon-rich circumstellar envelope CIT\,6 carried out between 90 and 116\,GHz with the Arizona Radio Observatory 12 m telescope. 
A total of 42 lines assigned to 10 molecular species and 4 isotopologues are detected. 
Despite the absence of any newly identified circumstellar molecules, several transitions are freshly reported for  this object. 
This work is a complement to our previous line survey toward CIT\,6 in the frequency ranges of 36--49\,GHz and 131--268\,GHz. 
Based on the measurements and in combination with previously published data, we perform a rotation-diagram analysis to determine the column densities, excitation temperatures, and fractional abundances of the molecules. 
The excitation temperature varies along the radius. 
The abundance pattern of CIT\,6 is broadly consistent with that of 
the prototype C-star IRC+10216 at the specified detection sensitivity, 
indicating that the molecular richness in IRC+10216 cannot 
be attributed to an interpretation of unusual chemistry. 
Nevertheless, subtle distinctions between the two C-stars are found. 
The higher abundance of carbon-chain molecules and 
lower $^{12}$C/$^{13}$C ratio in CIT\,6 compared to IRC+10216 
imply that the former is more massive or in a more evolved 
phase than the latter.

\end{abstract}


\section{Introduction}

When low- to intermediate-mass stars leave the main sequence and evolve to the asymptotic giant branch (AGB) phase, 
the products of nucleosynthesis are dredged up and ejected into the interstellar space, 
forming circumstellar envelopes (CSEs) \citep[see e.g.][]{2005ARA&A..43..435H}. 
The proto-planetary nebula (PPN) and planetary nebula (PN) phases follow the AGB phase in the CSE evolution. 
The physical characteristics of the CSE, 
such as temperature, density, radiation fields, and wind velocity, 
may drastically alter throughout the mass loss of the central star and the envelop expansion, 
leading to complicated circumstellar chemistry. 
CSEs can therefore act as an unconventional chemical laboratory \citep[see e.g.][]{2006PNAS..10312274Z}.
According to their C/O abundance ratio, AGB CSEs can be categorized into C-rich (C/O $>$ 1), O-rich (C/O $<$ 1), and S type (C/O $\sim$ 1) \citep[see e.g.][]{2018A&ARv..26....1H}. 
An O-rich CSE can be progressively transformed to an S-type and C-rich one through the third dredge-up, 
after which diverse carbon-bearing compounds manifest. 
More than 80 different molecular species have so far been detected in CSEs \citep{2018ApJS..239...17M, 2022ApJS..259...30M,2022A&A...658A..39P}. 
To better understand the chemical development of CSEs, 
we need to identify and analyze the lines from various molecular species in evolved stars at different phases. 
However, existing investigations have mostly concentrated on a small number of prominent molecules \citep[e.g.,][]{1992PASJ...44..469T, 2005ApJ...624..331H, 2014ApJ...790L..27A, 2007PASJ...59L..47K, 2017PASJ...69L..10S, 2019A&A...625A.101B, 2022PASJ...74..273A}. 
To this end, we have been working on a long-term project that involves wide-band systematic, unbiased spectral line surveys of a large sample of the CSEs surrounding evolved stars. 
This effort will give us an unbiased perspective on the circumstellar chemistry \citep[see][for a thorough review]{2011IAUS..280..237C}.
The present work is part of our series studies of gas-phase molecular species in CSEs \citep{2008ApJS..177..275H, 2008ApJ...678..328Z, 2009ApJ...691.1660Z, 2009ApJ...700.1262Z, 2013ApJ...773...71Z, 2020PASJ...72...46Z, 2012ApJ...760...66C, 2022ApJS..259...56Q, 2023A&A...669A.121Q}.

The majority of line surveys have been focused on the prototype C-rich CSE IRC+10216 
due to its closeness to {\bf the} Earth (110--135 pc, \citealt{1998A&A...338..491G}) and significant mass loss (3.25 $\times$ 10$^{-5}$ $M_\odot$\,yr$^{-1}$, \citealt{1997ApJ...483..913C}). 
Previous observations of IRC+10216 revealed unparalleled chemical complexity \citep[e.g.][]{2000A&AS..142..181C,2008Ap&SS.313..229A}, 
and more faint molecules are constantly have been found by deeper line surveys \citep{2022A&A...658A..39P}. 
IRC+10216 has become a common reference for research on circumstellar chemistry; however, it is crucial to determine whether it genuinely shares chemical characteristics with other C-rich stars.
This issue has to be resolved by doing line surveys toward more C-rich CSEs. 

CIT\,6, also named RW\,LMi, GL\,1403, IRC+30219, or IRAS\,10131+3049, 
is the second bright C-rich star after IRC+10216, making it a superb line survey target. 
In contrast to IRC+10216, 
unbiased line survey toward CIT\,6 is rather scarce 
(see table~\ref{table:1}). 
CIT\,6 is optically dim but infrared dazzling due to being shrouded in a thick dust envelope \citep{1966ApJ...146..288U}.
It has a variable star with a pulsation period of about 640 days, 
similar to IRC+10216, which has a variable period of about 630 days \citep{2003A&A...402..617W}. 
However, the distance of CIT\,6 was determined to be 400 $\pm$ 50 pc according to \cite{1996AJ....111..962C}, which is further away than the distance of IRC+10216.
There are studies showing that 
CIT\,6 seems to be slightly more evolved than IRC+10216 \citep[e.g.][]{1994ApJ...437..410F, 2009ApJ...691.1660Z}.

Our earlier line surveys demonstrate that the chemical makeup of CIT\,6 and IRC+10216 is strikingly comparable \citep{2009ApJ...691.1660Z, 2012ApJ...760...66C}. 
This essay can be used in conjunction with these publications. 
Section~\ref{Section2} describes the observations and data reductions. 
Section~\ref{Section3} describes the identifications and abundance. 
Section~\ref{Section4} discusses the implication of our findings for circumstellar chemistry. 
Section~\ref{Section5} finally provides a brief summary of this investigation.

\section{Observation and data reduction}
\label{Section2}

The $\lambda$ = 3 mm observations were made in 2015 between mid-May and early June 
using the Arizona Radio Observatory (ARO) 12 m telescope on Kitt Peak, southwest of Tucson, Arizona, in beam switching mode with an azimuthal beam-throw of about 2$\arcmin$. 
In 2013, a 12-meter prototype antenna for the Atacama Large Millimeter-submillimeter Array (ALMA) was installed at Kitt Peak to replace the deteriorating ARO 12-meter dish antenna. 
The on-source integration time can be cut in half to achieve the same spectrum sensitivity thanks to the increased sensitivity of the new antenna. 
The receivers were set to operate in single sideband (SSB) dual polarization mode with a typical rejection ratio of about 20 dB during the observations. 
The spectrometer back ends were equipped with two 256-channel filter banks (FBs) with a channel width of 1 MHz and a 3072-channel millimeter autocorrelator (MAC) with a channel width of 195 kHz.
The equipment settings closely resembled those in \citet{2008ApJS..177..275H} and \citet{2009ApJ...691.1660Z}, 
and thus the systematic errors could be reduced when comparing these spectra.
For the sky conditions and the frequency range, the system temperature ($T_{\rm sys}$) was typically between 100 and 250 K.
At each frequency setting, the on-source integration time took roughly 30 minutes. 
Pointing calibrations were carried out around every two hours by using accessible planets. 

\begin{figure*}
    \centering
    \includegraphics[width=1.0\textwidth]{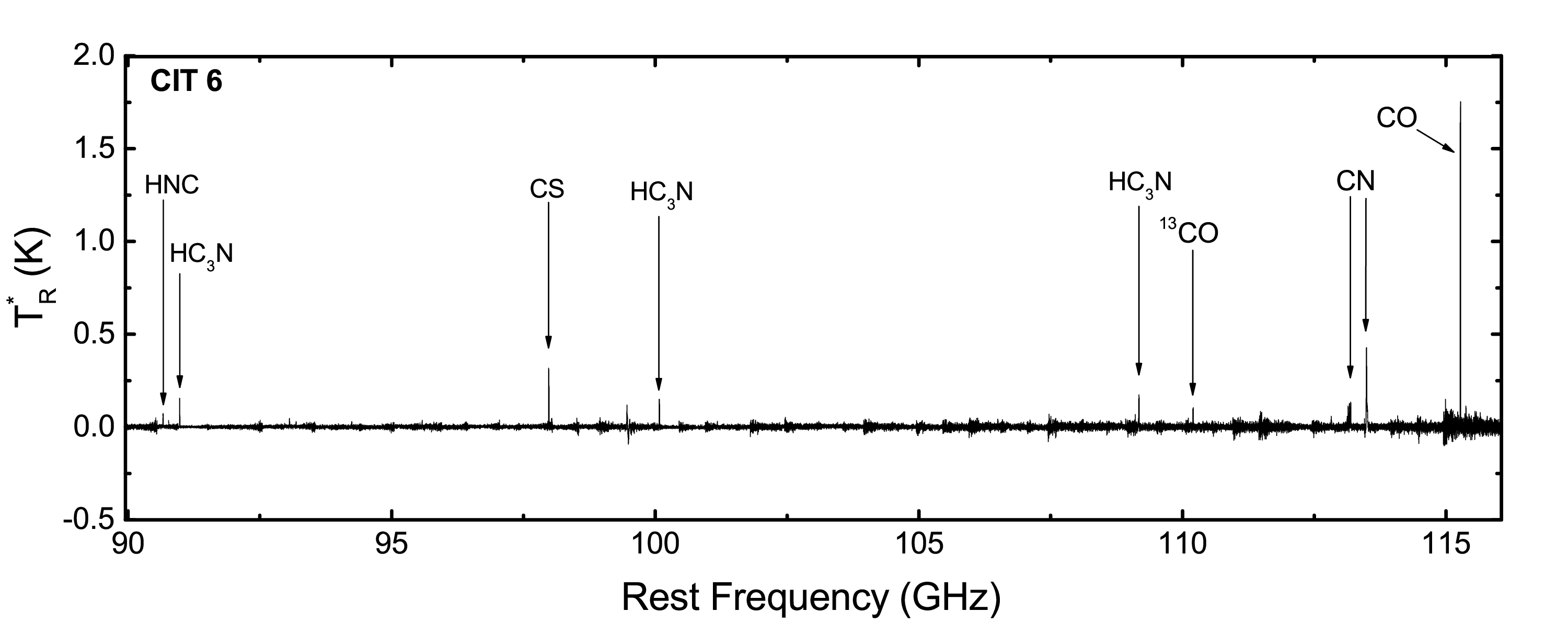}
    \caption{Overview of the full spectra of CIT\,6 obtained in current observations with strong lines marked.}
    \label{figure:1}
\end{figure*}

Data reduction was conducted using the CLASS package in Grenoble Image and Line Data Analysis Software (GILDAS) distributed by Institut
de Radioastronomie Millim\'{e}trique (IRAM). 
After identifying and eliminating the faulty scans that were severely impacted by bandpass anomalies, we co-added the individual scans to generate the spectra. 
A low-order polynomial was applied to fit the baseline of the spectra. 
The complete spectra derived from the current observations are displayed in figure~\ref{figure:1} with strong lines marked. 

The antenna temperatures ($T_{\rm R}^*$) of the line-free regions 
of the co-added spectra have a typical root mean square (RMS) level
of $<$10 mK at a spectral resolution of 600 kHz. 
$T_{\rm R}^*$ can be converted into the main beam temperature by $T_{\rm R}=T_{\rm R}^*/\eta^*$, where $\eta*$ is the main beam efficiency employed as 0.95 (as calculated by extrapolating the experimental data provided by ARO).

\section{Results}
\label{Section3}

\subsection{Line identification and measurement}

Line identifications are based on 
the Spalatalogue database for astronomical spectroscopy\footnote[1]{https://splatalogue.online}, 
which contains data from Jet Propulsion Laboratory Catalogue \citep{1998JQSRT..60..883P}, 
Cologne database for molecular spectroscopy \citep{2001A&A...370L..49M, 2005JMoSt.742..215M}, 
and National Institute of Standards and Technology (NIST) Recommended Rest Frequencies for Observed Interstellar Molecular Microwave Transitions \citep{2004JPCRD..33..177L}. 
We identified 42 emission lines that belong to 10 molecular species and four isotopologues. 
No circumstellar molecules are newly identified. 
We detect all strong lines reported in IRC+10216 within the same frequency coverage.
No lines invisible in IRC+10216 are discovered in CIT 6.
To our knowledge, 20 lines are discovered in this source for the first time. 
Figure~\ref{figure:2} shows the complete zoom-in spectra, which have been smoothed among four neighboring channels  
for ease of viewing. 
The line profiles and fittings are displayed in figure~\ref{figure:3}. 
Table~\ref{table:2} lists the measurements of the detected lines, where the
lines that are recently or previously discovered in this object
are also indicated.
The noise level nearby each line is listed in the 4th column of table~\ref{table:2}.
We discover three unidentified (U) emission features, as listed in table~\ref{table:3}. 
There are no records of any of these U lines in the NIST database. 
The line number per unit frequency interval is about 1.7 GHz$^{-1}$, which is comparable with those of our earlier observations of CIT\,6 (see table~\ref{table:1}). 

\textbf{CO.} 
The $J=1-0$ transitions of CO and its isotopologue $^{13}$CO are shown in figure~\ref{figure:3}. 
The CO $J=1-0$ line exhibits an optically thick parabolic profile, 
while the $^{13}$CO $J=1-0$ line has a double-peaked profile 
and thus is optically thin. 
The two lines have also been detected by \citet{1992A&A...256..235K} using the IRAM 30m telescope, 
who report the velocity-integrated intensities of 176 and 8.5 K\,km\,s$^{-1}$ 
for the CO and $^{13}$CO lines, respectively.
When the beam dilution effect is taken into account, 
our results are consistent with their measurements, 
therefore we do not observe a substantial change in the intensities of the two lines over the previous 30 years.

\textbf{CS.} 
The $J=2-1$ transitions of CS and its isotopologue C$^{34}$S are detected, as shown in figure~\ref{figure:4}. 
Since the line profile of CS $J=2-1$ transition has a parabolic shape and the C$^{34}$S $J=2-1$ line appears a double-peaked shape, 
the former line is optically thick, and the latter is optically thin. 
The CS $J=2-1$ line had been detected by \cite{2003A&A...402..617W} using the Onsala Space Observatory (OSO) 20 m telescope, 
while the C$^{34}$S $J=2-1$ line is a new detection.  
Less than 1/20 of the CS $J=2-1$ emission line's intensity is found in the C$^{34}$S $J=2-1$ emission line.
After taking the impact of the different beam sizes into account, 
the observed intensity of the CS $J=2-1$ line is reasonably consistent with that reported by \citet{2003A&A...402..617W}.

\textbf{SiS.} 
The $J=5-4$ and $J=6-5$ transitions of SiS are detected in this survey, as shown in figure~\ref{figure:5}.
Both lines have been detected by \cite{1985A&A...147..143H} and \cite{2003A&A...402..617W} with the NRAO 11 m, Bell Laboratories 7 m, and OSO 20 m telescopes. 
The integrated intensity ratio between the $J=6-5$ and $J=5-4$ lines is 1.3, 
which is substantially lower than the value of 6.1 reported by \cite{2003A&A...402..617W}. 
Using the Plateau de Bure IRAM (PdBI) interferometer, 
\citet{2000A&A...361.1036L} carried out a spatially resolved observation in the SiS $J=5-4$ line, 
which displays a somewhat different profile from our observations. 
This is presumably because the interferometer observation poorly samples the extended emission regions. 
The SiS $J=5-4$ line has also been detected by \citet{2007A&A...473..871S} with the OSO 20 m telescope. 
Their measurement of integrated intensity matches ours within a 16\% error.

\textbf{SiC$_{2}$.}
The $N_{J,F} = 4_{0,4}-3_{0,3}$, $4_{2,3}-3_{2,2}$, $4_{2,2}-3_{2,1}$, and $5_{0,5}-4_{0,4}$ transitions of SiC$_{2}$ are detected, as shown in figure~\ref{figure:6}. 
The average width of the four SiC$_{2}$ lines is 26.0 $\pm$ 0.4 km\,s$^{-1}$.
The $^{29}$SiC$_{2}$ $5_{0,5}-4_{0,4}$ line is marginally detected with a signal level of 2$\sigma$.
Its velocity-integrated intensity is about 4.3 times lower than the intensity of the main line. 
Its width of 29.0 $\pm$ 3.2 km\,s$^{-1}$ is comparable with those of the main lines, and thus we consider it as a positive detection.
With the exception of the SiC$_{2}$ $N_{J,F}=5_{0,5}-4_{0,4}$ transition, which has been discovered by \citet{2003A&A...402..617W} using the OSO 20 m telescope, 
we do not find any earlier report on the detection of the other four SiC$_{2}$ lines.

\textbf{CN.} 
There are nine fine structures of the CN $N=1-0$ transition lying in the frequency coverage of our observations.
Except for the weak $N_{J,F} = 1_{3/2, 1/2}-0_{1/2, 3/2}$ component at 113.5 GHz, all are visible, as shown in figure~\ref{figure:7}.
The integrated intensities are consistent with those reported by \citet{2009ApJ...690..837M}.

\textbf{HNC.}
The HNC $J=1-0$ transition is detected with a line width of 27.5 $\pm$ 0.3 km\,s$^{-1}$,{\bf as shown in figure~\ref{figure:8}}. 
The integrated intensity of this line is in agreement with that reported by \citet{2003A&A...402..617W} using the OSO 20 m telescope.

\textbf{C$_{3}$N.} 
Four C$_{3}$N transitions are definitely detected, as shown in figure~\ref{figure:9}. 
The $N_J=11_{23/2}-10_{21/2}$ and $11_{21/2}-10_{19/2}$ transitions have been detected by \citet{2003A&A...402..617W} using the OSO 20 m telescope, 
while the $N_J=10_{21/2}-9_{19/2}$ and $10_{19/2}-9_{17/2}$ transitions are new detection for CIT\,6.

\textbf{HC$_{3}$N.} 
The HC$_{3}$N $J=10-9$, $11-10$, and $12-11$ are detected, as shown in figure~\ref{figure:10}. 
The $J=10-9$ and $12-11$ lines have been observed by \citet{1982ApJ...255L..69J} and \citet{1984ApJ...278..176J}, respectively. 
The H$^{13}$CCCN $J=11-10$ and $13-12$ lines are marginally detected with a signal-to-noise ratio of about 1.5$\sigma$.

\textbf{HC$_{5}$N.}
Eight rotational transitions of HC$_{5}$N from the $J=34-33$ to $41-40$ are detected, as shown in figure~\ref{figure:11}. 
Apart from the $J=34-33$ line detected by \citet{2000A&A...361.1036L} and \citet{2003A&A...402..617W}, 
the others are new detection in this source.

\textbf{C$_{4}$H.}  
Five rotational transitions of C$_{4}$H, including one in an excited vibration state, are detected, as shown in figure~\ref{figure:12}.
The C$_{4}$H $N_J=10_{19/2}-9_{17/2}$ and $10_{21/2}-9_{19/2}$ lines have been detected by \cite{2003A&A...402..617W}  using the OSO 20 m telescope.
The other lines are new detection in this object.

\subsection{Rotation-diagram analysis and fractional abundances}

Using the traditional ``rotation-diagram" method, 
we determine the excitation (rotational) temperatures ($T_{\rm ex}$) and the column densities ($N$) of the molecular species identified in recent observations. 
This method makes the assumption that the emission lines are optically thin and the level populations are in thermodynamical equilibrium (LTE). 
A line's integrated intensity and the energy of the transition to the upper level ($E_{\rm u}$) satisfy the expression
\begin{equation}
   \ln\frac{3k\int {T_{\rm S}\,{\rm d}V}}{8\pi \nu S \mu ^2}= \ln\frac{N_{\rm u}}{g_{\rm u}}=\ln\frac{N}{Q(T _{\rm ex})}-\frac{E_{\rm u}}{kT_{\rm ex}}, \label{eq1}
\end{equation}
where $N_{\rm u}$ and $g_{\rm u}$ are the population and degeneracy of the upper level,
$\nu$ the rest frequency of the line, 
$S\mu^2$ the line strength multiplied by the square of the dipole moment, 
and Q($T_{\rm ex}$) the rotational partition function, 
The source brightness temperature $T_{\rm S}$ is obtained after correcting for the beam-dilution effect, 
\begin{equation}
     T_{\rm S}=\frac{\theta^2_{\rm b}+\theta^2_{\rm S}}{\theta^2_{\rm S}} T_{\rm R},
     \label{eq2}
\end{equation}
where $\theta _{\rm b}$ is the half-power beam width (HPBW = 55$\arcsec$--90$\arcsec$ corresponding to 115--90 GHz). 
We simply assume that the source size ($\theta _{\rm S}$) is 20$\arcsec$, 
in line with the assumptions made in \citet{2009ApJ...691.1660Z} and \citet{2012ApJ...760...66C}.
{\bf Equation~\ref{eq1}} implies a linear relation between $ \ln({N_{\rm u}}/{g_{\rm u}})$ and $E_{\rm u}/k$ 
and that $T_{\rm ex}$ and $N$ can be derived by fitting a straight line to observational values. 
Linearity may deviate due to the non-thermal excitation process or misidentification. 
The rotation diagrams, which were constructed by combining the observations of this work and our earlier researches \citep{2009ApJ...691.1660Z,2012ApJ...760...66C}, 
are shown in figure~\ref{figure:13}.

The fractional abundances of molecular X with respect to H$_{2}$ are determined using the same method in \citet{1997IAUS..178..457O} with the expression of 
\begin{center}
    \begin{equation}
        f_{\rm X}=1.7\times10^{-28} \frac{V_{\rm exp}\theta_{\rm b}D}{\dot{M}_{\rm H_2}}\frac{Q(T_{\rm ex})\nu^2}{g_{\rm u}A_{\rm ul}}\frac{e^{\frac{E_{\rm l}}{kT_{\rm ex}}}\int{T_{\rm R}\,{\rm d}V}}{\int_{x_{\rm i}}^{x_{\rm e}} e^{-4\ln 2x^2}\,{\rm d}x}
    \end{equation}
\end{center}
where $V_{\rm exp}$ is the CSE's expansion velocity in km\,s$^{-1}$ taken from the measured line width, 
the distance $D$ is in pc \citep[400 pc,][]{1996AJ....111..962C}, 
$\int{T_{\rm R}\,{\rm d}V}$ is in the scale of K\,km\,s$^{-1}$, 
and the H$_2$ mass loss rate $\dot{M}_{\rm H_2}$ is 3.2 $\times$ 10$^{-6}$ $M_{\odot}$\,yr$^{-1}$ \citep{2009ApJ...691.1660Z}. 
$A_{\rm ul}$ is the Einstein's A-coefficient in $s^{-1}$.  
$E_{\rm l}$ is the energy of the lower level. 
The limits of the integration are given in $x_{\rm i,e}=R_{\rm i,e}/(\theta_{\rm b}D)$, 
where the inner and outer boundaries $R_{\rm i,e}$ are taken from \citet{2003A&A...402..617W} and vary depending on the type of molecule.

The results are summarized in table~\ref{table:4}, 
which demonstrates a good degree of agreement with the earlier findings.

\subsection{Isotopic ratios}

Isotopic abundance ratios reveal the nucleosynthesis and mixing processes of the star. 
Based on the fractional abundances, 
we compute the isotopic ratios (or their lower limits) of carbon, silicon, and sulfur in CIT 6, as presented in table~\ref{table:5}. 
For comparison, we also list the results of CIT\,6 from our earlier work \citep{2009ApJ...691.1660Z} 
and the values of IRC+10216 \citep{2000A&AS..142..181C, 2017A&A...606A..74Z}, CRL\,2688 \citep{2022ApJS..259...56Q}, and the Sun \citep{2003ApJ...591.1220L}.

One of the most useful tracers of the relative degree of primary and secondary processing of stellar nucleosynthesis is the $^{12}$C/$^{13}$C ratio, 
which is the most studied isotope ratio in evolved stars \citep{2005ARA&A..43..435H}.
During the red giant phase, the $^{12}$C/$^{13}$C ratio drops as $^{13}$C, a by-product of the CNO cycle, is brought to the surface of the star by the ``first dredge-up". 
In the AGB phase, the main product of the 3$\alpha$ process, $^{12}$C, is continuously transported to the surface with the periodic ``third dredge-up", 
and thus the $^{12}$C/$^{13}$C ratio gradually \textbf{increases} \citep[e.g.][]{2009ApJ...690..837M}. 
However, extensive observations show that the $^{12}$C/$^{13}$C ratios in CSEs are far lower than those predicted by standard stellar models \citep[e.g.][]{1998A&A...336..915C}. 
This is attributed to an extra mixing process known as ``cool bottom processing" \citep{1995ApJ...453L..41C}, 
which may reduce the $^{12}$C/$^{13}$C ratio of a low-mass AGB star to roughly 4 \citep{1999ApJ...510..232B,1999ApJ...510..217S}. 
For massive AGB stars ($>$4 $M_\odot$), a hot bottom burning process occurs, 
bringing the $^{12}$C/$^{13}$C ratio to an equilibrium value of approximately 3.5 \citep{1998A&A...332L..17F}. 
The optically-thin line of HC$_{3}$N detected suggests a $^{12}$C/$^{13}$C ratio of 17.8 $\pm$ 5.5, 
which is in good agreement with the results of 25 $\pm$ 10 and 12.1 $\pm$ 1.3 obtained by \citet{1989A&A...210...78S} and \citet{2009ApJ...691.1660Z}, respectively. 
CIT\,6 and IRC+10216 have lower $^{12}$C/$^{13}$C ratios compared to the solar value, 
which is an indication of the drudge-up process of AGB stars. 
The fact that CIT\,6 has a lower $^{12}$C/$^{13}$C ratio than IRC+10216 suggests that it is either more massive or more evolved.

The elements in the third row of the periodic table, including Si and S, are barely affected by the nucleosynthesis process of AGB stars. 
According to \citet{2000A&AS..142..181C}, the isotopic ratios of Si and S in IRC+10216 are roughly the same as the solar values. 
Our results show a $^{28}$SiC$_{2}$/$^{29}$SiC$_{2}$ ratio of $5.1\pm1.5$, 
which is in good accord with the prior value of \cite{2009ApJ...691.1660Z} but is lower than the solar value. 
This probably indicates that SiC$_{2}$ has undertaken a chemical fractionation. 
We obtain a C$^{32}$S/C$^{34}$S ratio of 20.4 $\pm$ 4.8, 
which is in accordance with the solar S isotopic ratio.

\section{Discussion}
\label{Section4}

Similarly to IRC+10216, the spectrum of CIT\,6 is characterized by abundant linear carbon-bearing compounds, 
such as CO, CS, SiC$_{2}$, CN, HNC, C$_{3}$N, HC$_{3}$N, HC$_{5}$N, and C$_{4}$H. 
As shown in figure~\ref{figure:13}, the data points of 
HC$_{3}$N and SiC$_{2}$ do not follow a linear relationship
with the lower- and higher-$J$ transitions denoting 
lower and higher excitation temperatures, respectively.
This clearly indicates a stratified temperature structure.
Presumably, the circumstellar temperature decreases with increasing 
radius. Table~\ref{table:2} shows a rough trend that the higher-$J$ transitions have narrower line widths. This is consistent with the idea
that the higher-$J$ transitions primarily arise from the inner layer
and and tend to be spatially unresolved.

In figure~\ref{figure:14}, we compare the fractional abundances relative to HC$_{3}$N for CIT\,6, IRC+10216, and CRL\,2688. 
CRL\,2688 is a PPN that has just finished the AGB phase. 
Although the abundance patterns of the three sources are generally similar, 
the molecular abundances in CIT 6 are typically between those in IRC+10216 and CRL 2688, 
despite the three sources having generally similar patterns of molecular abundance. 
This likely indicates an evolutionary consequence from IRC+10216 through CIT\,6 to CRL\,2688.

The abundance of C$_{4}$H is an exception:
it is significantly lower in CIT\,6 than those in IRC+10216 and CRL\,2688. 
The amount of C$_{4}$H in IRC+10216 is found to be five times more abundant than predicted by photochemical models \citep{1993ApJ...407L..37D}.
The enhancement of C$_{4}$H in CRL\,2688 has been discovered by \citet{1986A&A...154L..12L}. 
This molecule in CRL\,2688 appears to dominantly be located in cold regions, 
with low-$N$ transitions being enhanced and high-$N$ transitions being strongly suppressed \citep{2013ApJ...773...71Z}. 
A speculative conjecture is that C$_{4}$H is destructed during the expansion of AGB envelope but can be synthesized again in cold regions.

The most striking difference between the abundance patterns is the abundance of CN. 
The apparent augmentation of CN in CIT\,6 over IRC+10216 can be attributed to the increasing photodissociation by interstellar radiation field 
via the reaction $ \rm HCN/HNC + {\it h\nu} \rightarrow CN + H$ although other
routes cannot be ruled out.
Because the PPN has a hotter central star emitting more intense ultraviolet photons, 
CRL\,2688 exhibits substantially more plentiful CN than CIT\,6, and IRC+10216. 
CN might be enhanced even further when the CSE transitions into the PN phase. 
This has been verified through the observations of NGC 7027 \citep{2008ApJ...678..328Z}.

On the other hand, CN can be reprocessed to produce cyanogenic polynylene molecules (HC$_{\rm 2n-1}$N, n $>$ 1), 
which are of astronomical interest due to their large dipole moments. 
In line with earlier results \citep{2003A&A...402..617W,2009ApJ...691.1660Z,2012ApJ...760...66C}, 
we find that the fractional abundances of HC$_{3}$N and HC$_{5}$N in CIT\,6 are larger than those in IRC+10216. 
This is in accordance with the idea that cyanopolyynes are formed from CN, and it suggests that the longer cyanopolyyne molecules are then synthesized in a sequential fashion.

\section{Summary}
\label{Section5}

In this paper, we present an unbiased line survey toward CIT\,6 at the 3 mm band with frequency coverage of 90--116 GHz. 
The spectral data can be taken as a supplement to our prior investigations on the molecular species in CIT\,6.
Forty-two emission lines belonging to 10 molecular species and four isotopologues are detected; 
to our knowledge, some of these lines are seen in this object for the first time. 
Three lines remain unidentified. 

Given our detection limit and frequency coverage, there is not much of a chemical difference between IRC+10216 and CIT\,6. 
We determine the column densities and fractional abundances of the molecules in CIT\,6. 
The outcomes match our earlier results from observations made at different wavelength windows quite well. 
The temperature of the envelope is stratified. 
Through a comparison between the molecular abundances of different CSEs, 
we infer that CIT 6 is at an evolutionary stage between IRC+10216 and CRL\,2688. 

\section*{Funding}
The work was supported by the National Science Foundation of China (NSFC, Grant Nos. 11973099 and 12003080) and the science research grants from the China Manned Space Project (NO. CMS-CSST-2021-A09 and CMS-CSST-2021-A10). JJQ is supported by the Guangdong Basic and Applied Basic Research Foundation (No. 2019A1515110588) and the Fundamental Research Funds for the Central Universities (Sun Yat-sen University, No. 22qntd3101). 

\begin{ack} 
YZ thanks Xinjiang Uygur Autonomous Region
of China for support from the Tianchi Talent Program.
XL acknowledges support from the Xinjiang Tianchi project (2019).
\end{ack}

\clearpage

\newgeometry{top=2cm,bottom=3cm}

\clearpage

\begin{figure*}
    \centering
    \includegraphics[width=0.9\textwidth]{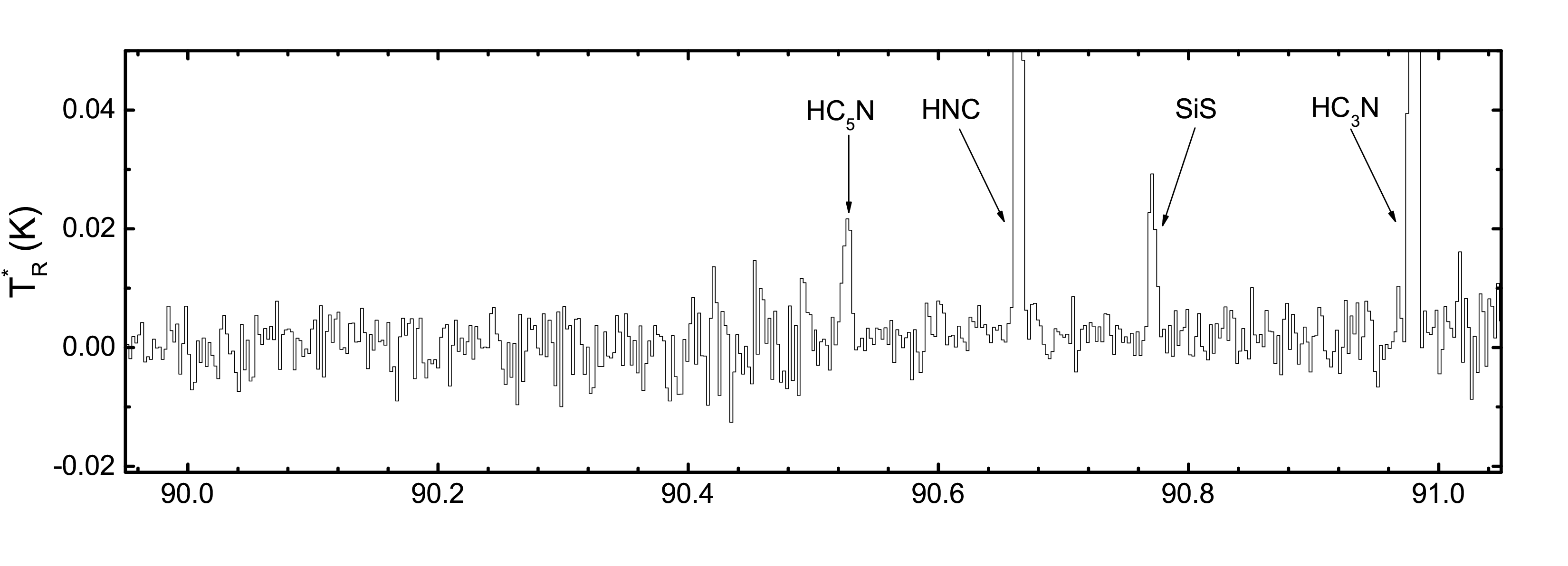}
    \includegraphics[width=0.9\textwidth]{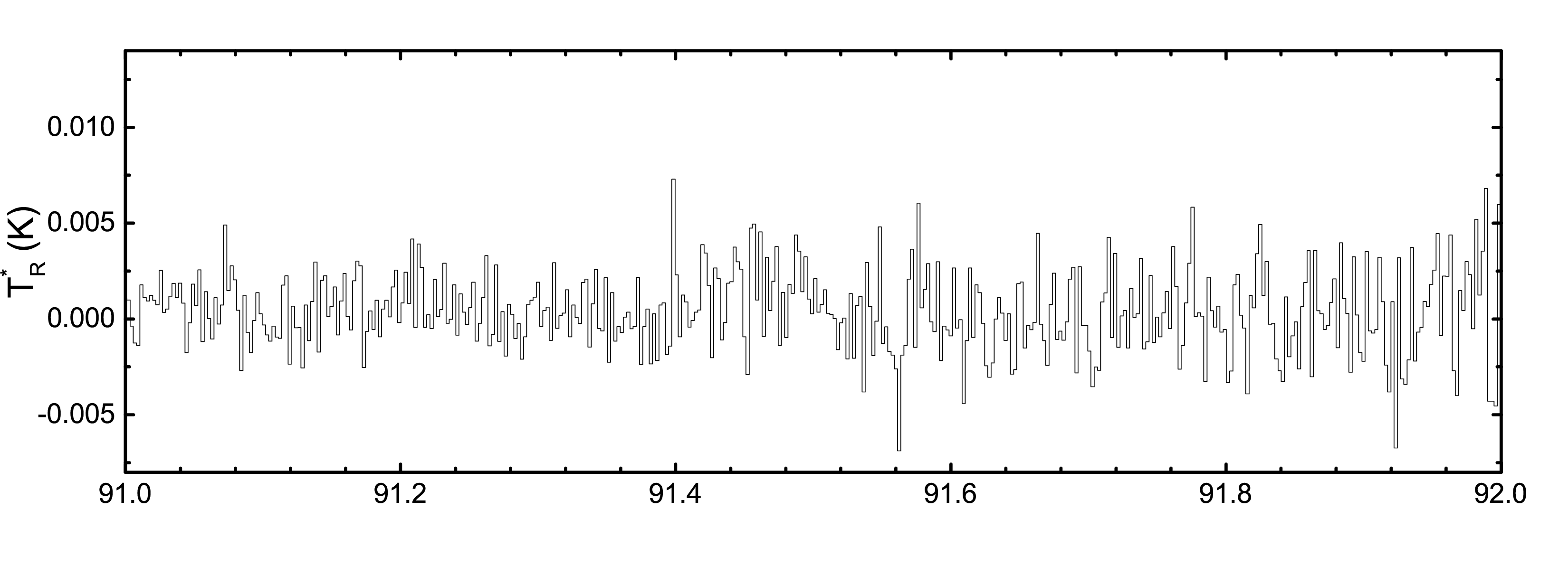}
    \includegraphics[width=0.9\textwidth]{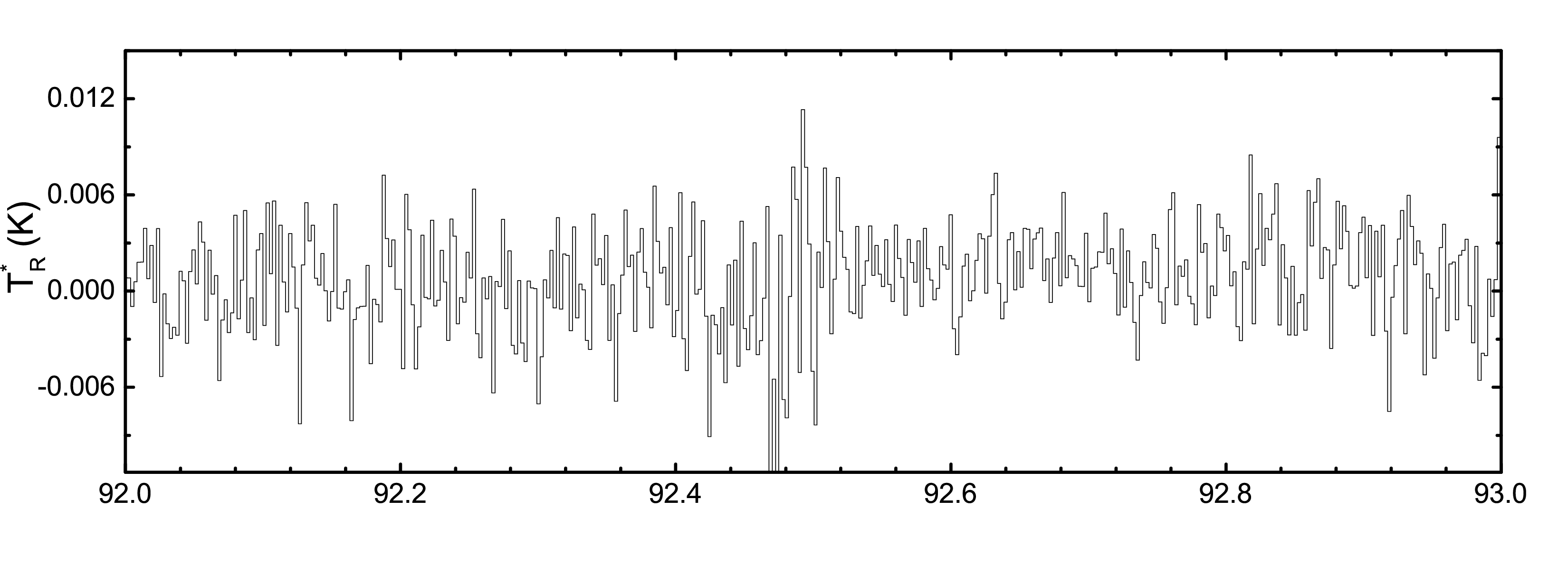}
    \includegraphics[width=0.9\textwidth]{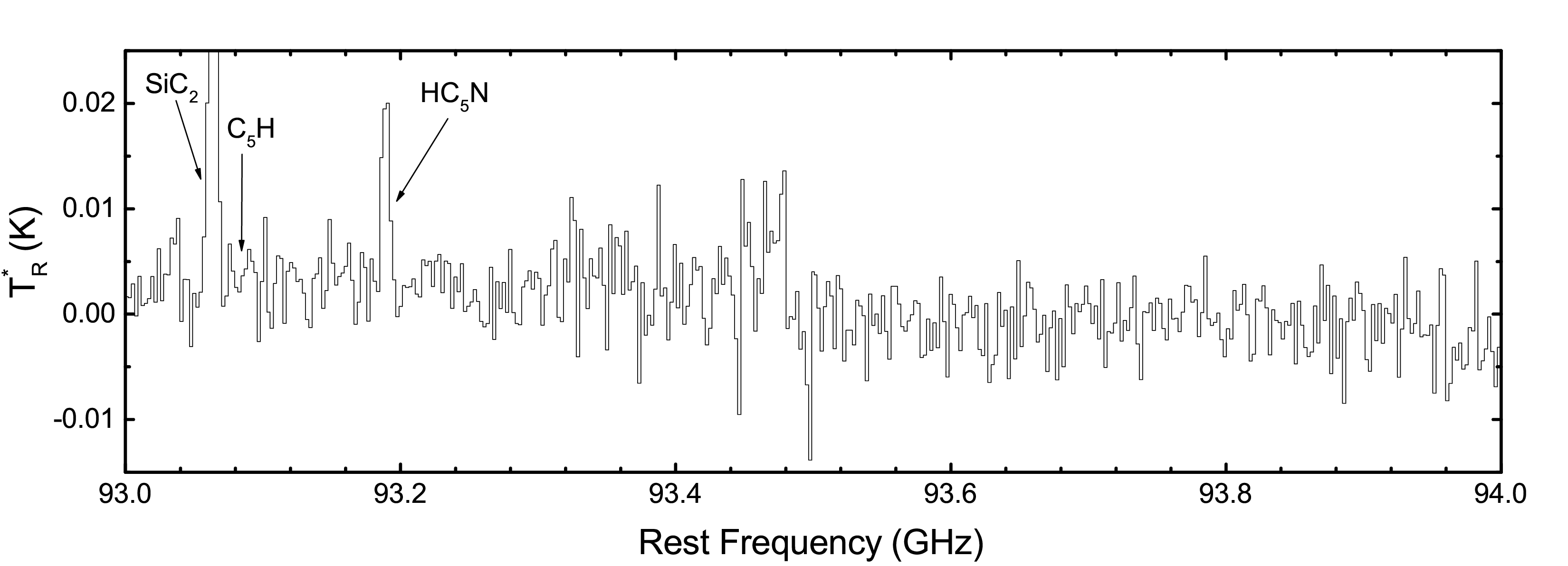}
    \caption{Zoom-in spectra of CIT\,6 in the frequency range of 90--116 GHz with all lines marked.   Note that the spectra
    have been smoothed over four adjacent channels for the sake of view.}
    \label{figure:2}
\end{figure*}
 
 \clearpage

\addtocounter{figure}{-1}
\begin{figure*}
    \addtocounter{figure}{0}
    \centering
    \includegraphics[width=0.9\textwidth]{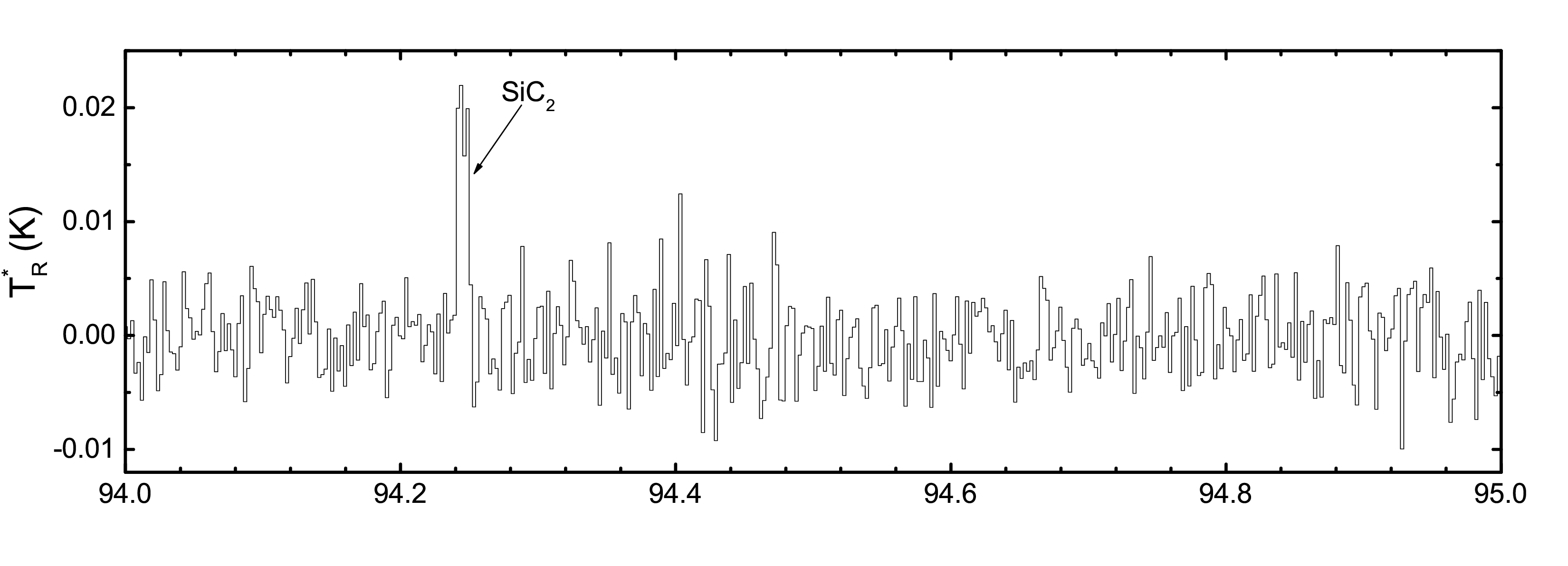}
    \includegraphics[width=0.9\textwidth]{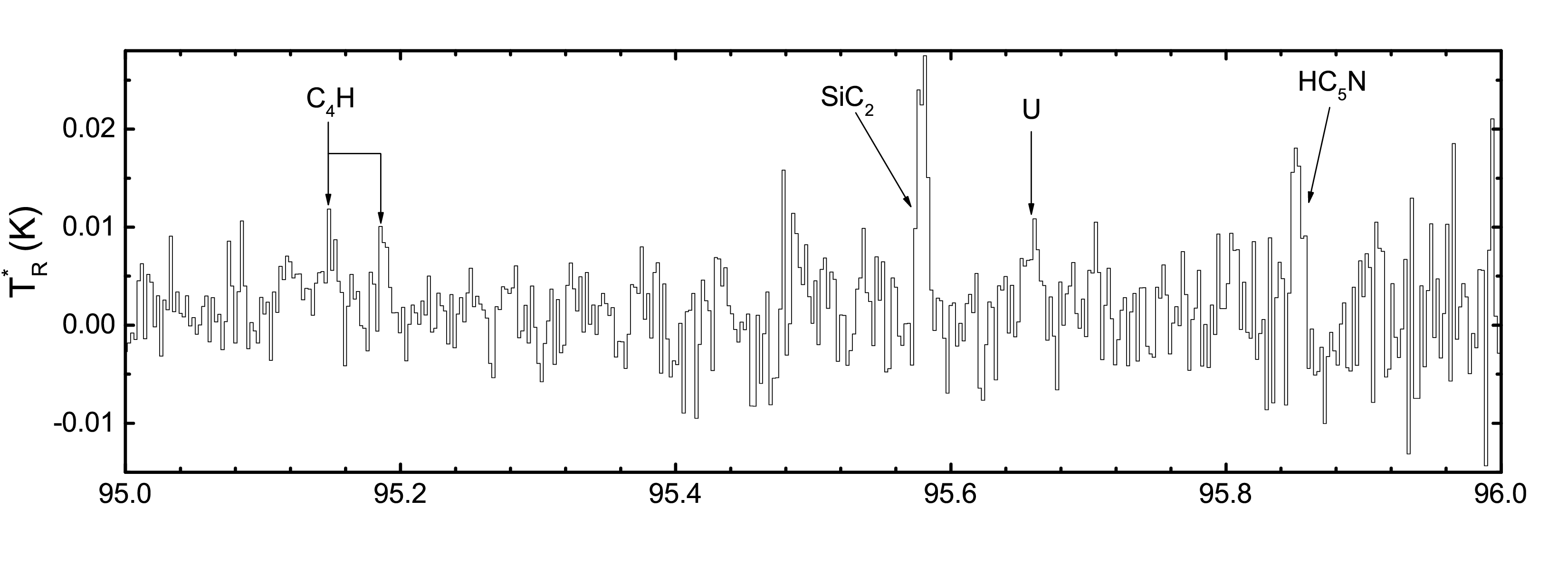}
    \includegraphics[width=0.9\textwidth]{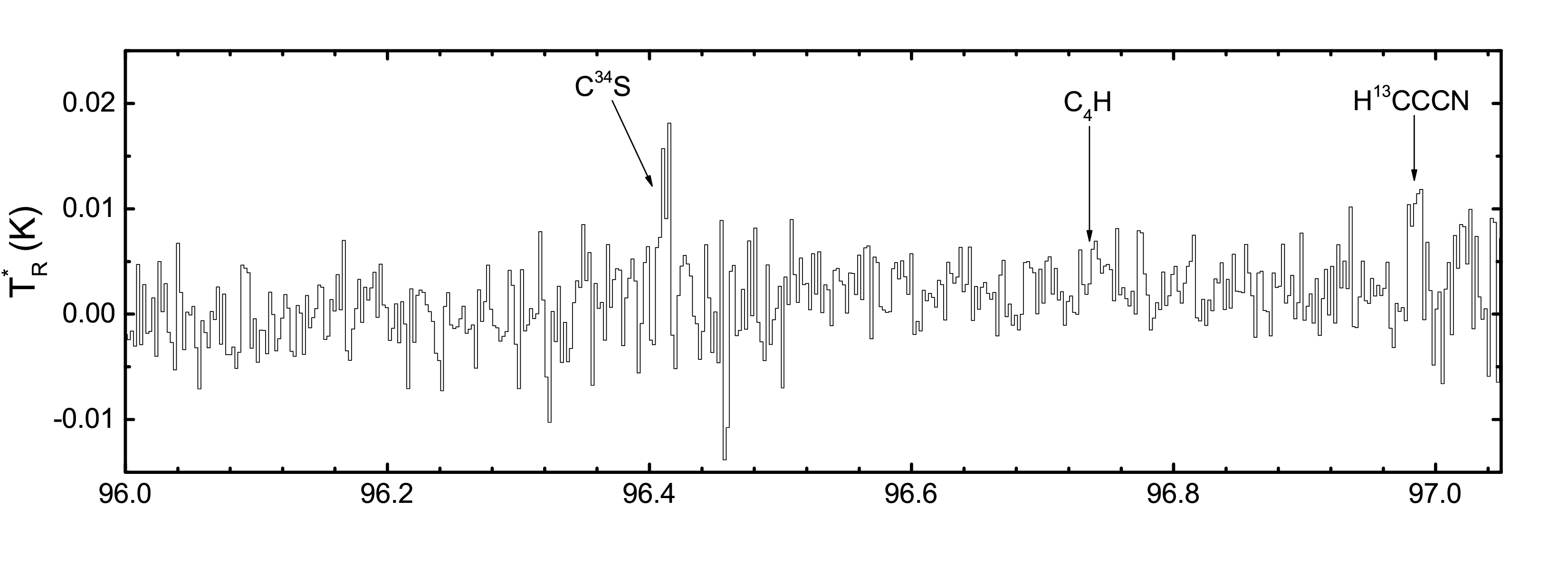}
    \includegraphics[width=0.9\textwidth]{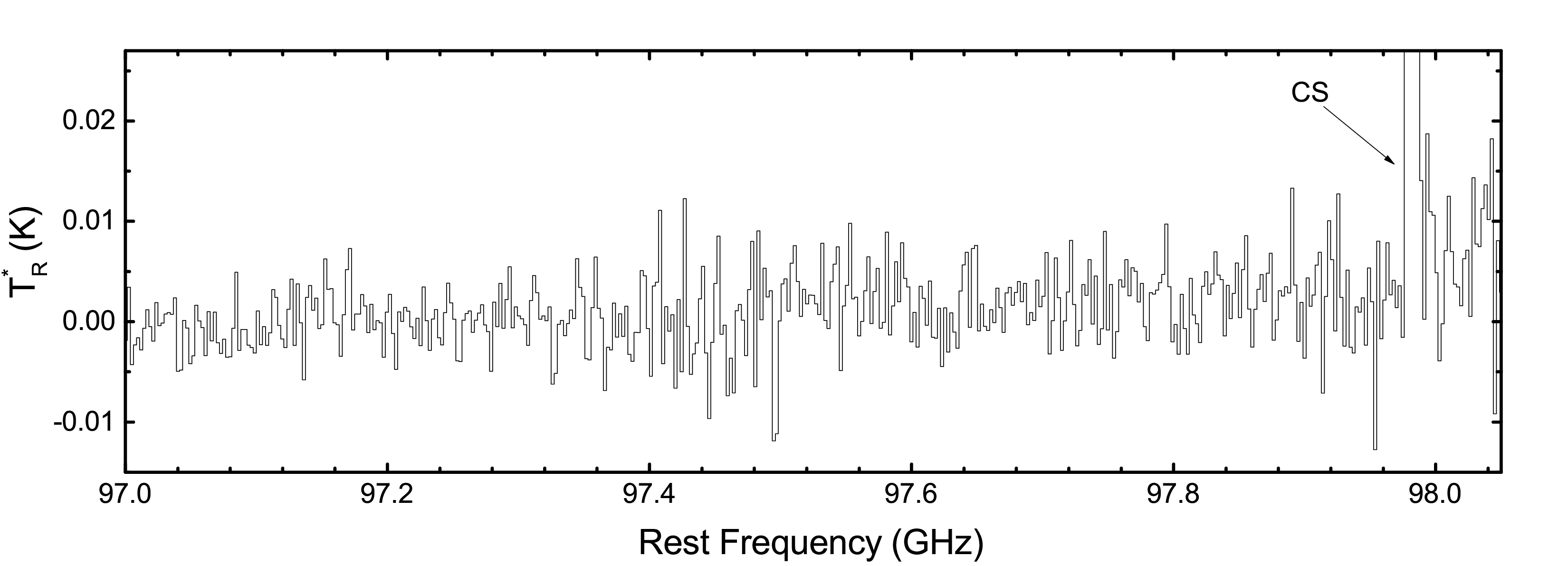}
    \caption{(Continued)}
   
\end{figure*}

\clearpage

\addtocounter{figure}{-1}
\begin{figure*}
    \addtocounter{figure}{0}
    \centering
    \includegraphics[width=0.9\textwidth]{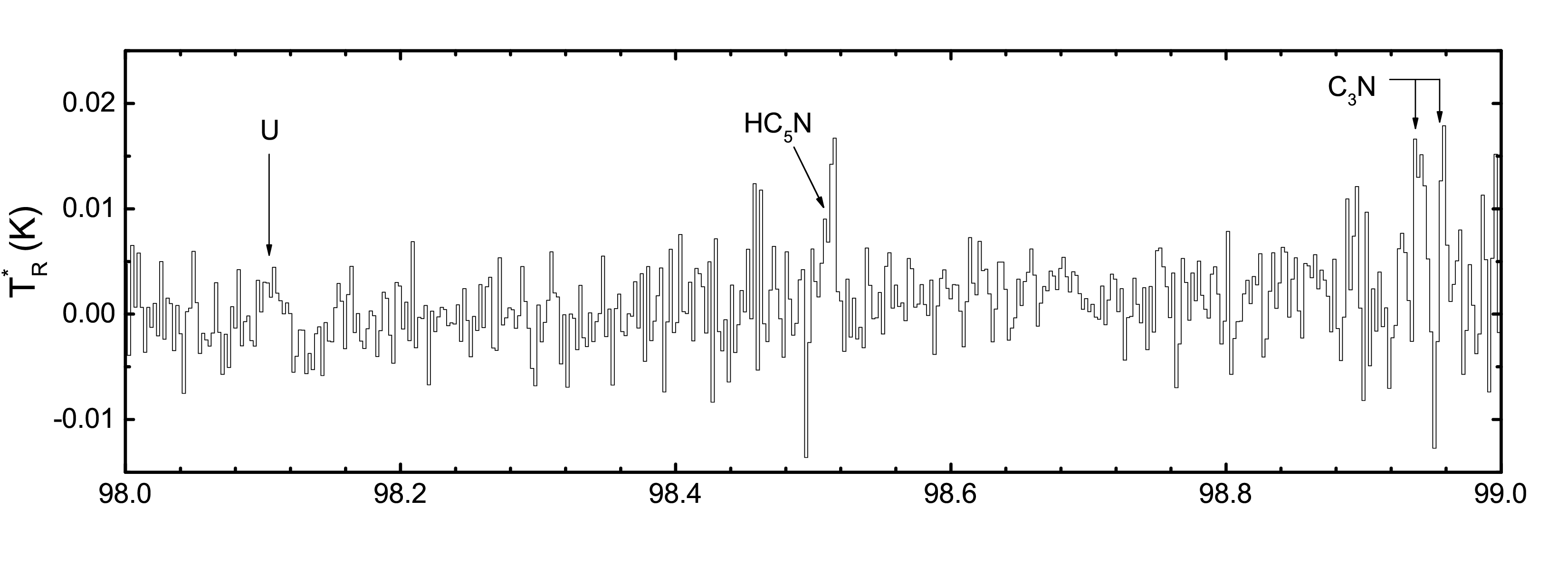}
    \includegraphics[width=0.9\textwidth]{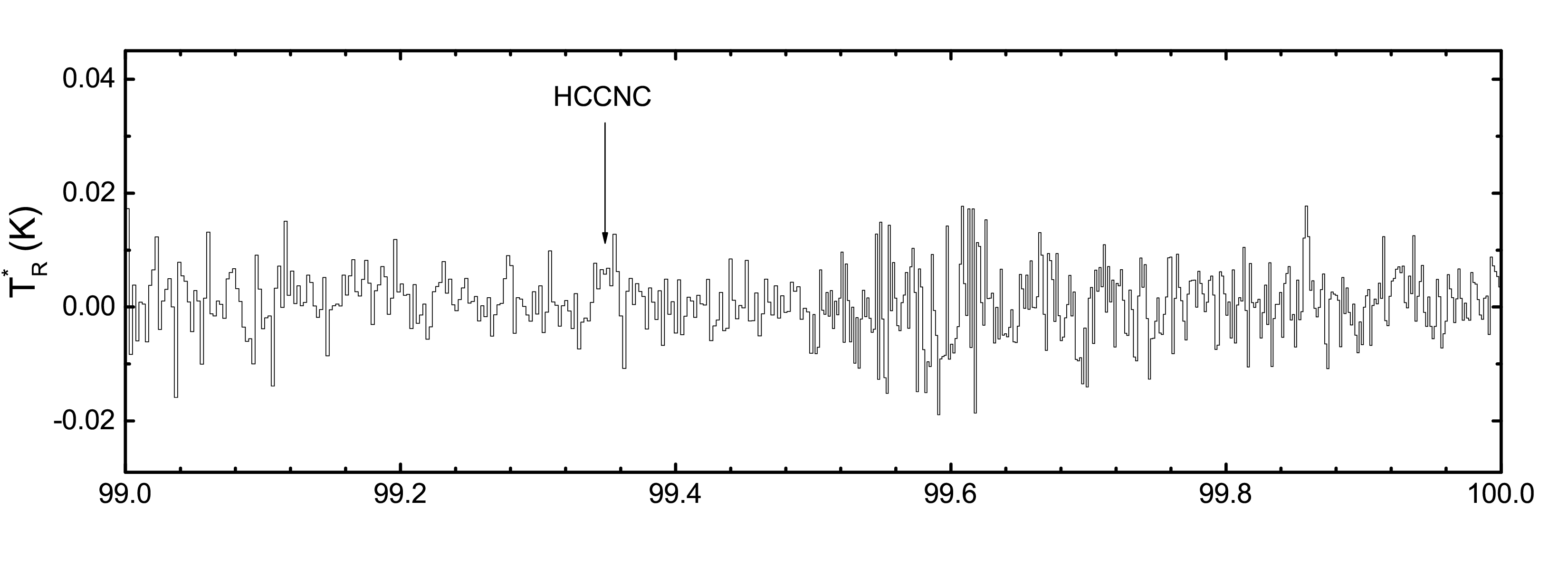}
    \includegraphics[width=0.9\textwidth]{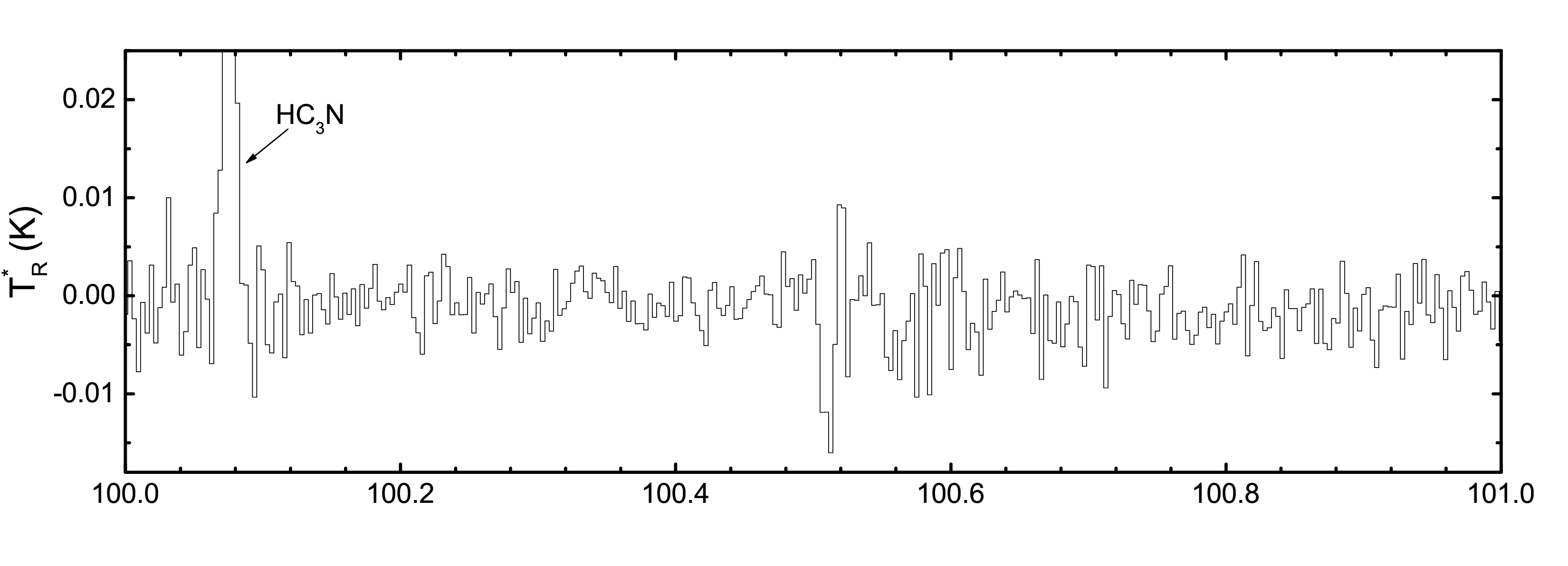}
    \includegraphics[width=0.9\textwidth]{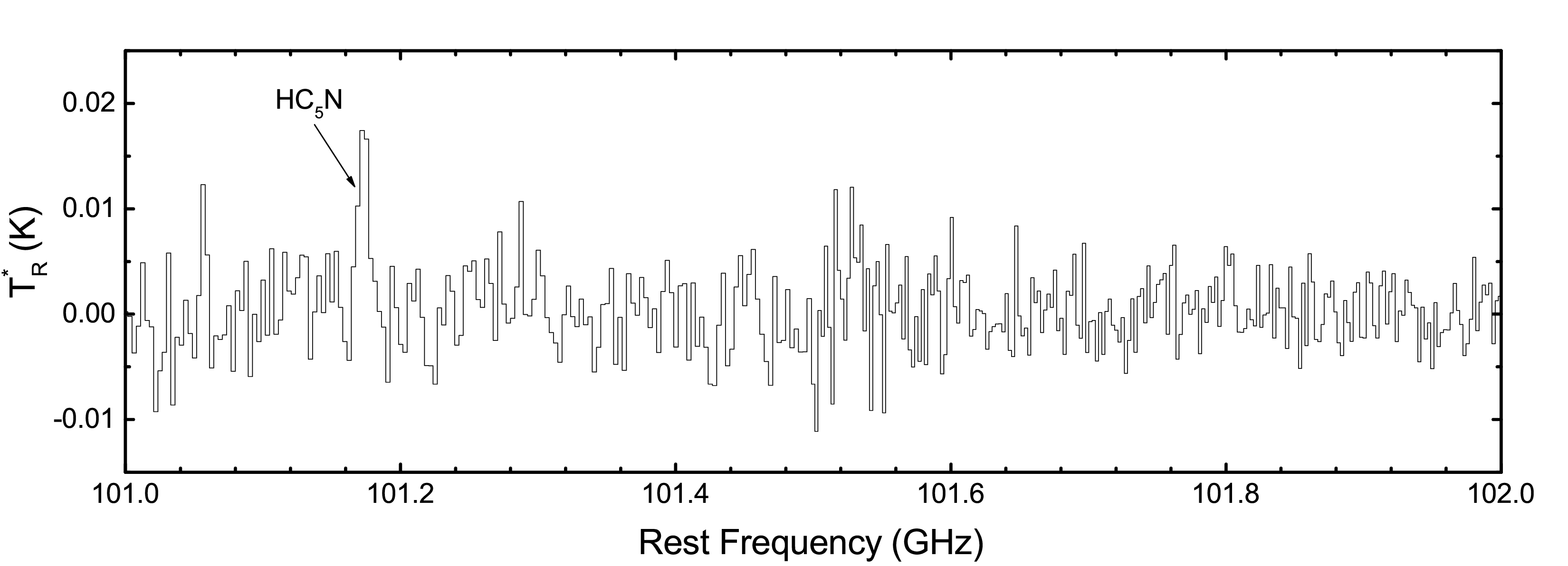}
    \caption{(Continued)}
\end{figure*}

\clearpage

\addtocounter{figure}{-1}
\begin{figure*}
    \addtocounter{figure}{0}
    \centering
    \includegraphics[width=0.9\textwidth]{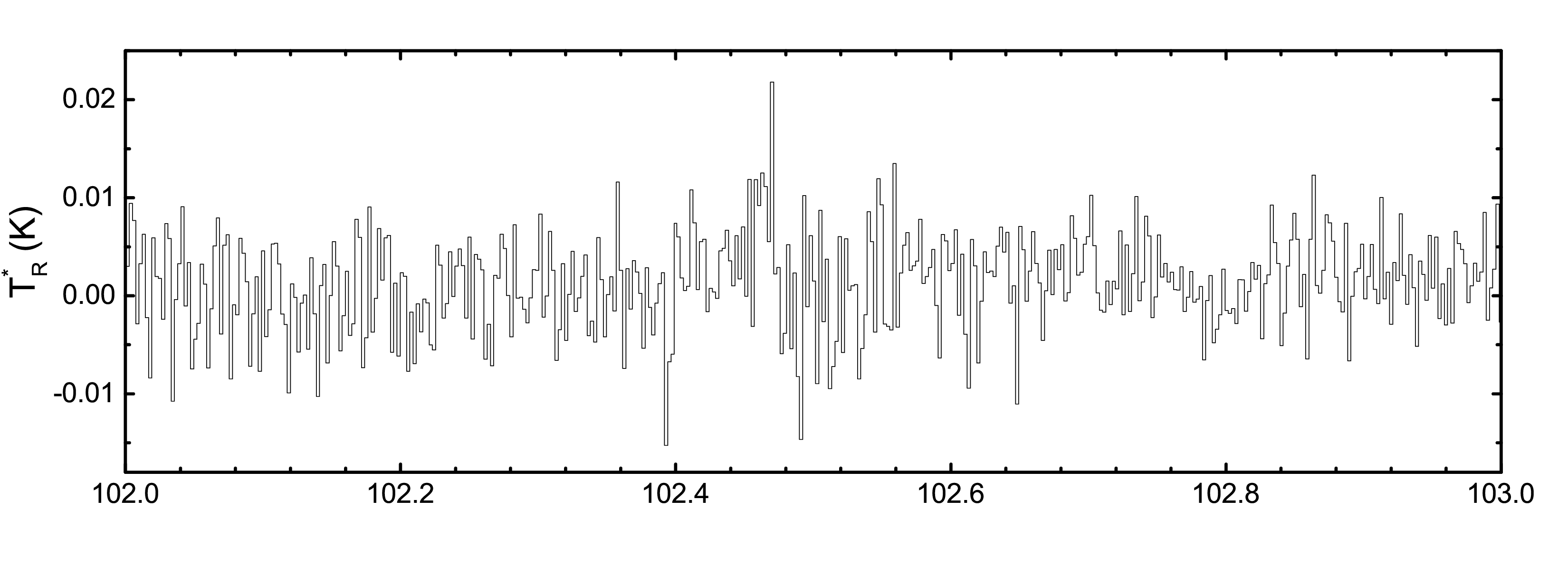}
    \includegraphics[width=0.9\textwidth]{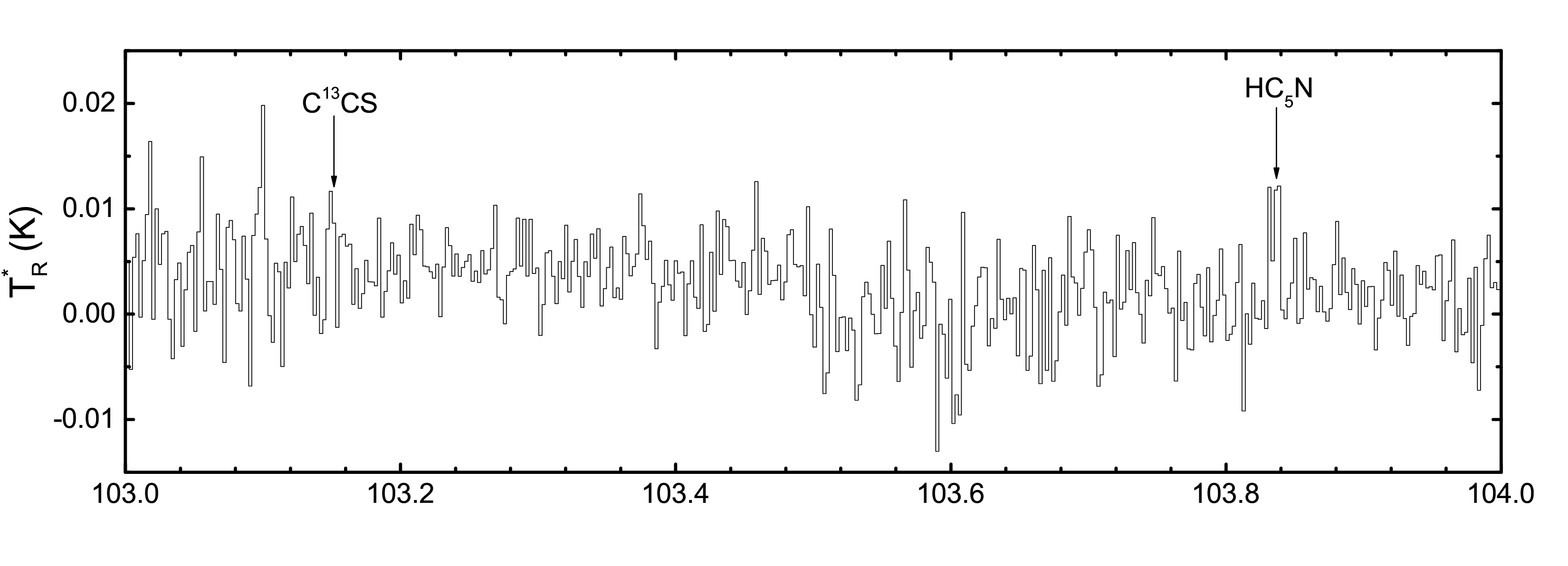}
    \includegraphics[width=0.9\textwidth]{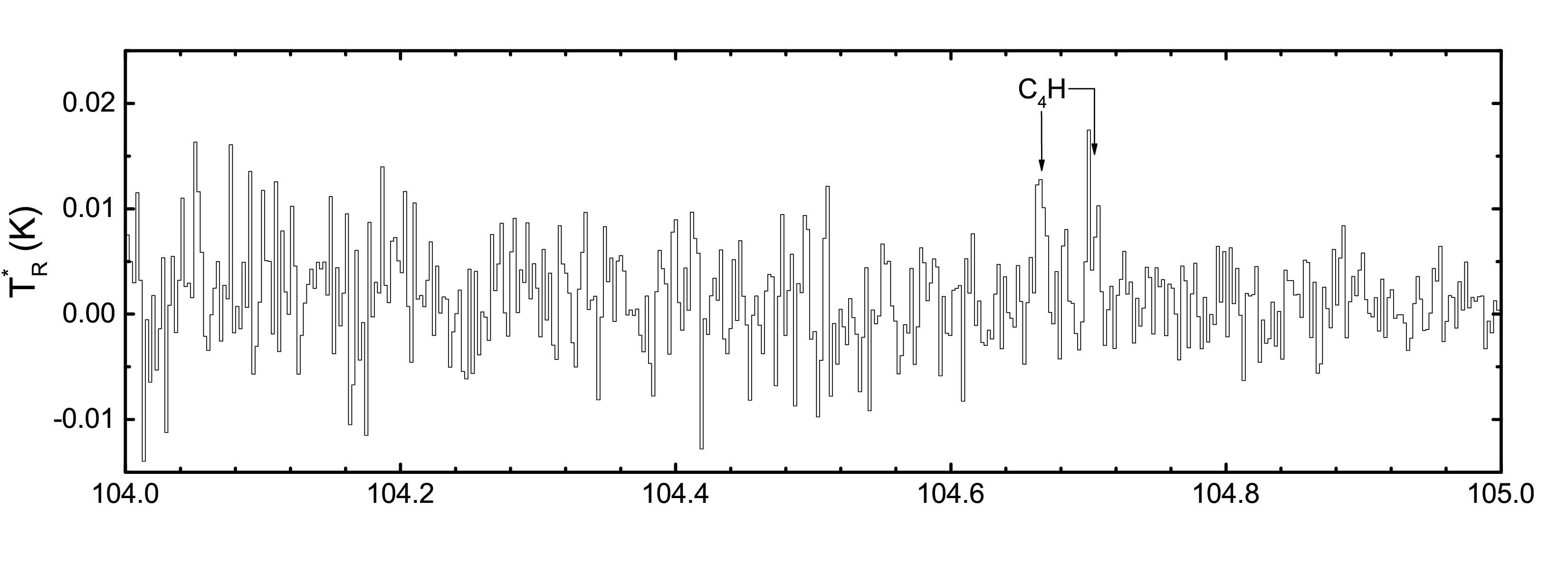}
    \includegraphics[width=0.9\textwidth]{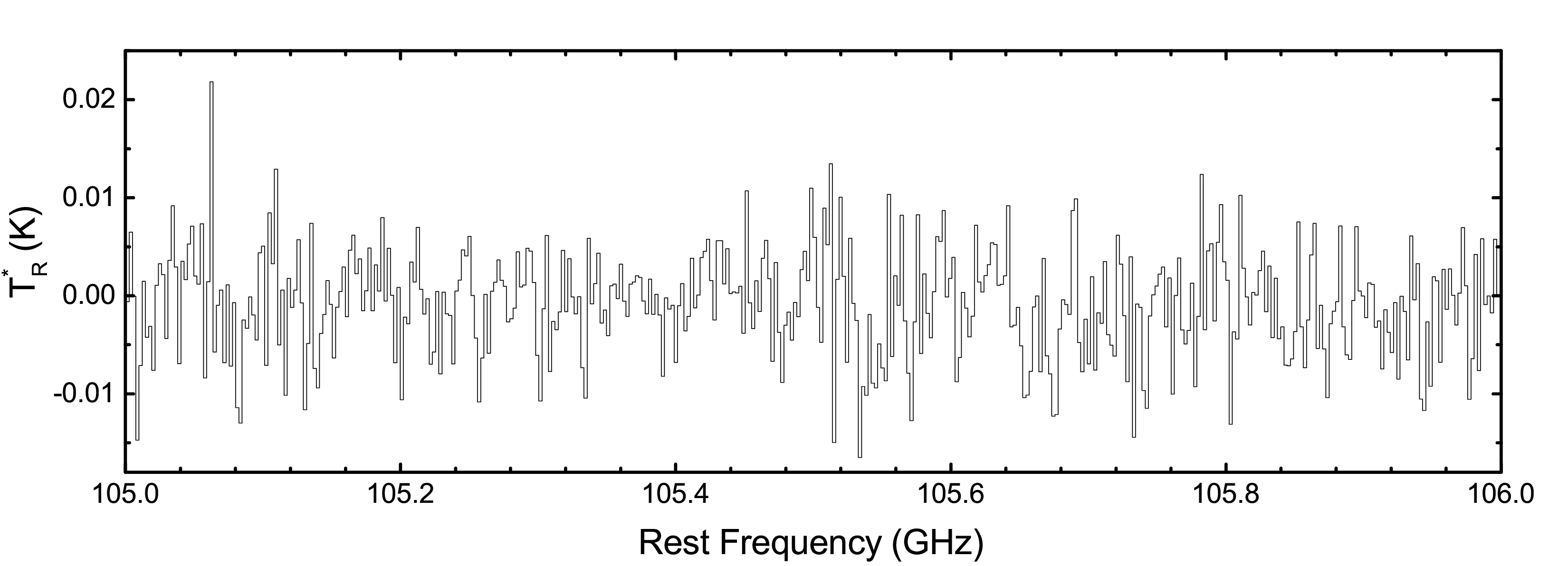}
    \caption{(Continued)}
   
\end{figure*}

\clearpage

\addtocounter{figure}{-1}
\begin{figure*}
    \addtocounter{figure}{0}
    \centering
    \includegraphics[width=0.9\textwidth]{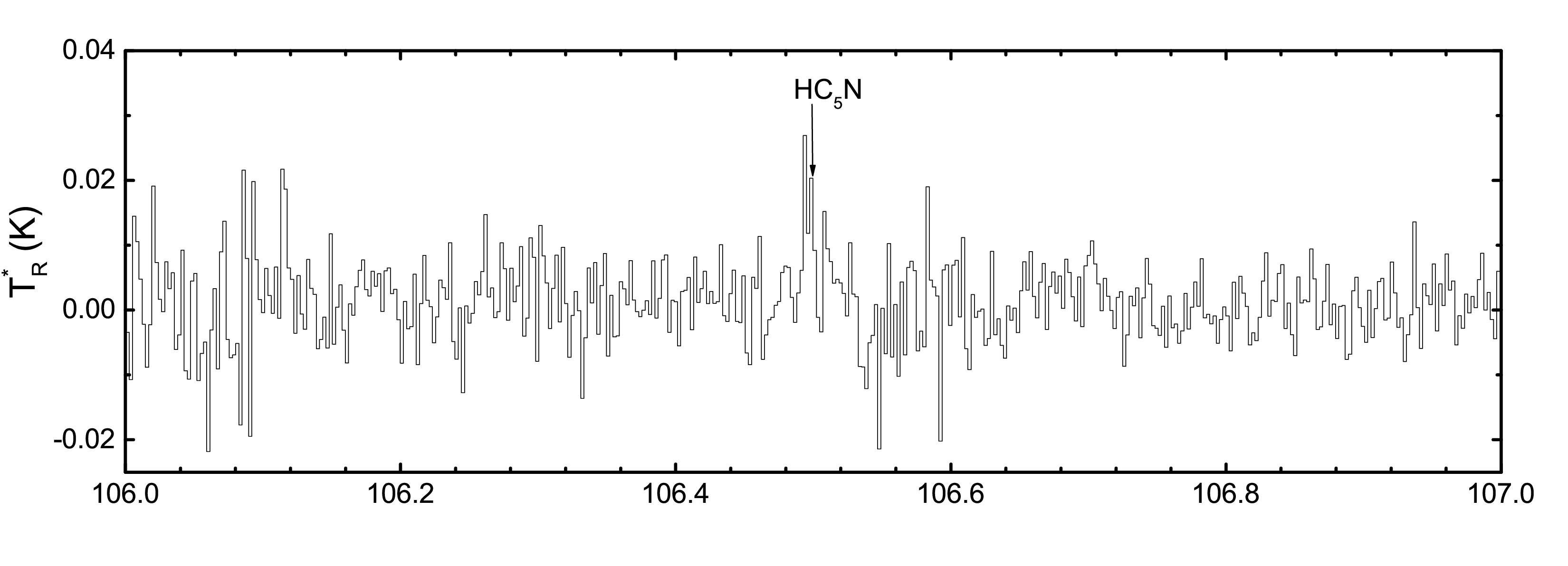}
    \includegraphics[width=0.9\textwidth]{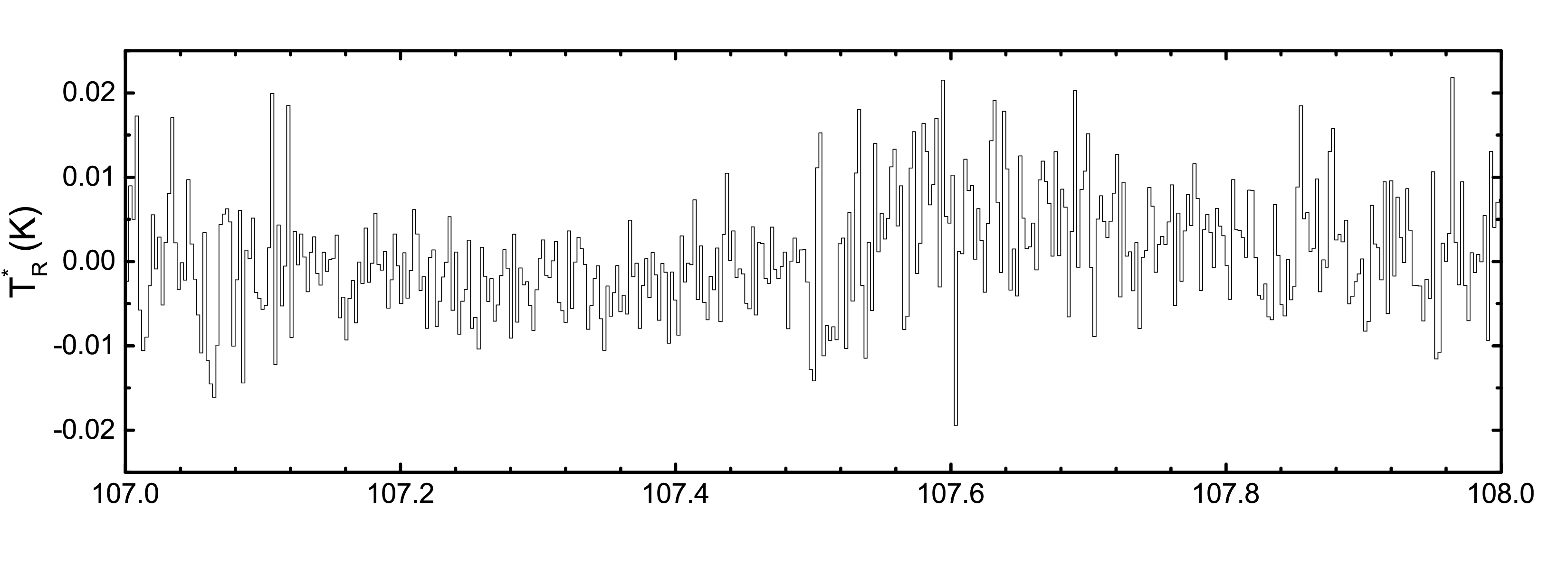}
    \includegraphics[width=0.9\textwidth]{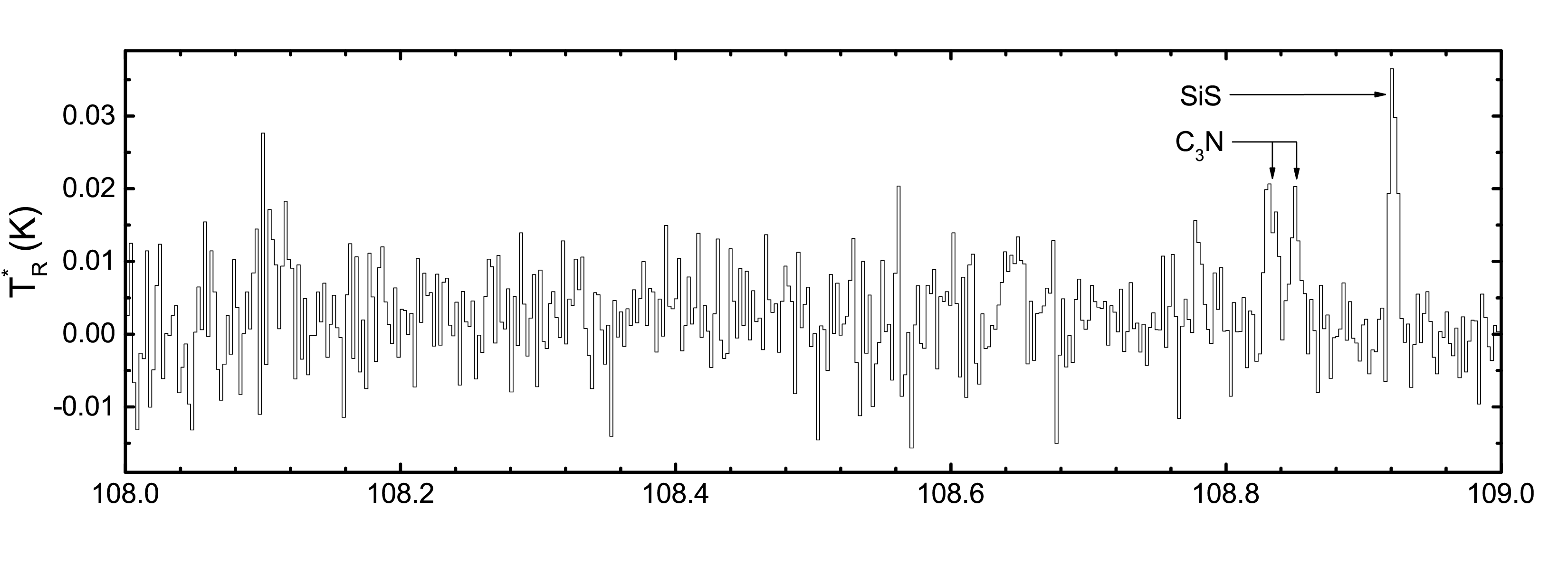}
    \includegraphics[width=0.9\textwidth]{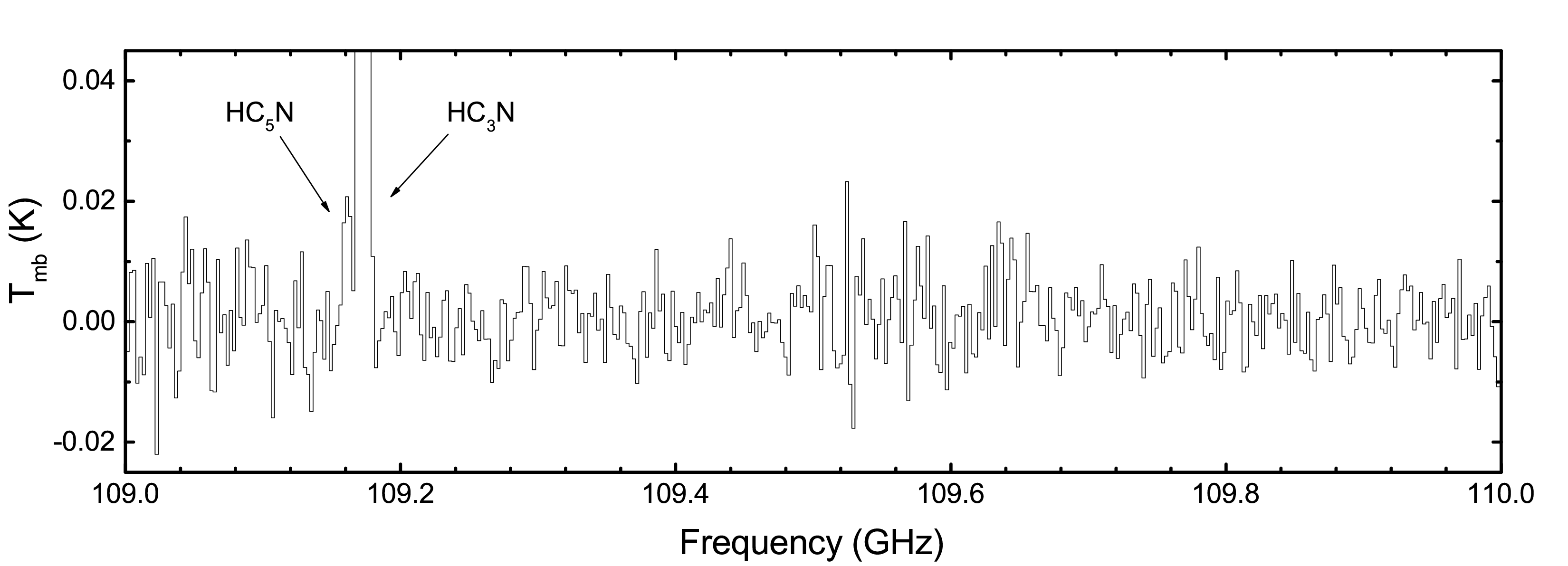}
    \caption{(Continued)}
    
\end{figure*}

\clearpage

\addtocounter{figure}{-1}
\begin{figure*}
    \addtocounter{figure}{0}
    \centering
    \includegraphics[width=0.9\textwidth]{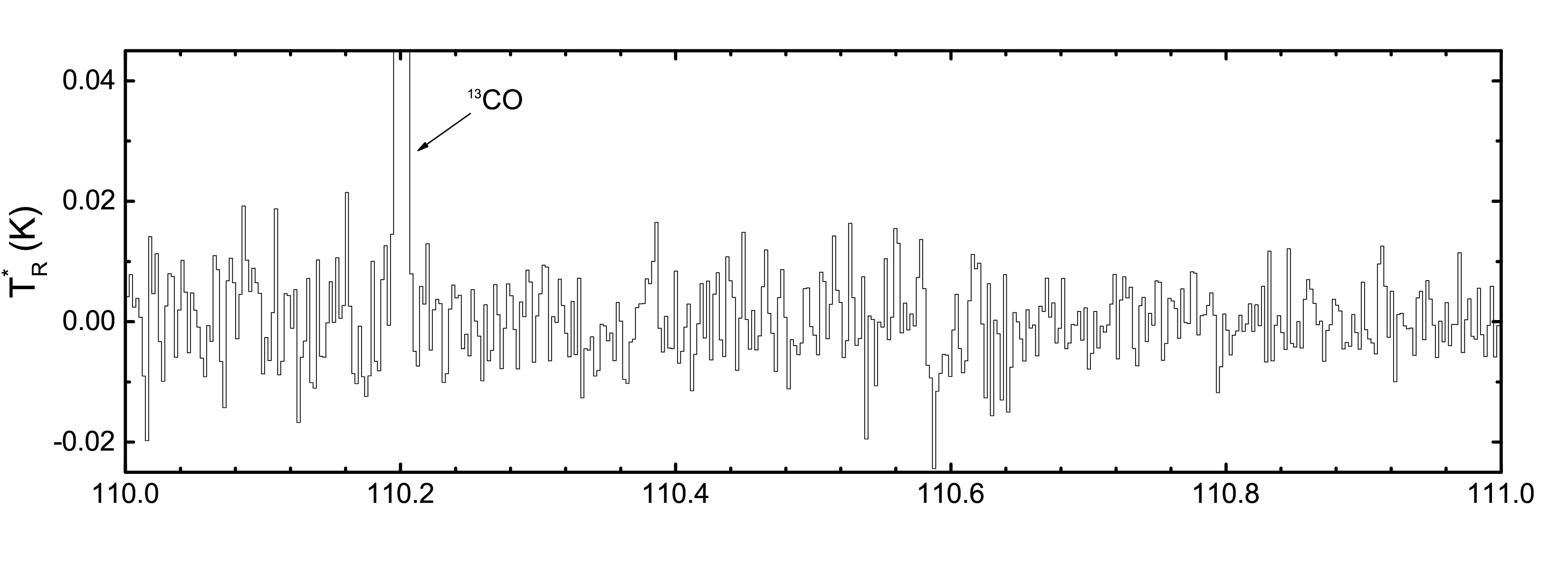}
    \includegraphics[width=0.9\textwidth]{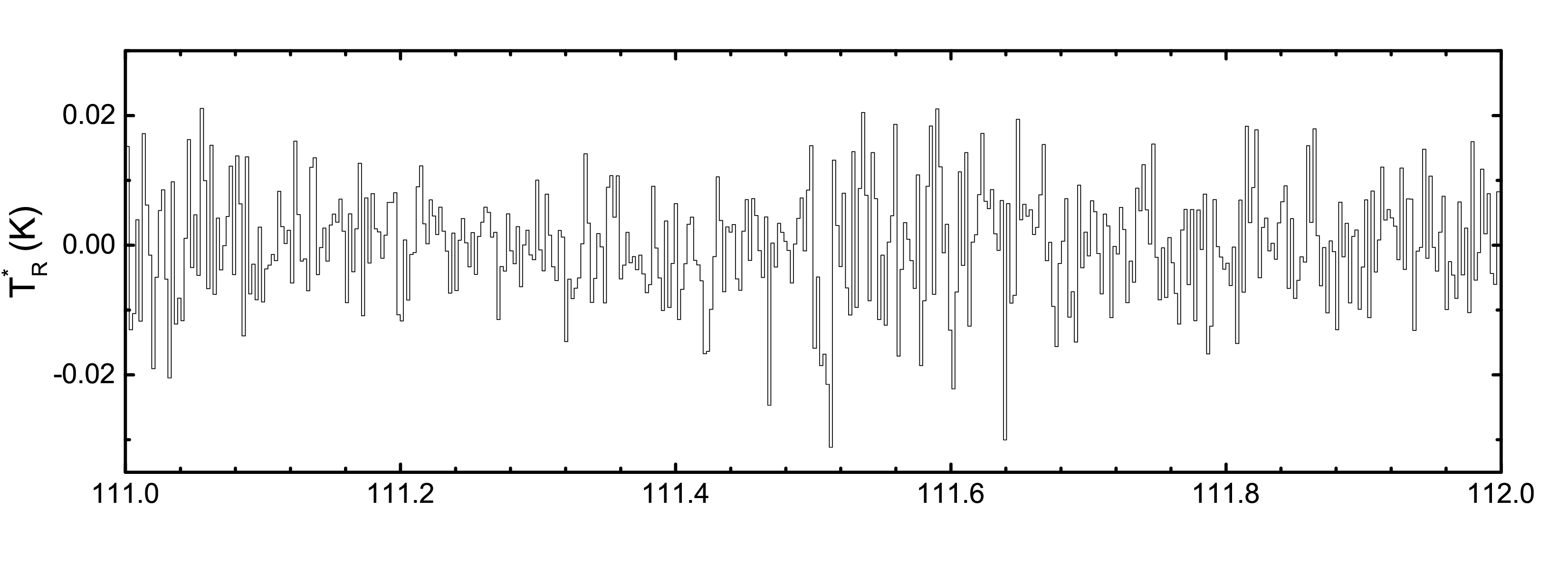}
    \includegraphics[width=0.9\textwidth]{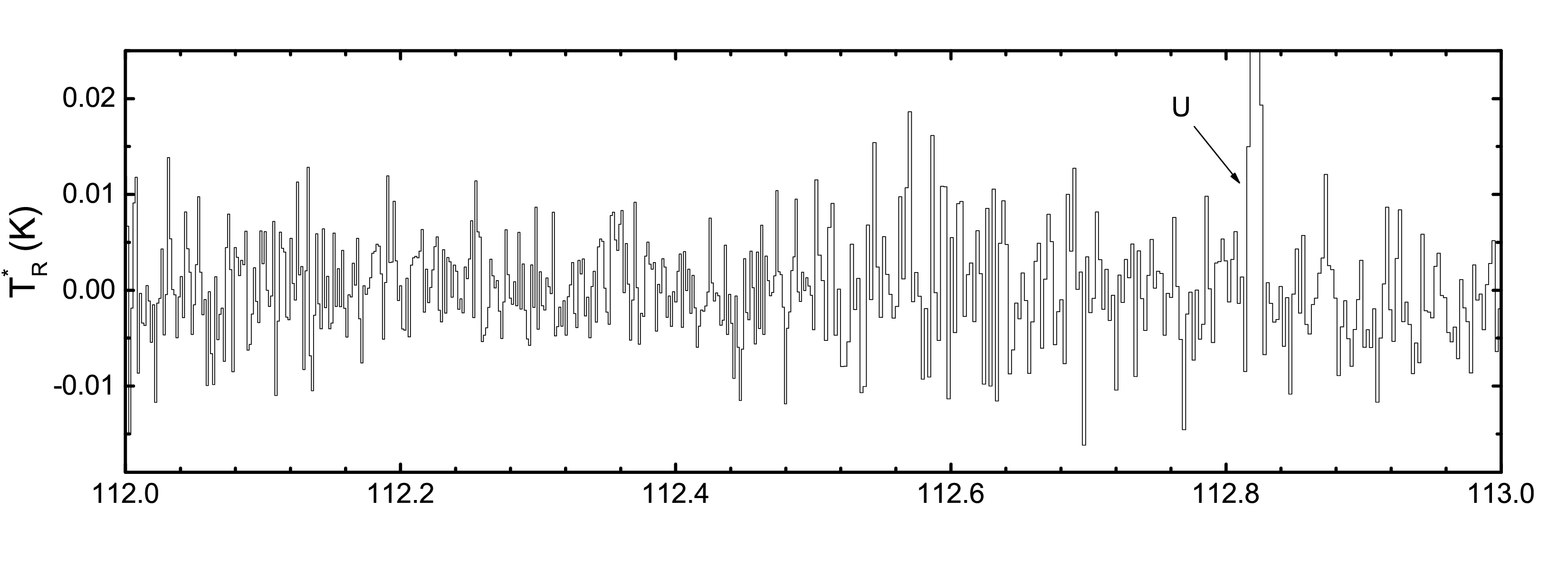}
    \includegraphics[width=0.9\textwidth]{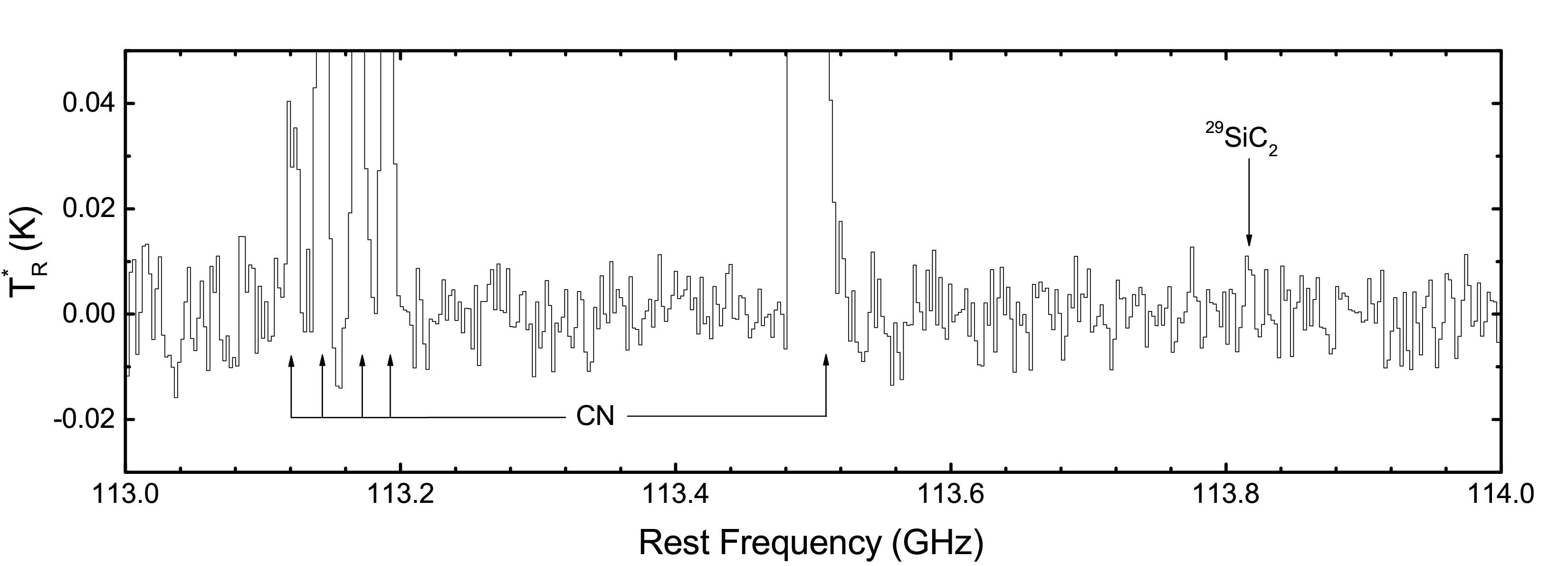}
    \caption{(Continued)}
    
\end{figure*}
    
\clearpage

\addtocounter{figure}{-1}
\begin{figure*}
    \addtocounter{figure}{0}
    \centering
    \includegraphics[width=0.9\textwidth]{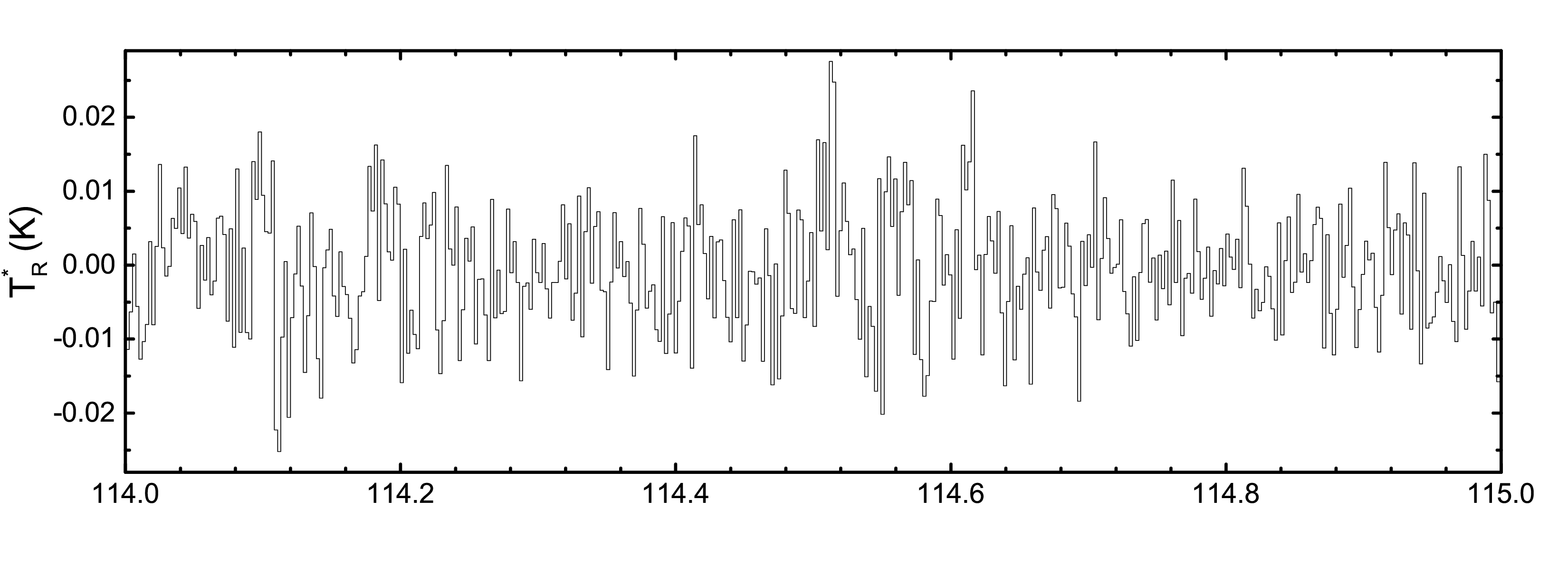}
    \includegraphics[width=0.9\textwidth]{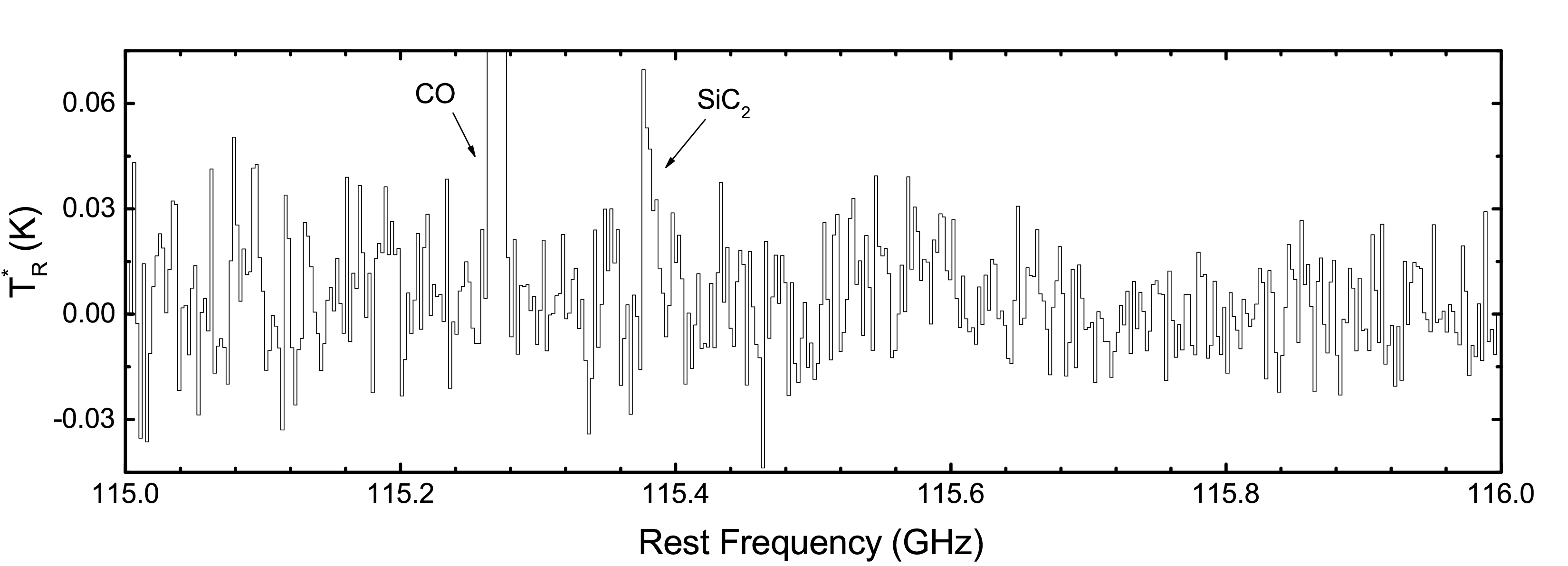}
    \caption{(Continued)}
    
\end{figure*}

\clearpage

\begin{figure*}
    \centering
    \includegraphics[width=0.45\textwidth]{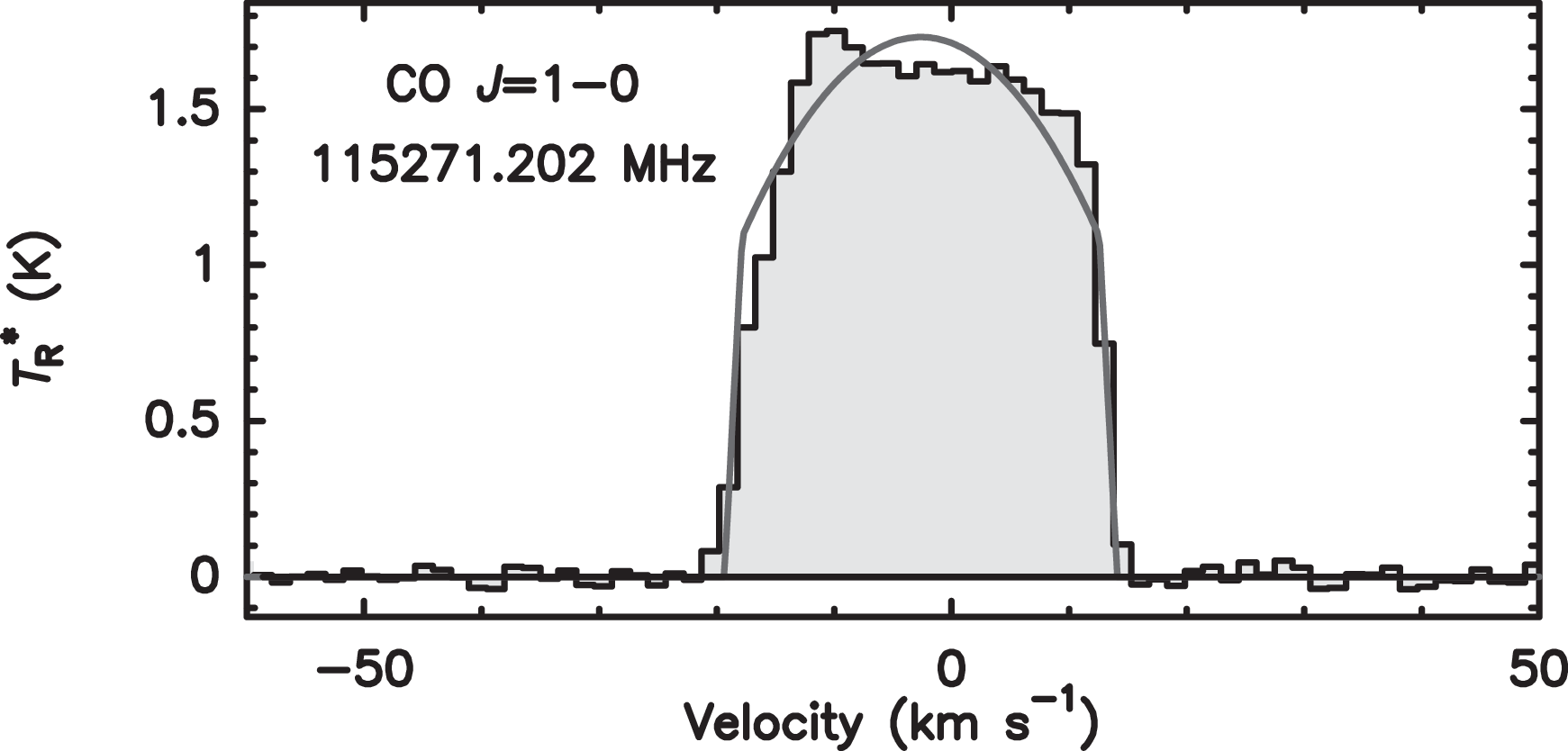}
    \includegraphics[width=0.45\textwidth]{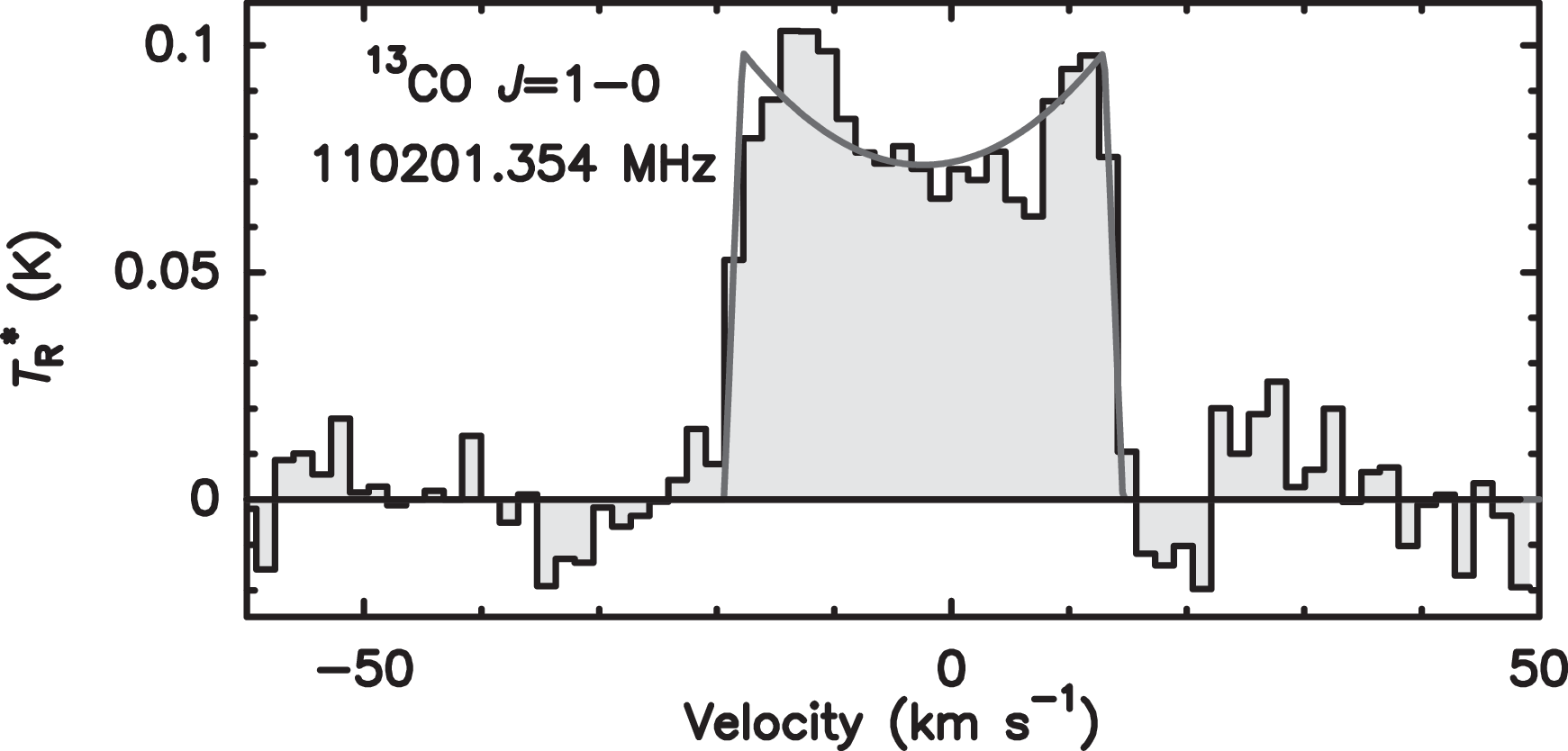}
    \caption{The CO and $^{13}$CO lines. The light grey curves represent the stellar-shell fitting.}
    \label{figure:3}
\end{figure*}

\clearpage

\begin{figure*}
    \centering
    \includegraphics[width=0.45\textwidth]{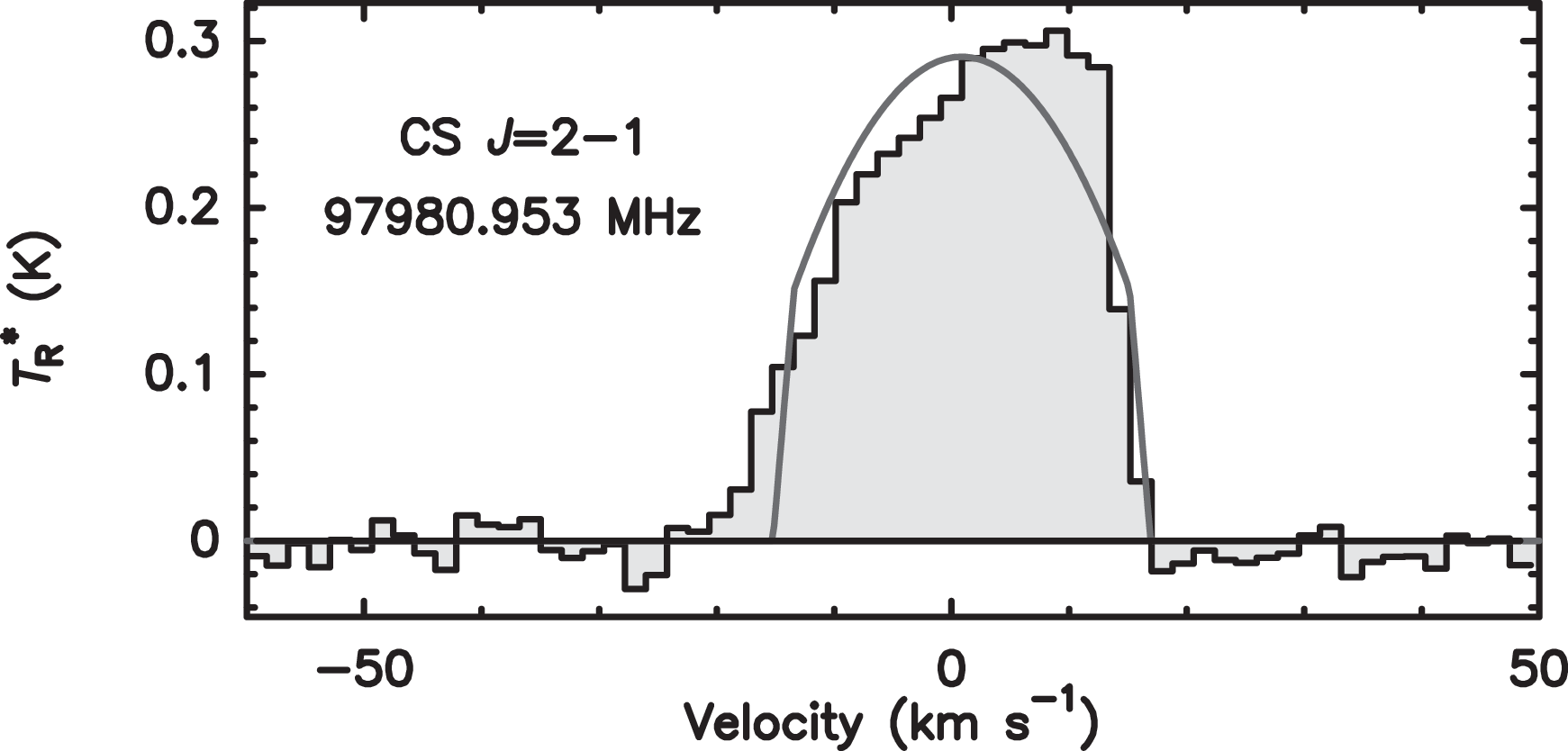}
    \includegraphics[width=0.45\textwidth]{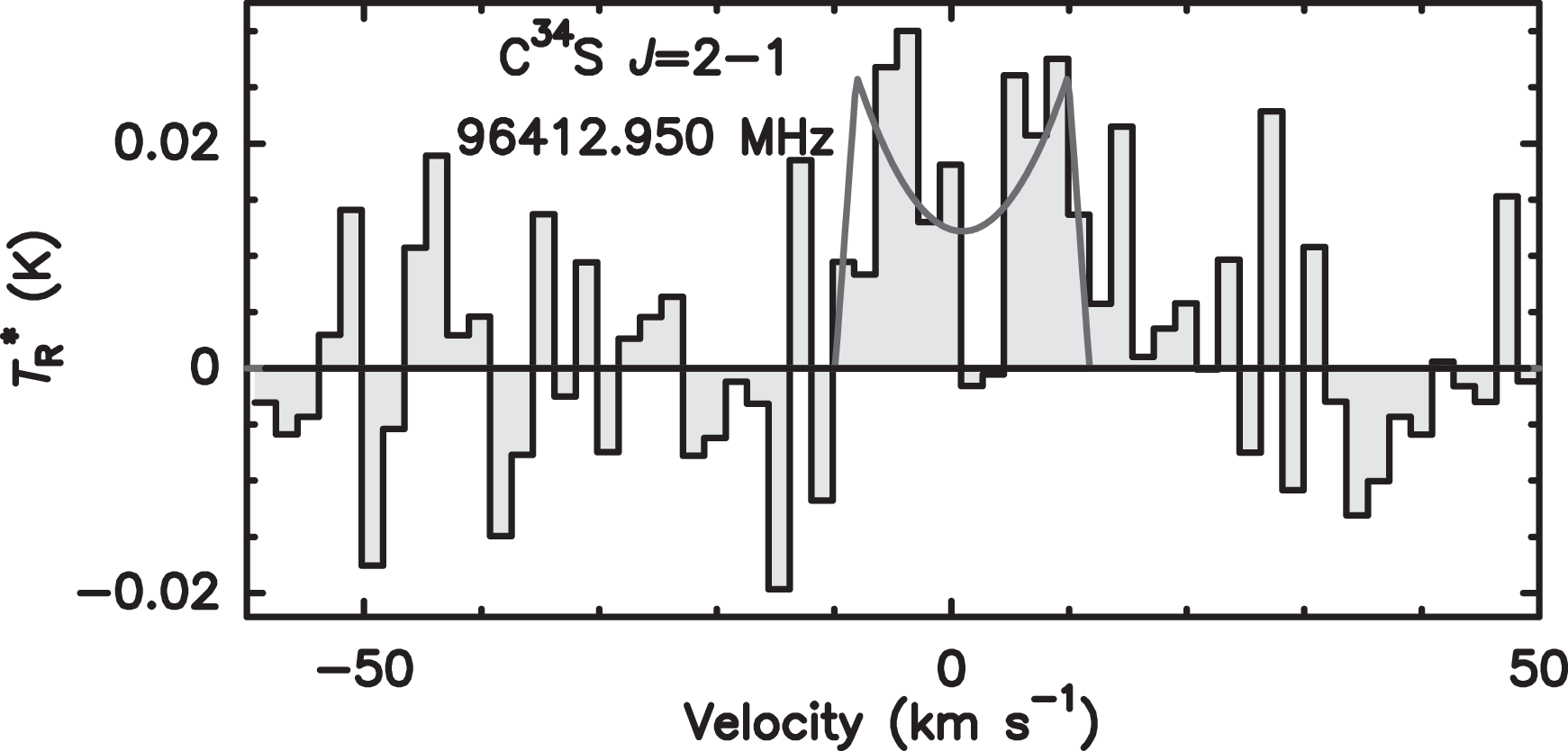}
    \caption{Same as figure~\ref{figure:3}, but for CS and C$^{34}$S.}
    \label{figure:4}
\end{figure*}

\clearpage

\begin{figure*}
    \centering
    \includegraphics[width=0.45\textwidth]{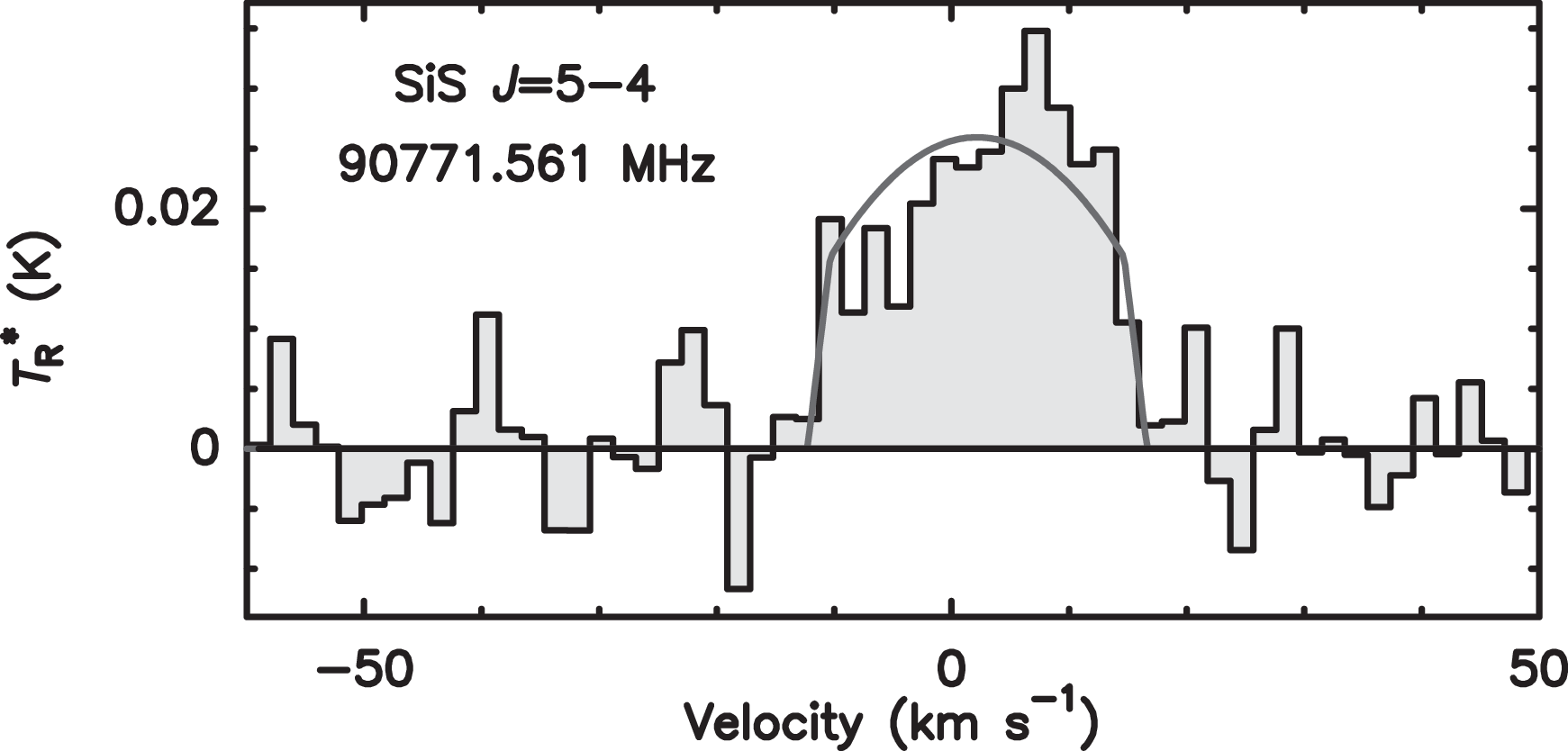}
    \includegraphics[width=0.45\textwidth]{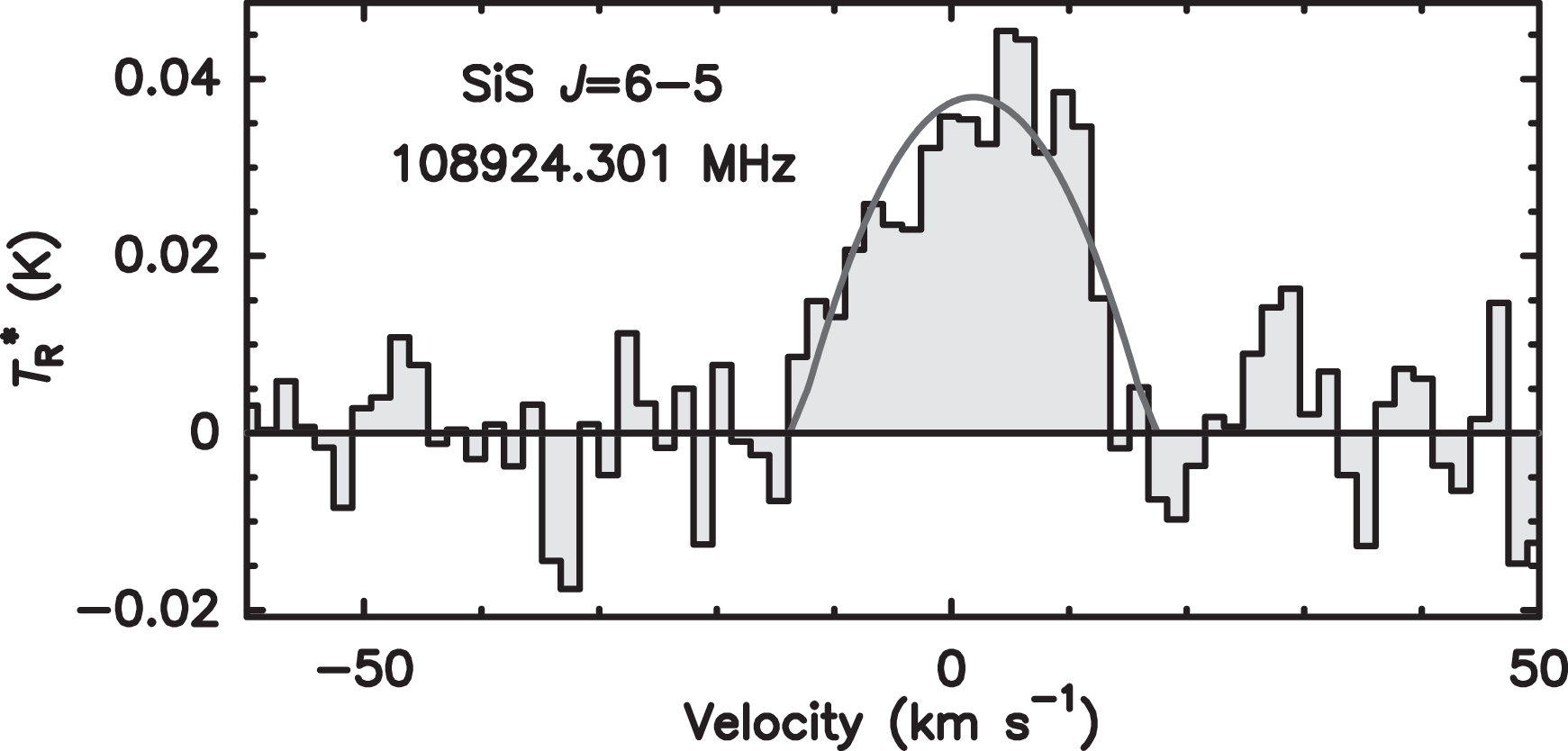}
    \caption{Same as figure~\ref{figure:3}, but for SiS.}
    \label{figure:5}
\end{figure*}

\clearpage

\begin{figure*}
    \centering
    \includegraphics[width=0.45\textwidth]{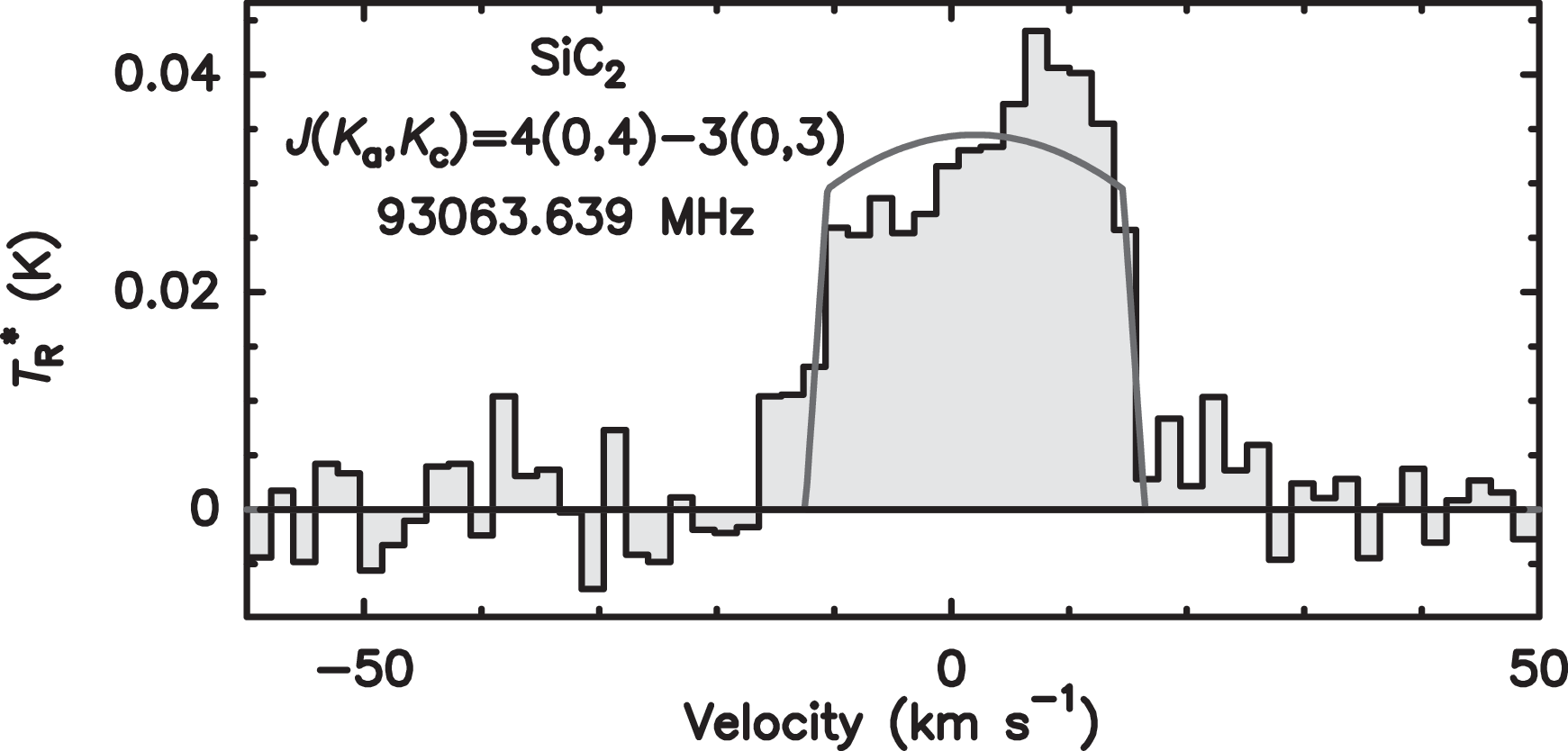}
    \includegraphics[width=0.45\textwidth]{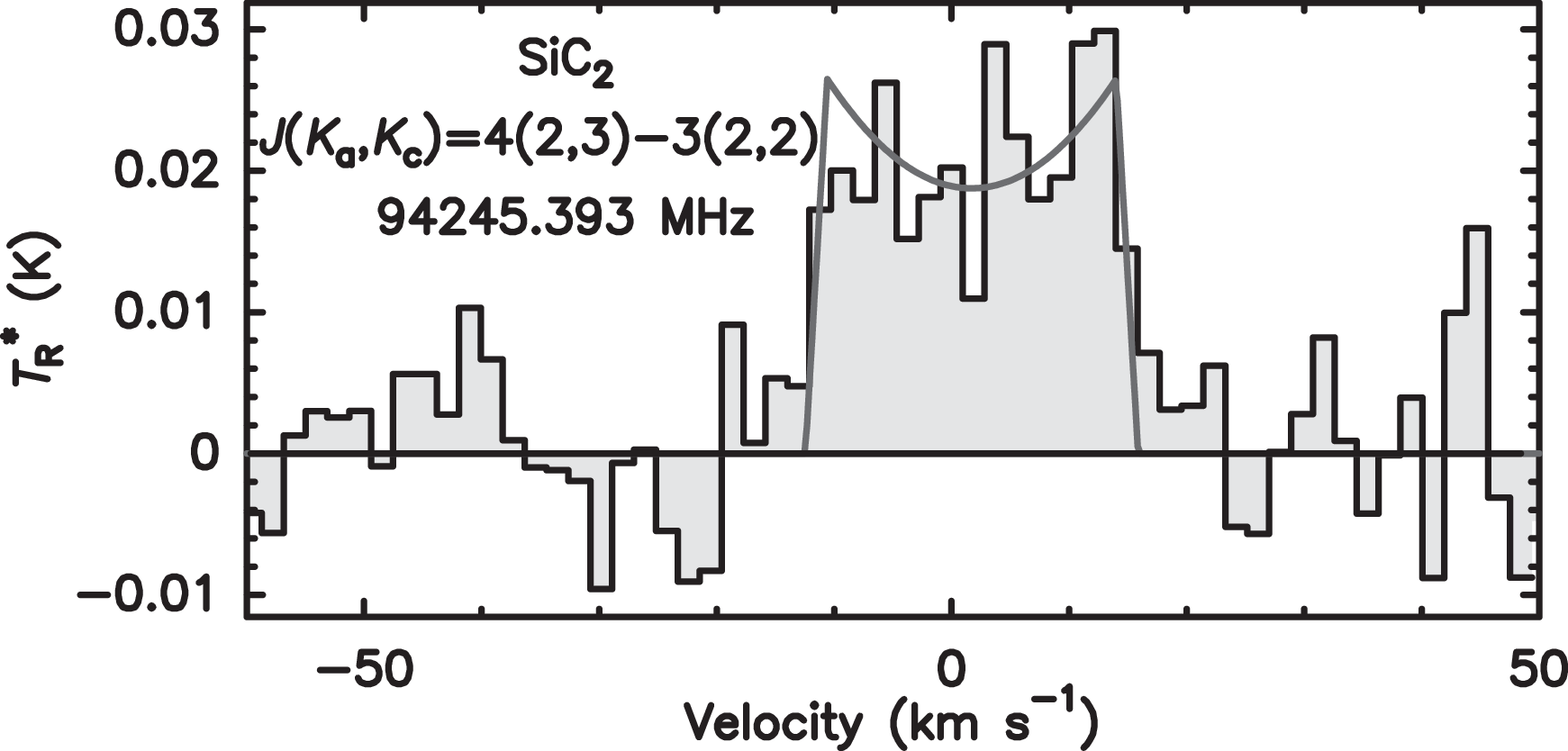}
    \includegraphics[width=0.45\textwidth]{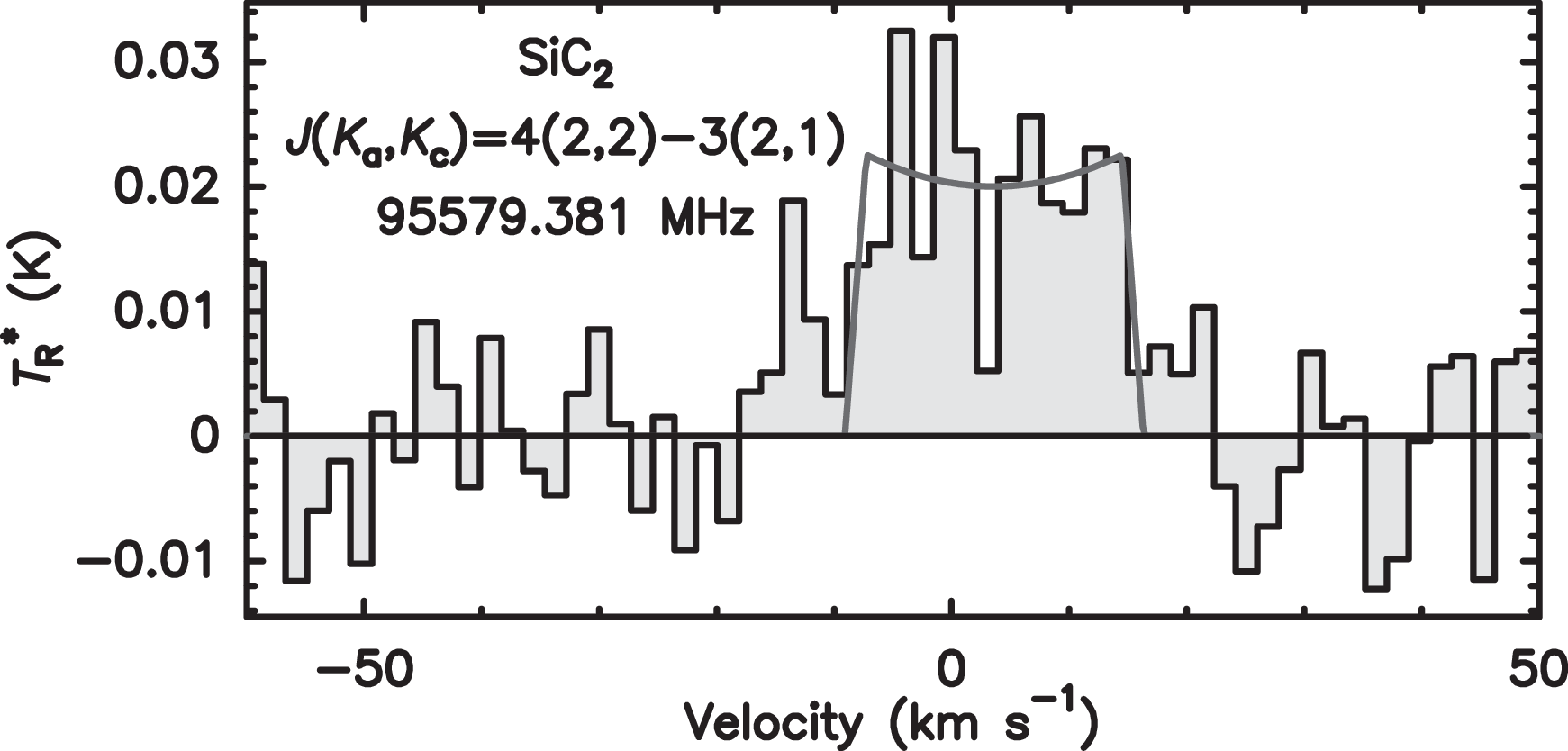}
    \includegraphics[width=0.45\textwidth]{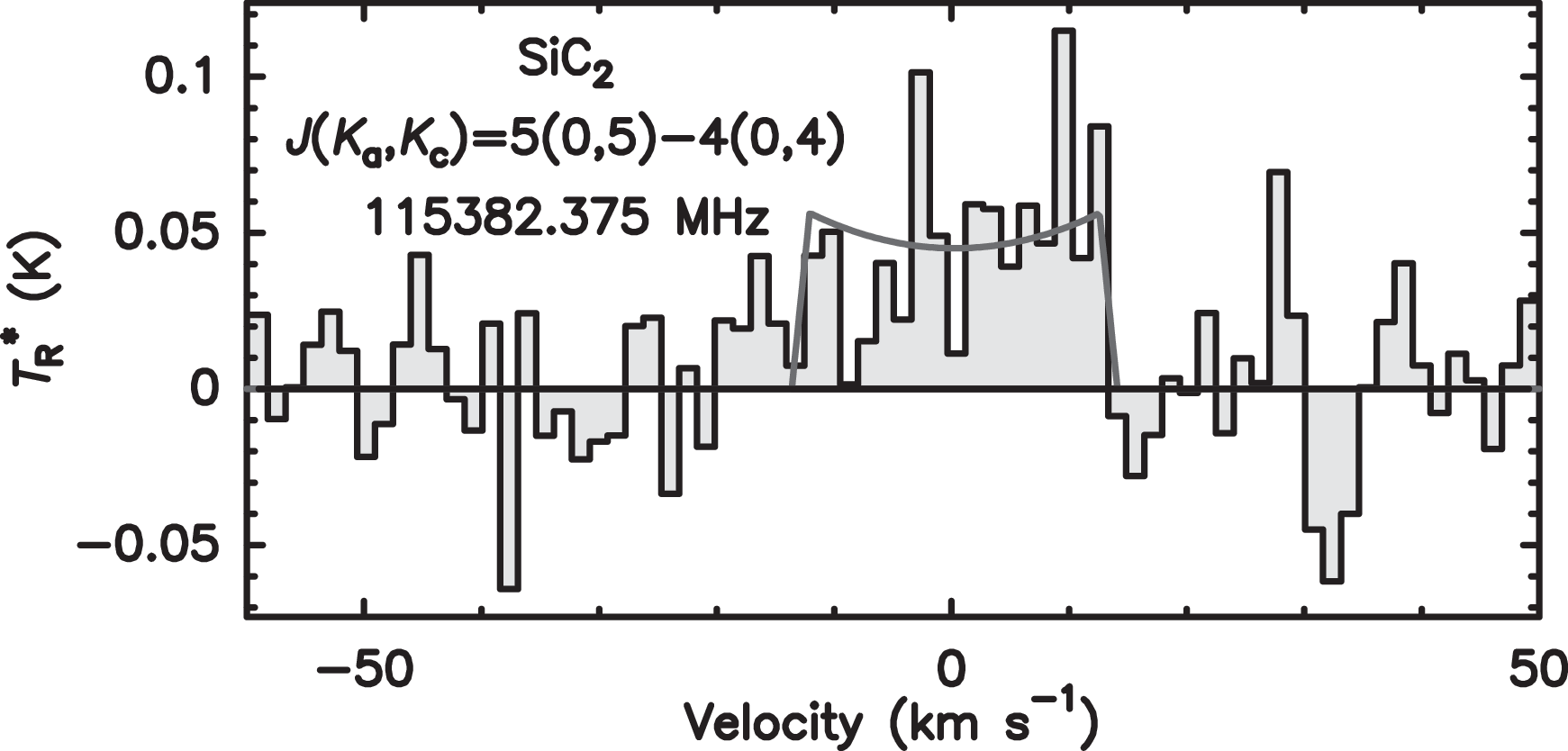}
    \includegraphics[width=0.45\textwidth]{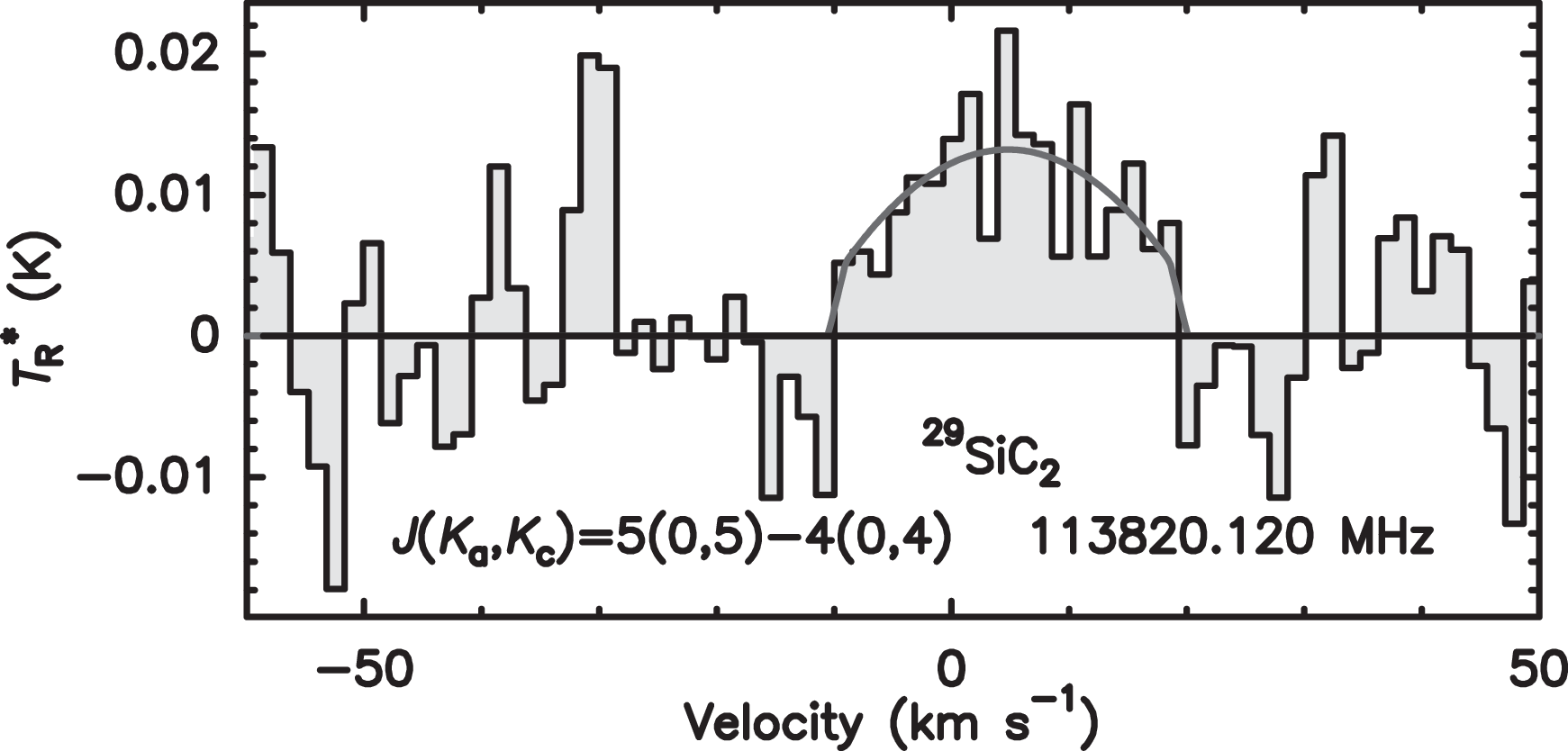}
    \caption{Same as figure~\ref{figure:3}, but for SiC$_{2}$ and $^{29}$SiC$_{2}$.}
    \label{figure:6}
\end{figure*}

\clearpage

\begin{figure*}
    \centering
    \includegraphics[width=0.45\textwidth]{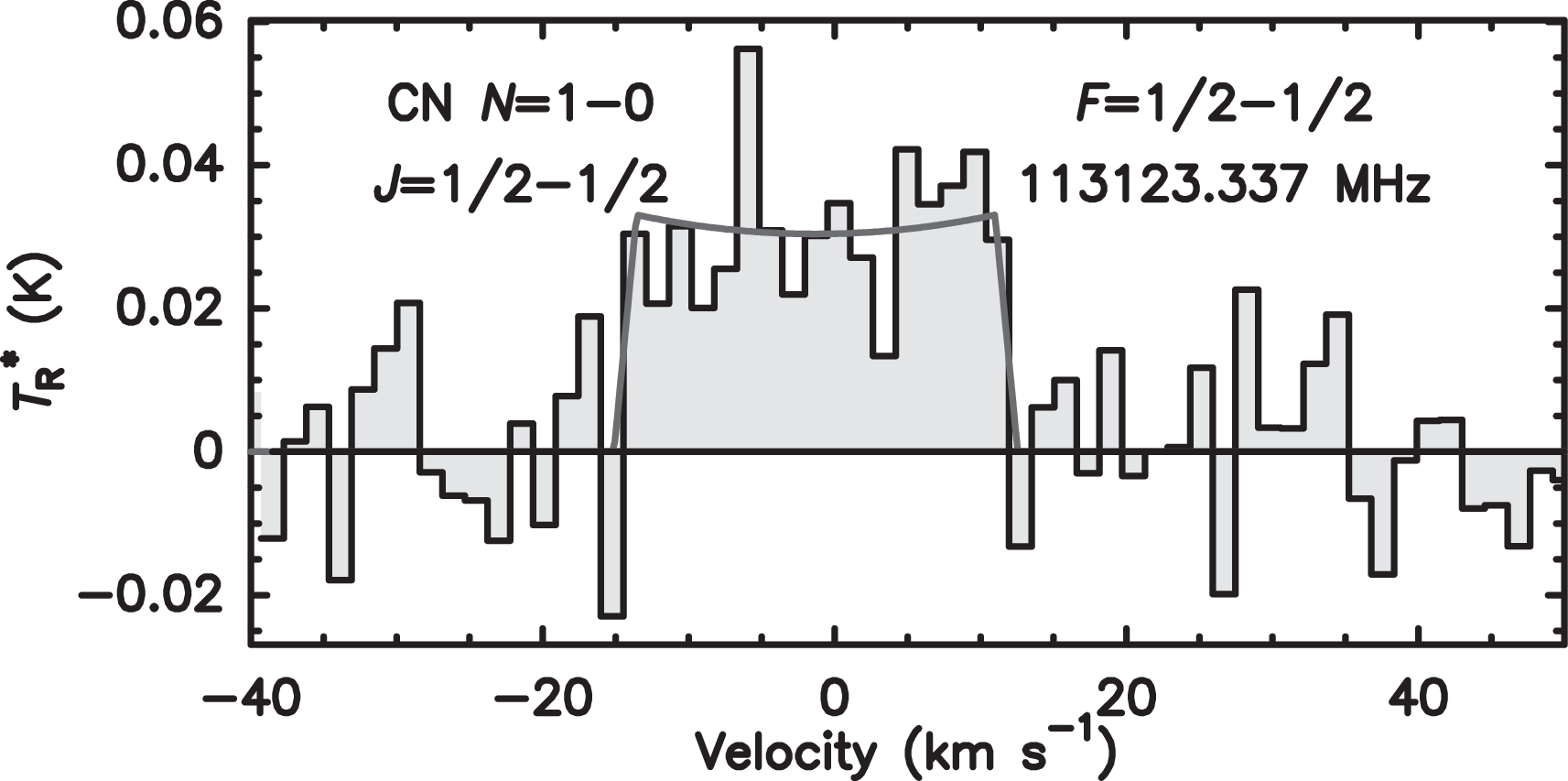}
    \includegraphics[width=0.45\textwidth]{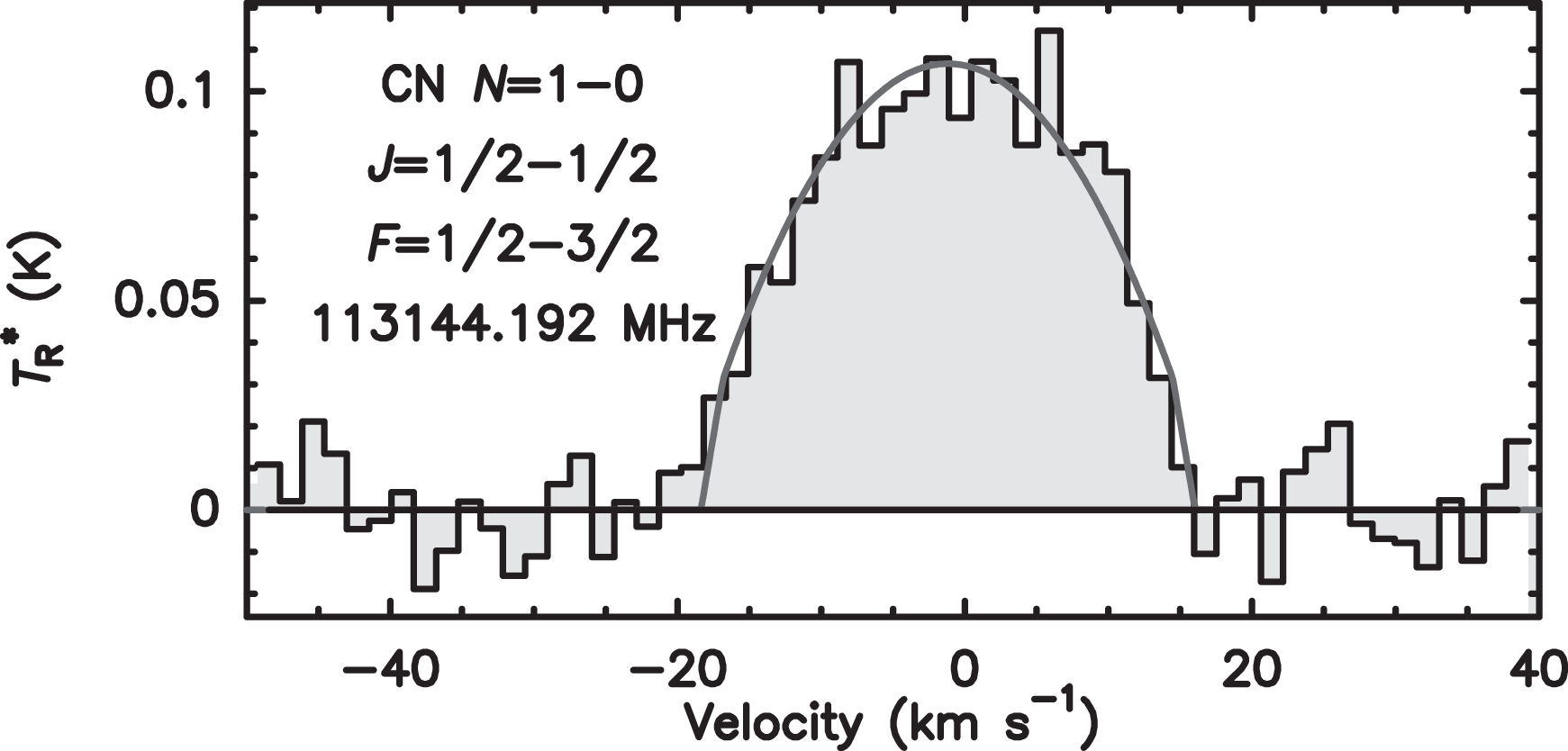}
    \includegraphics[width=0.45\textwidth]{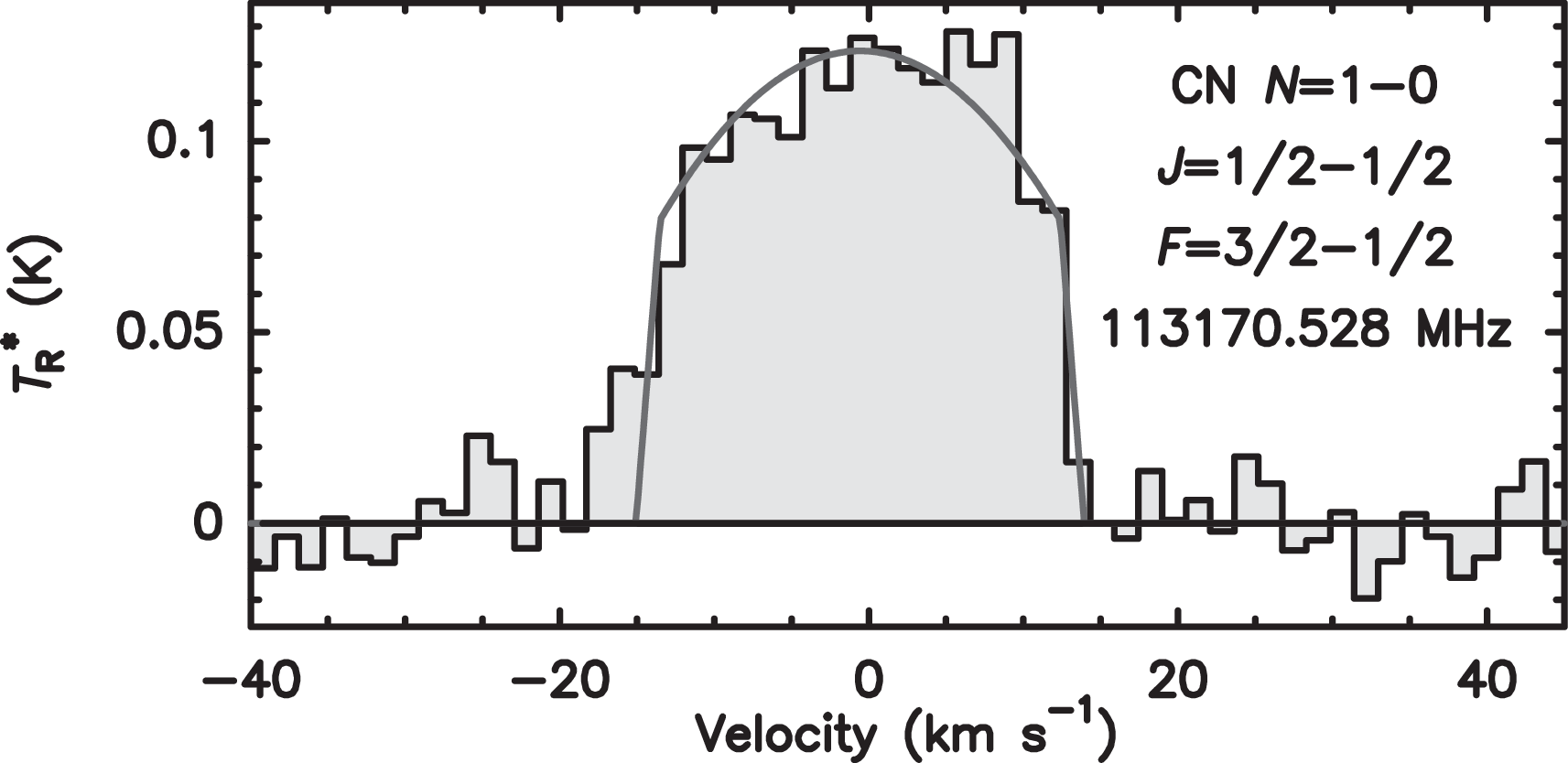}
    \includegraphics[width=0.45\textwidth]{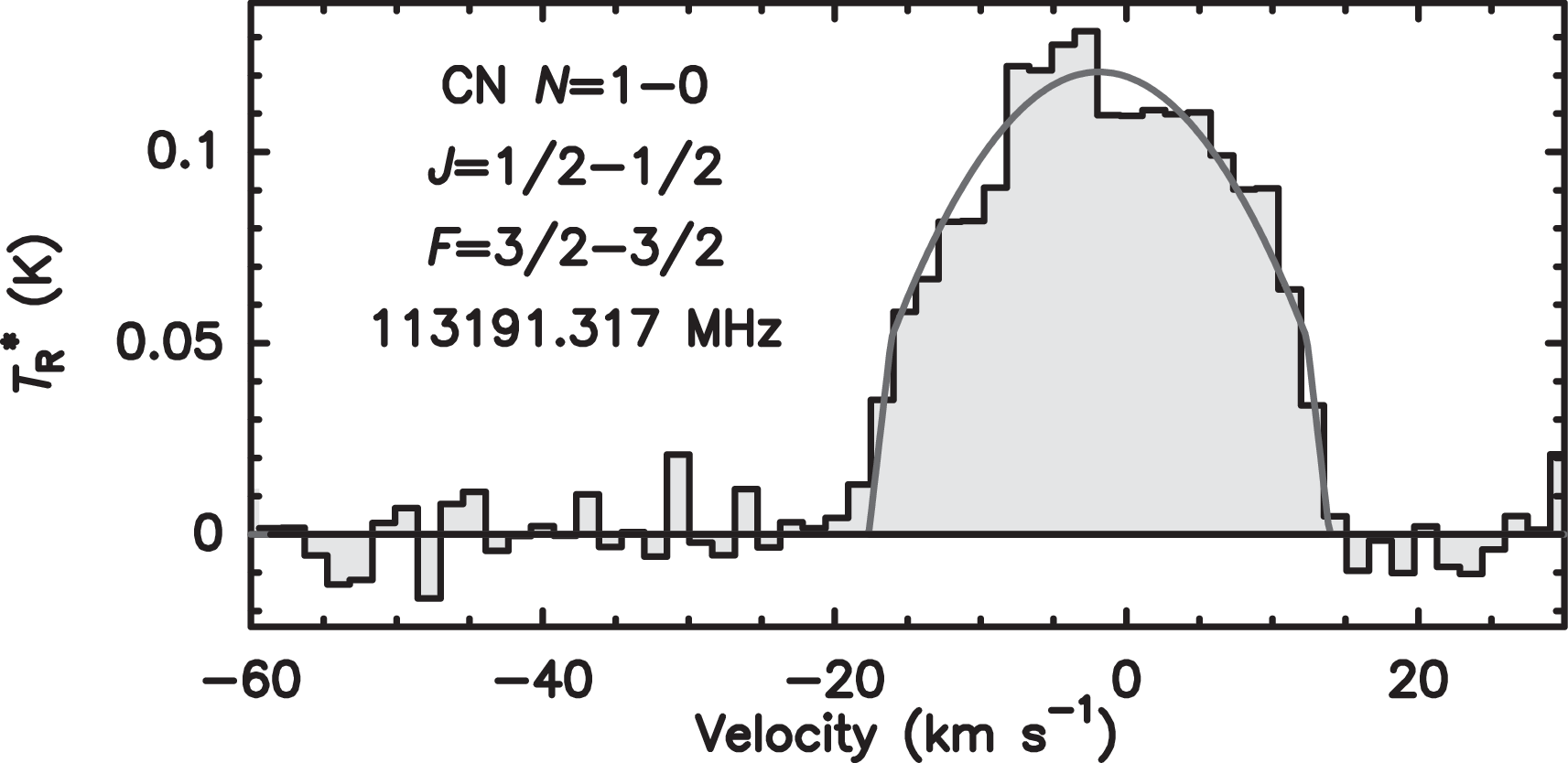}
    \includegraphics[width=0.45\textwidth]{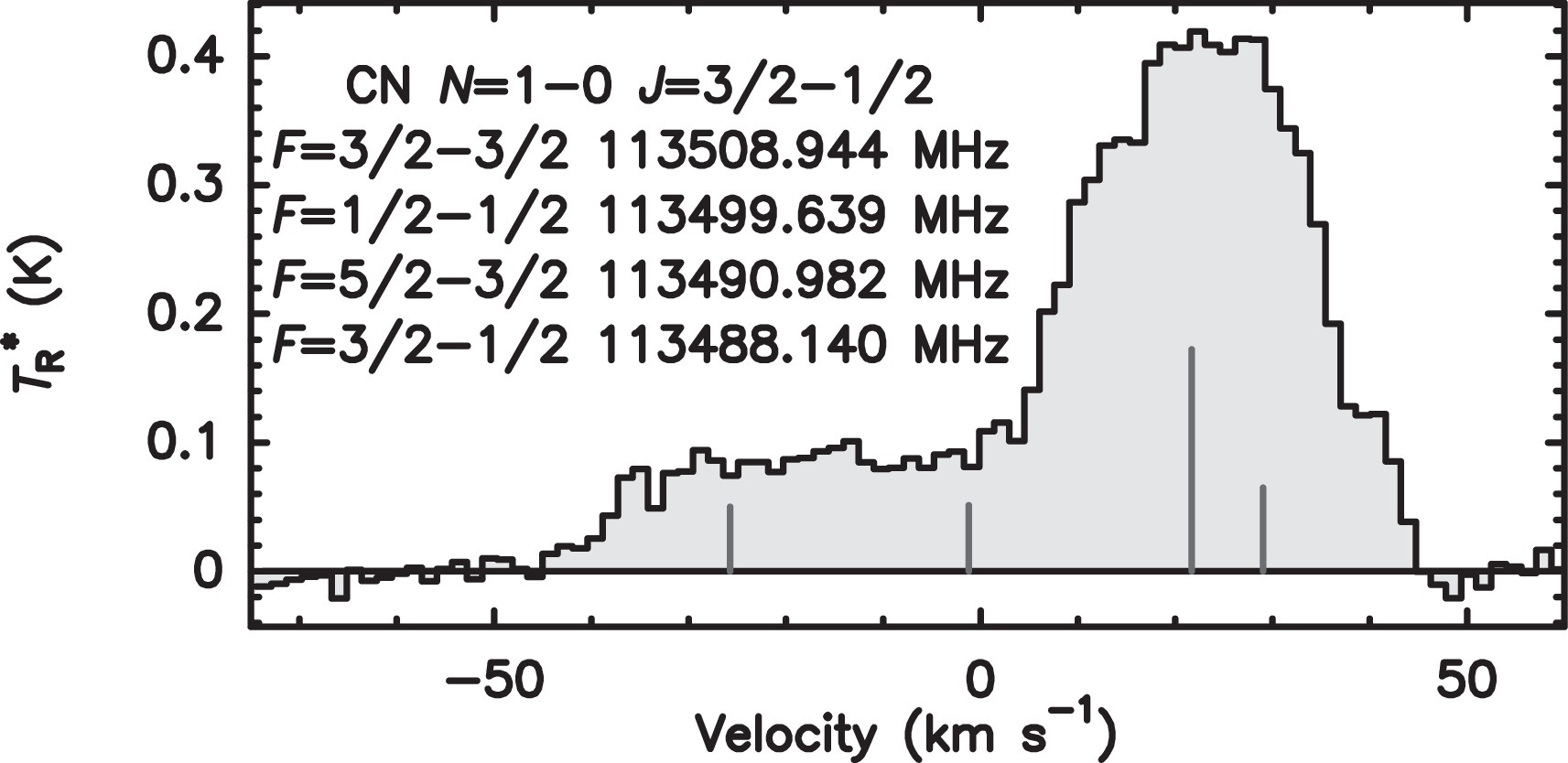}
    \caption{Same as figure~\ref{figure:3}, but for CN. The vertical lines show
    the relative strengths of blended fine-structure lines.}
    \label{figure:7}
\end{figure*}

\clearpage

\begin{figure*}
    \centering
    \includegraphics[width=0.45\textwidth]{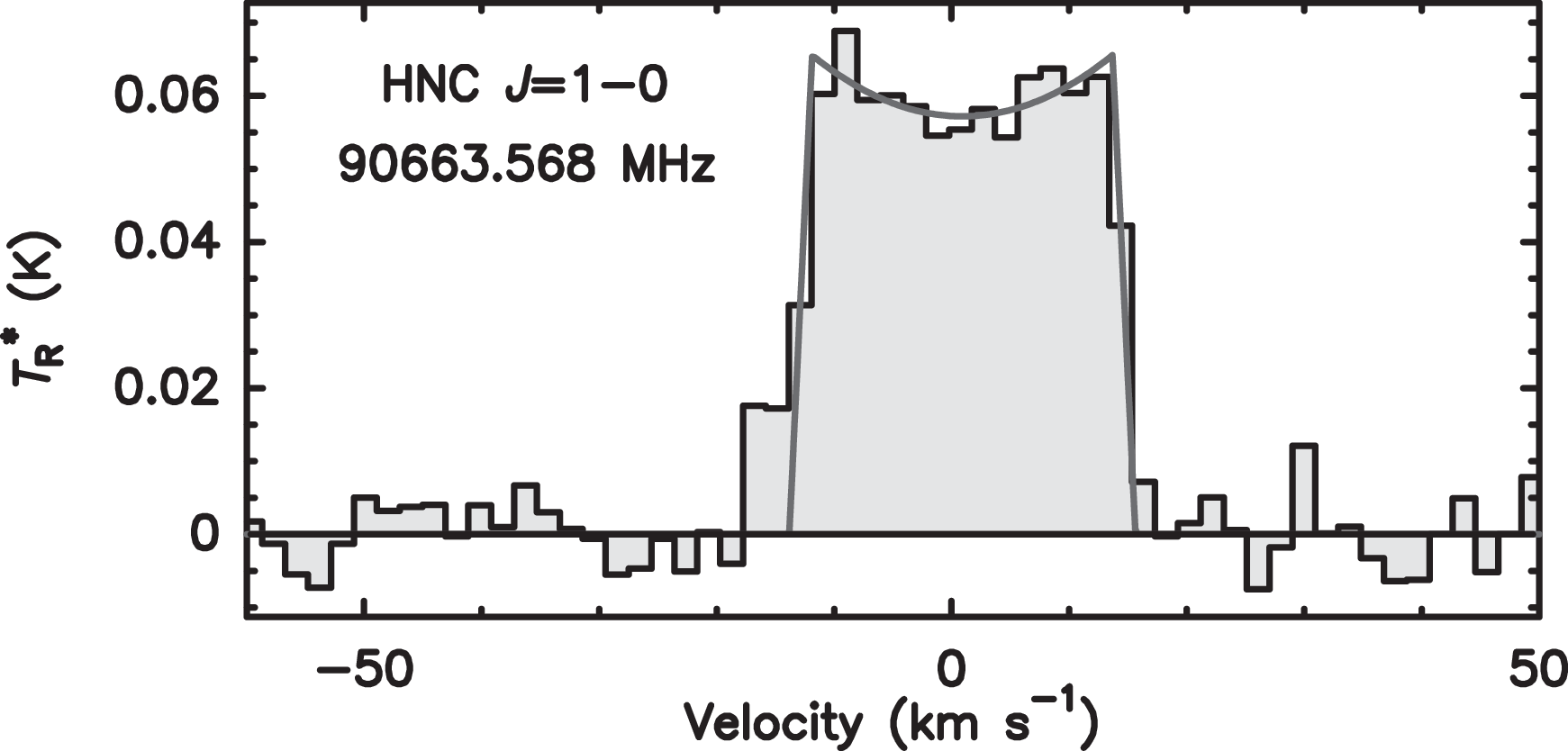}
    \caption{Same as figure~\ref{figure:3}, but for HNC.}
    \label{figure:8}
\end{figure*}

\clearpage

\begin{figure*}
    \centering
    \includegraphics[width=0.45\textwidth]{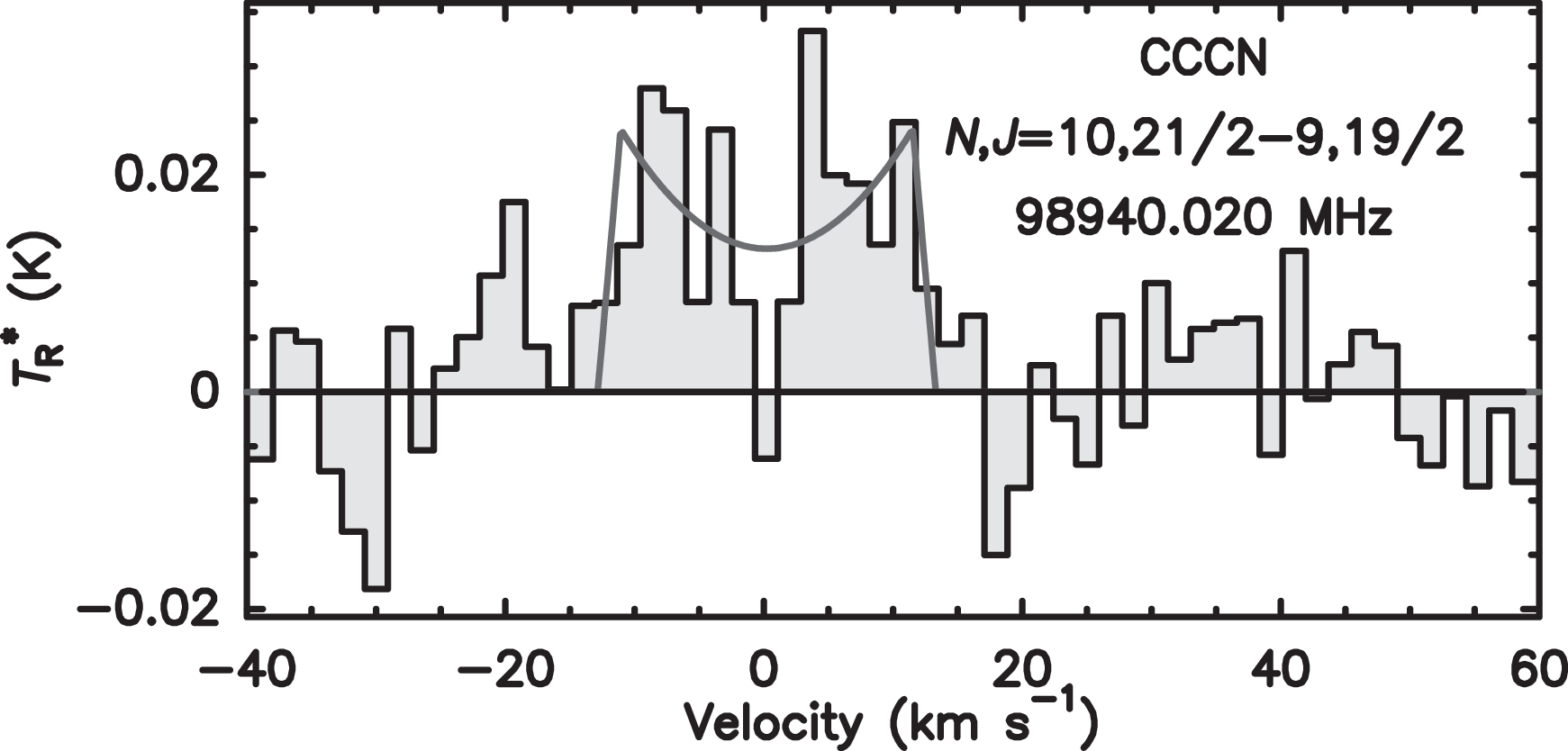}
    \includegraphics[width=0.45\textwidth]{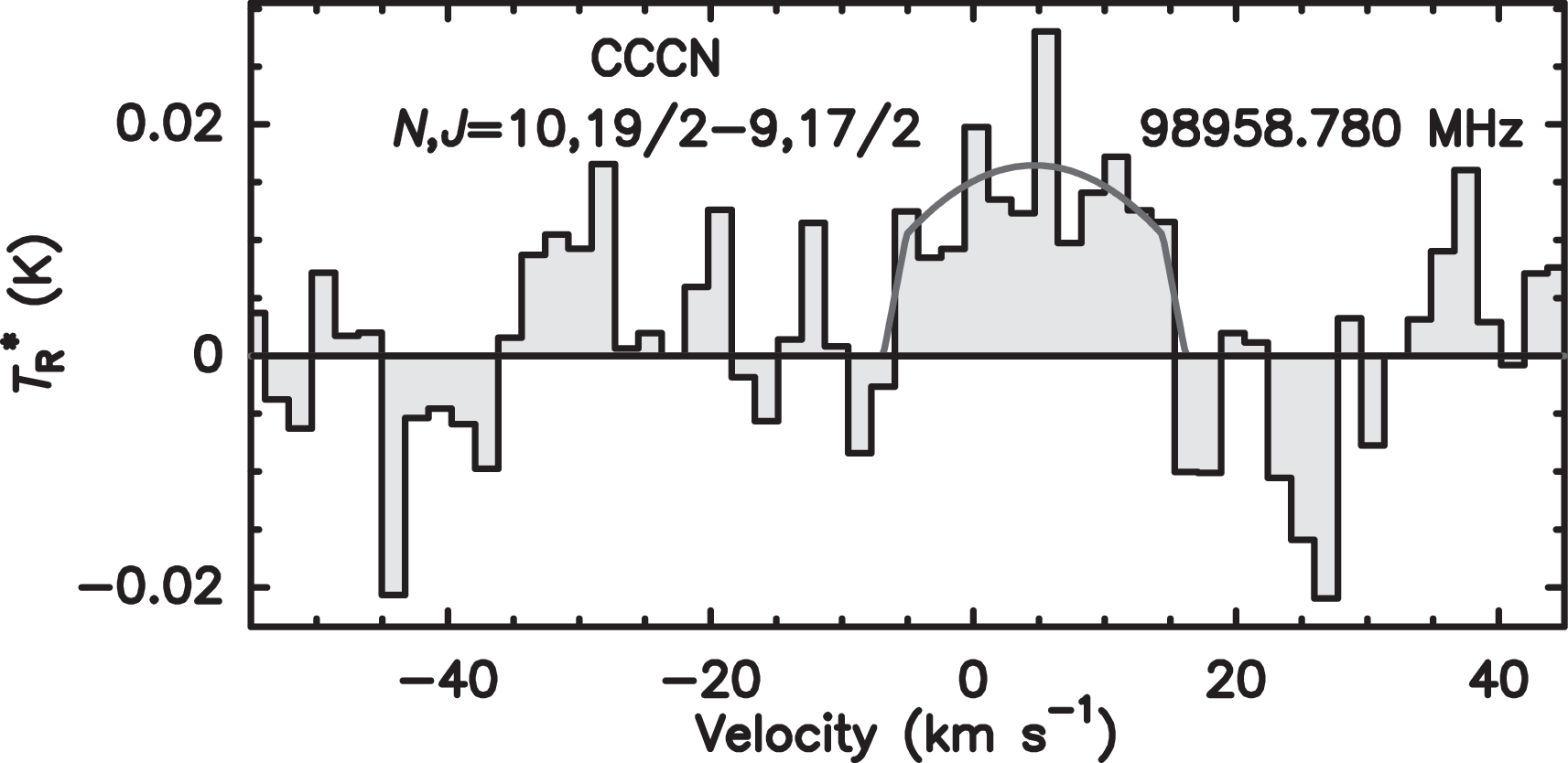}
    \includegraphics[width=0.45\textwidth]{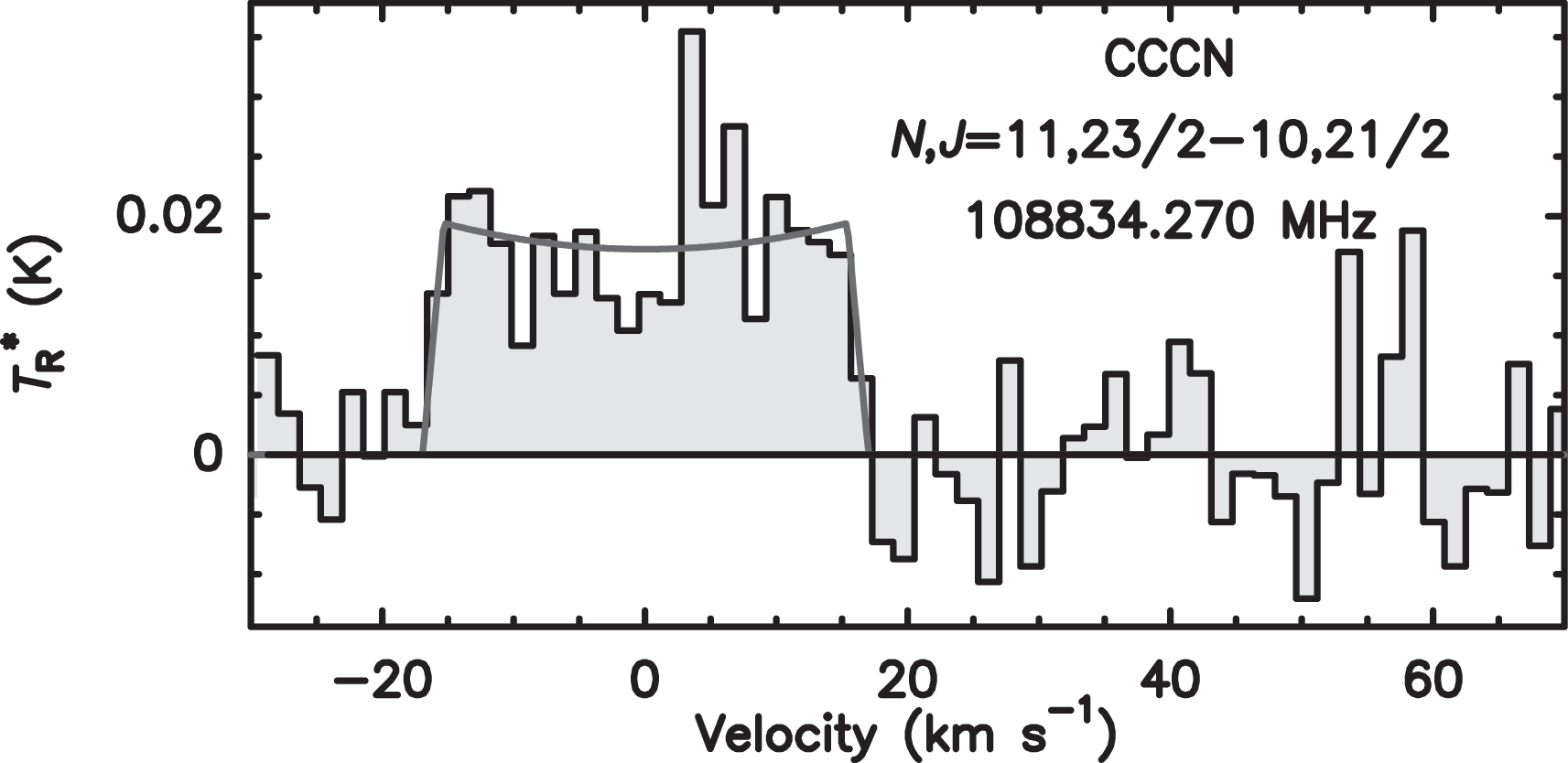}
    \includegraphics[width=0.45\textwidth]{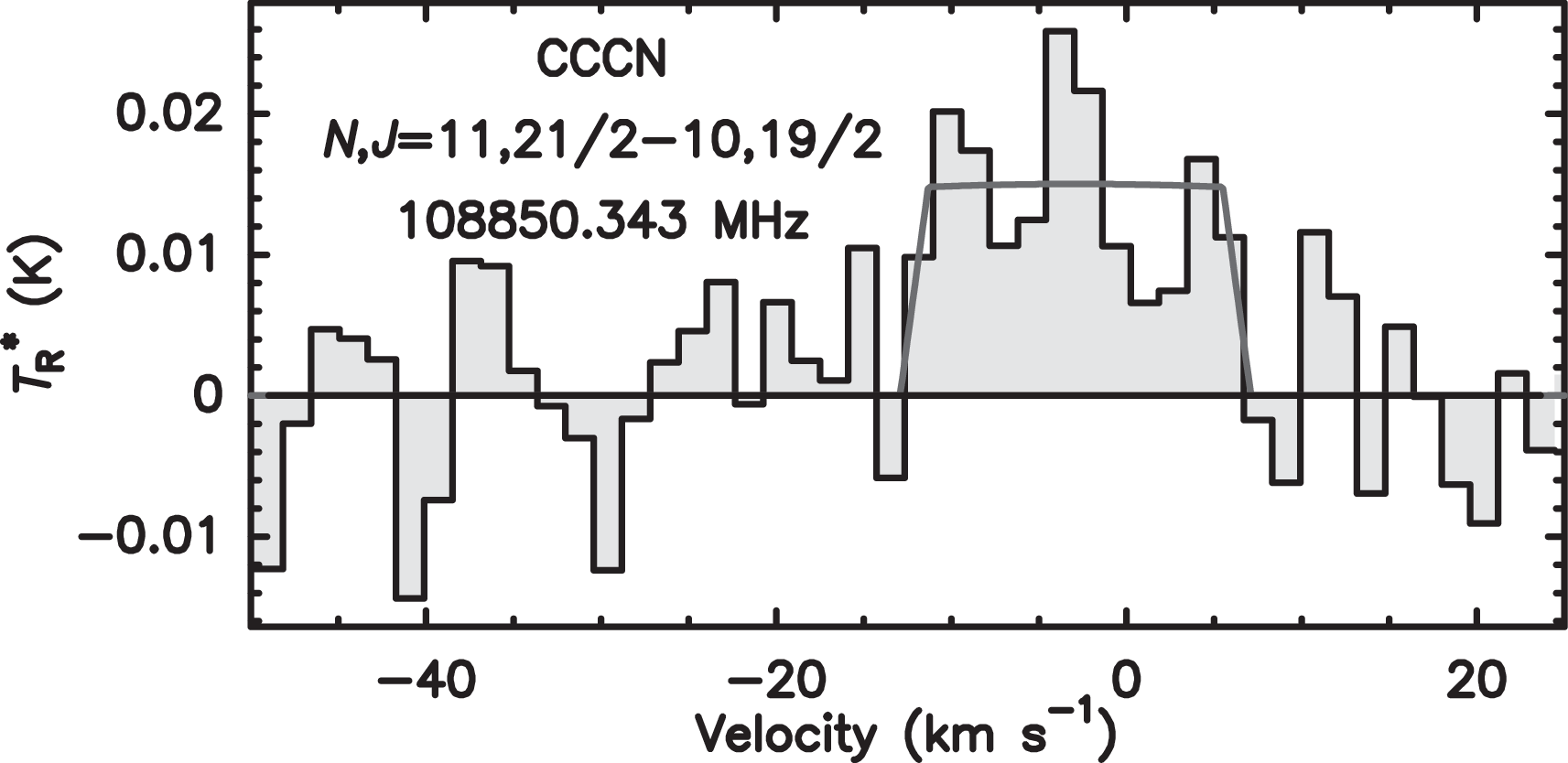}
    \caption{Same as figure~\ref{figure:3}, but for C$_{3}$N.}
    \label{figure:9}
\end{figure*}

\clearpage

\begin{figure*}
    \centering
    \includegraphics[width=0.45\textwidth]{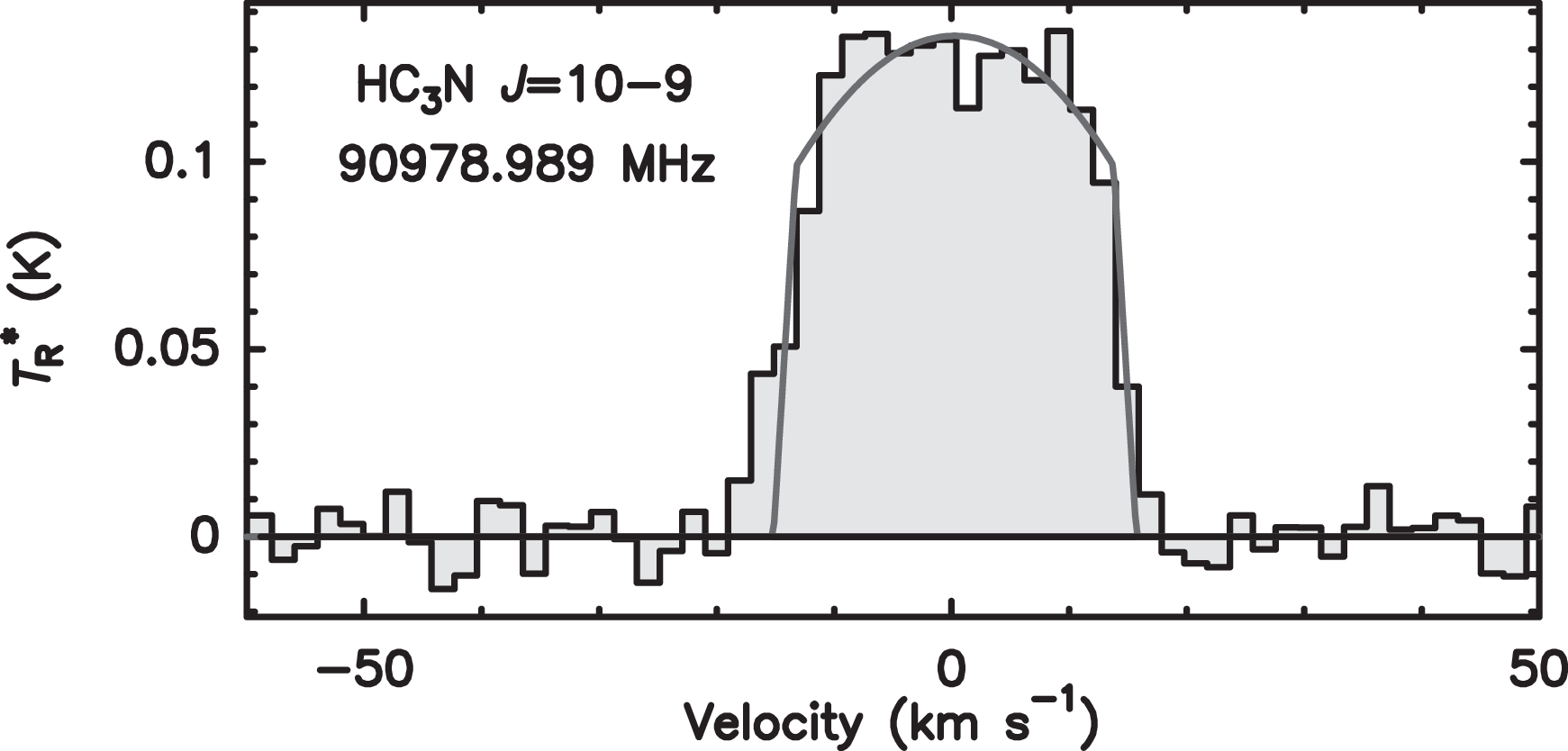}
    \includegraphics[width=0.45\textwidth]{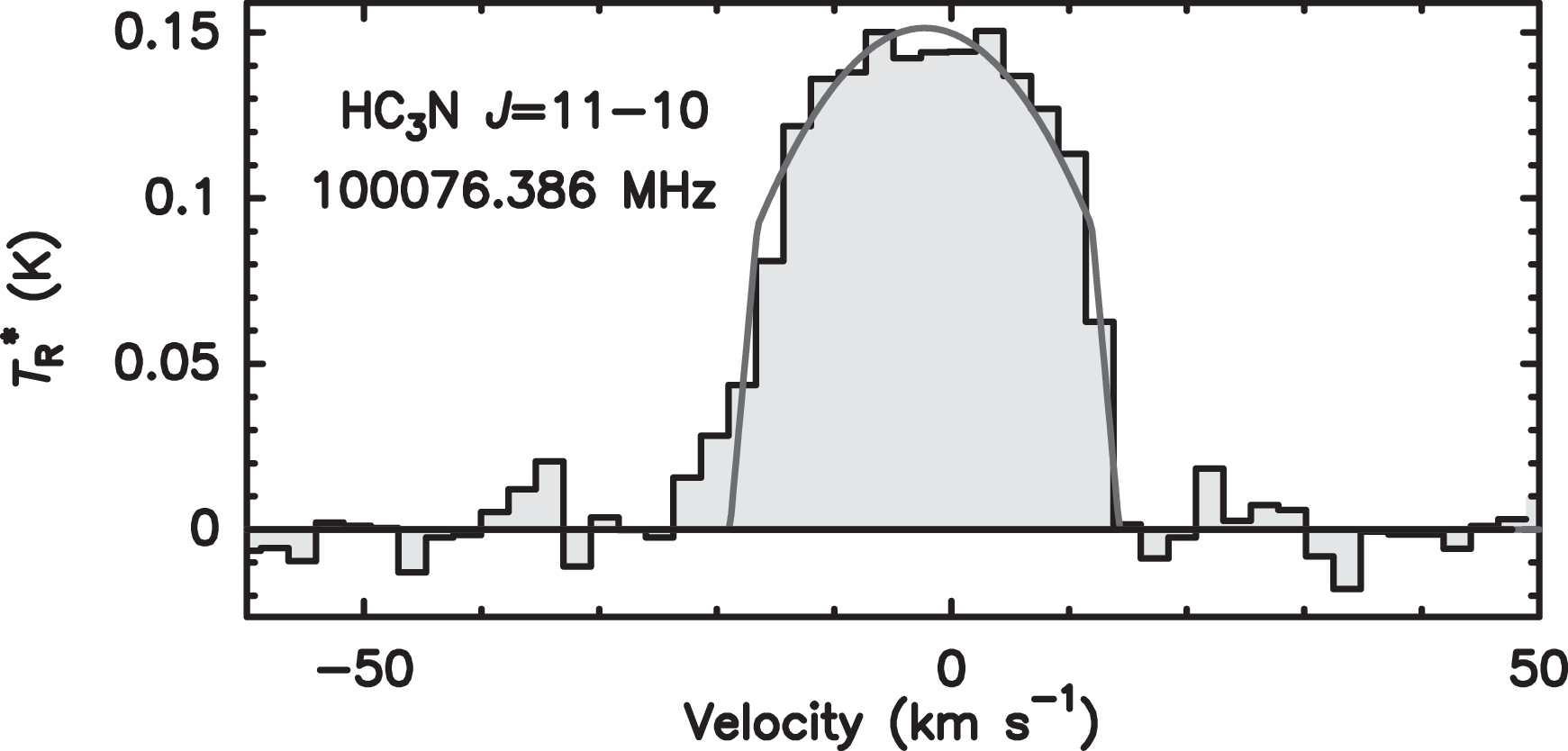}
    \includegraphics[width=0.45\textwidth]{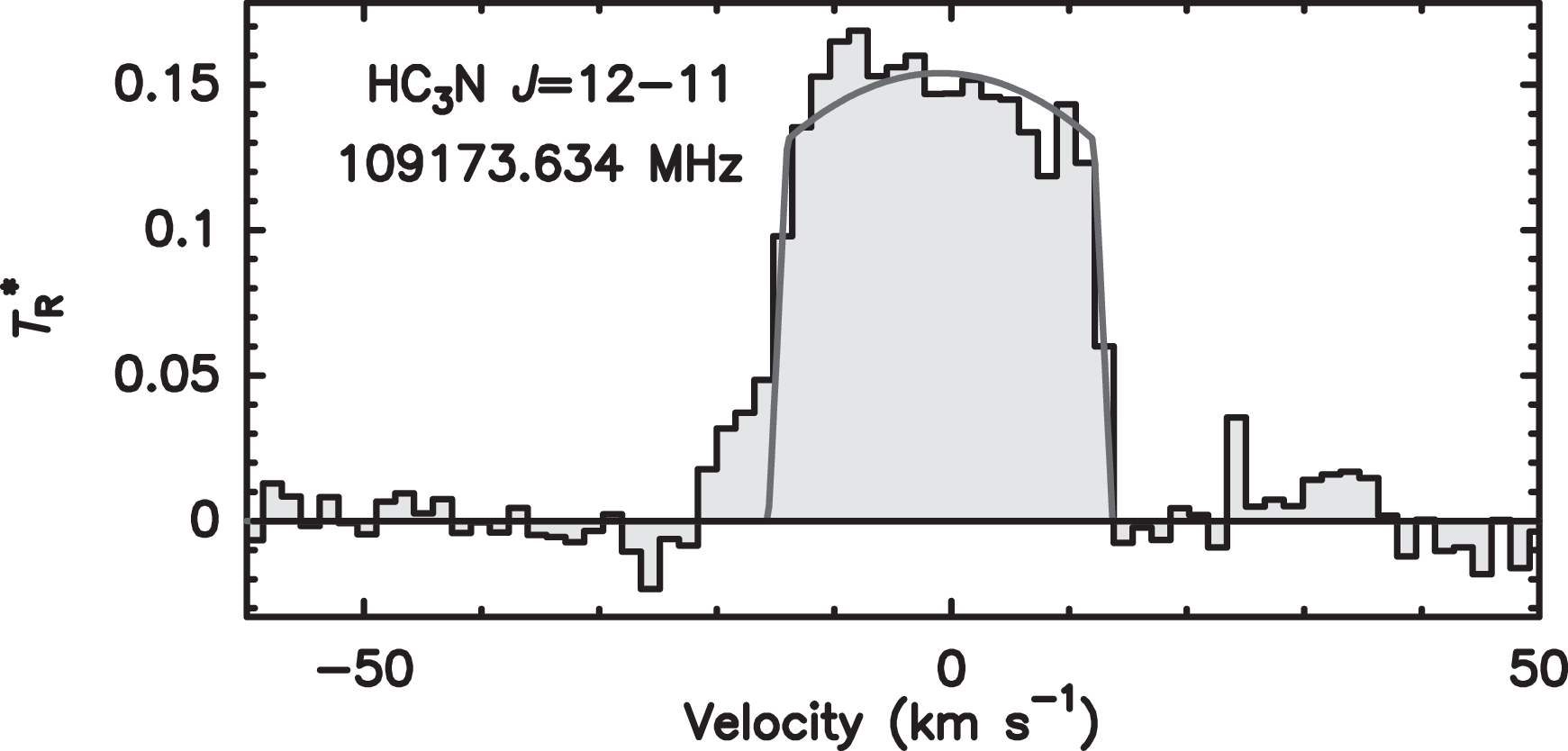}
    \includegraphics[width=0.45\textwidth]{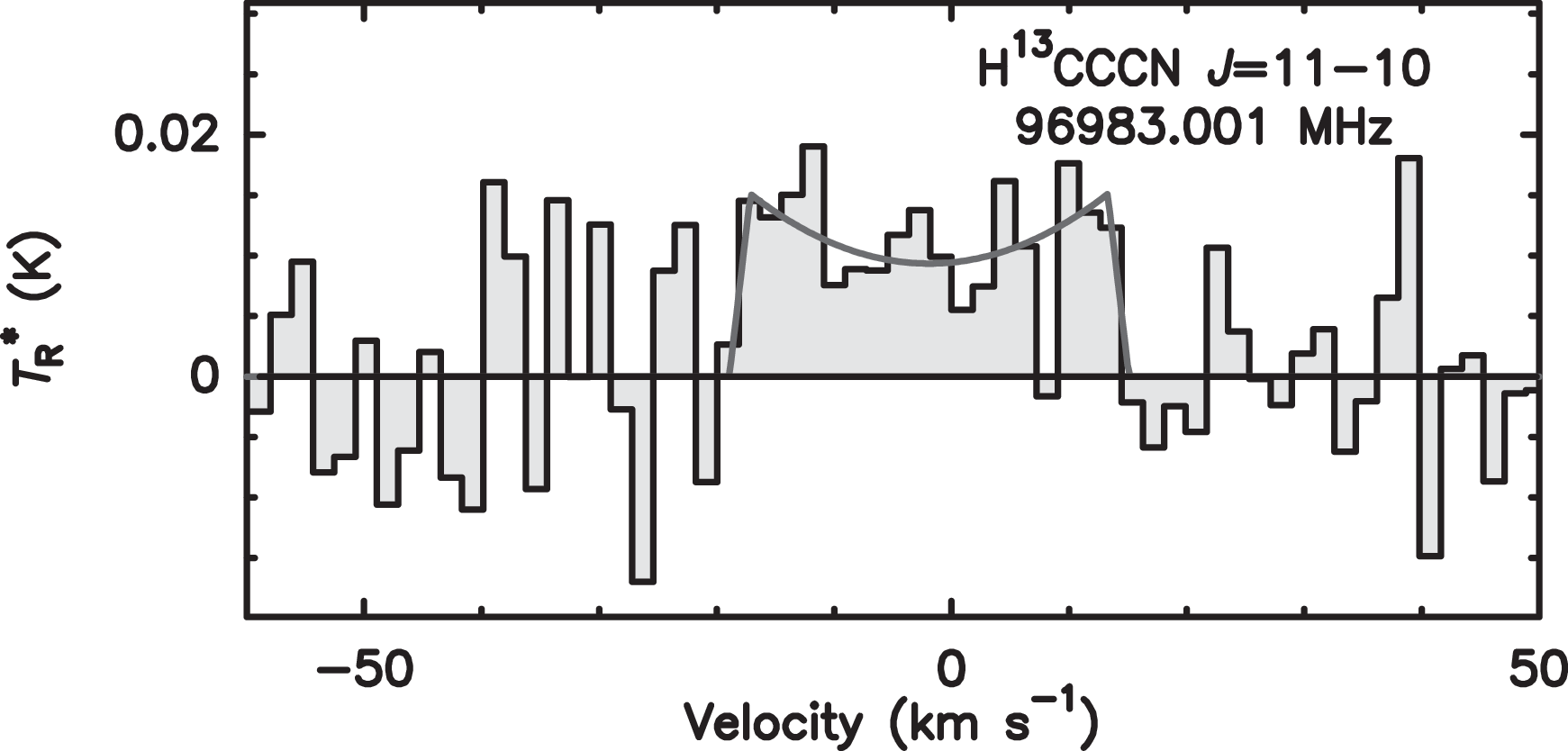}
    \includegraphics[width=0.45\textwidth]{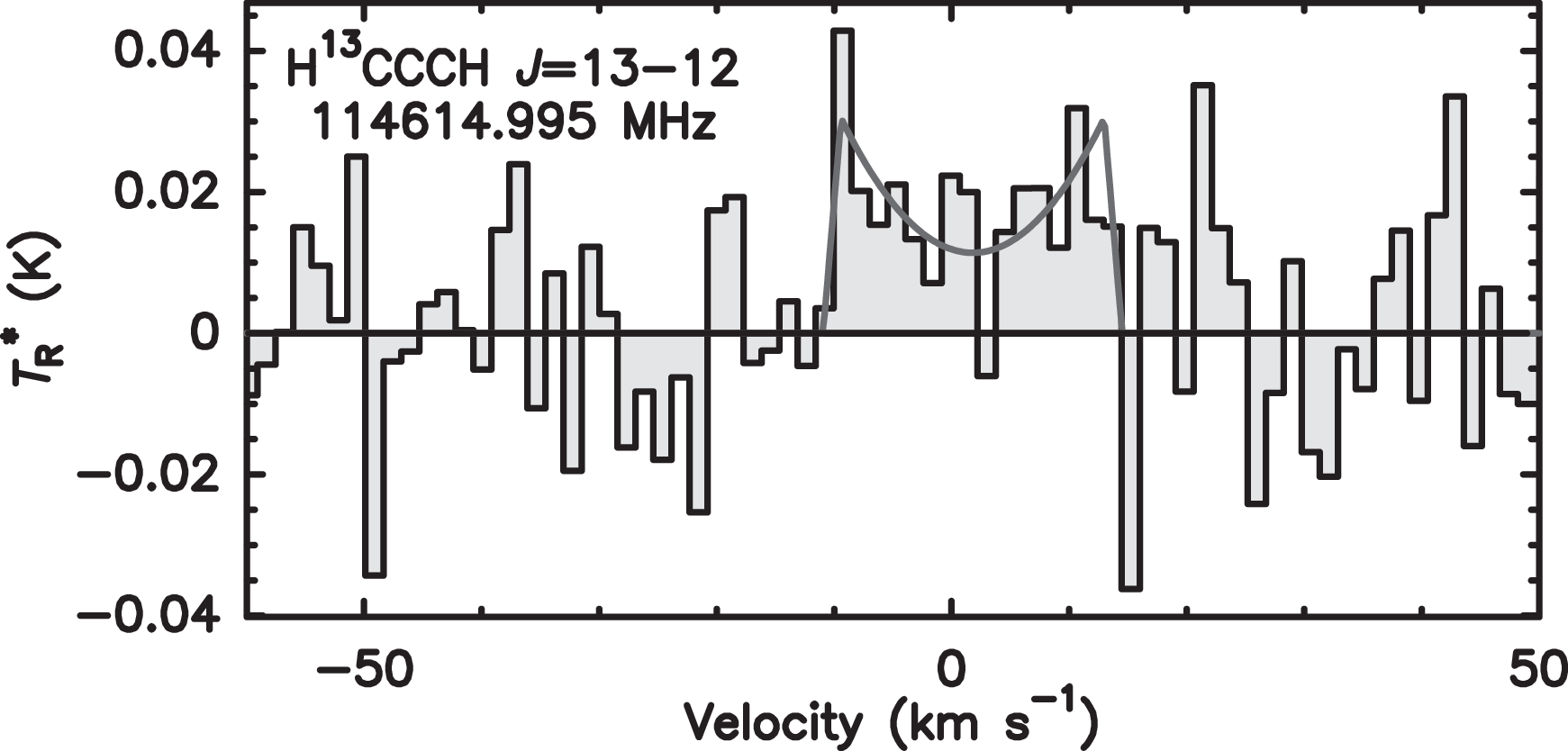}
    \caption{Same as figure~\ref{figure:3}, but for HC$_{3}$N.}
    \label{figure:10}
\end{figure*}

\clearpage

\begin{figure*}
    \centering
    \includegraphics[width=0.45\textwidth]{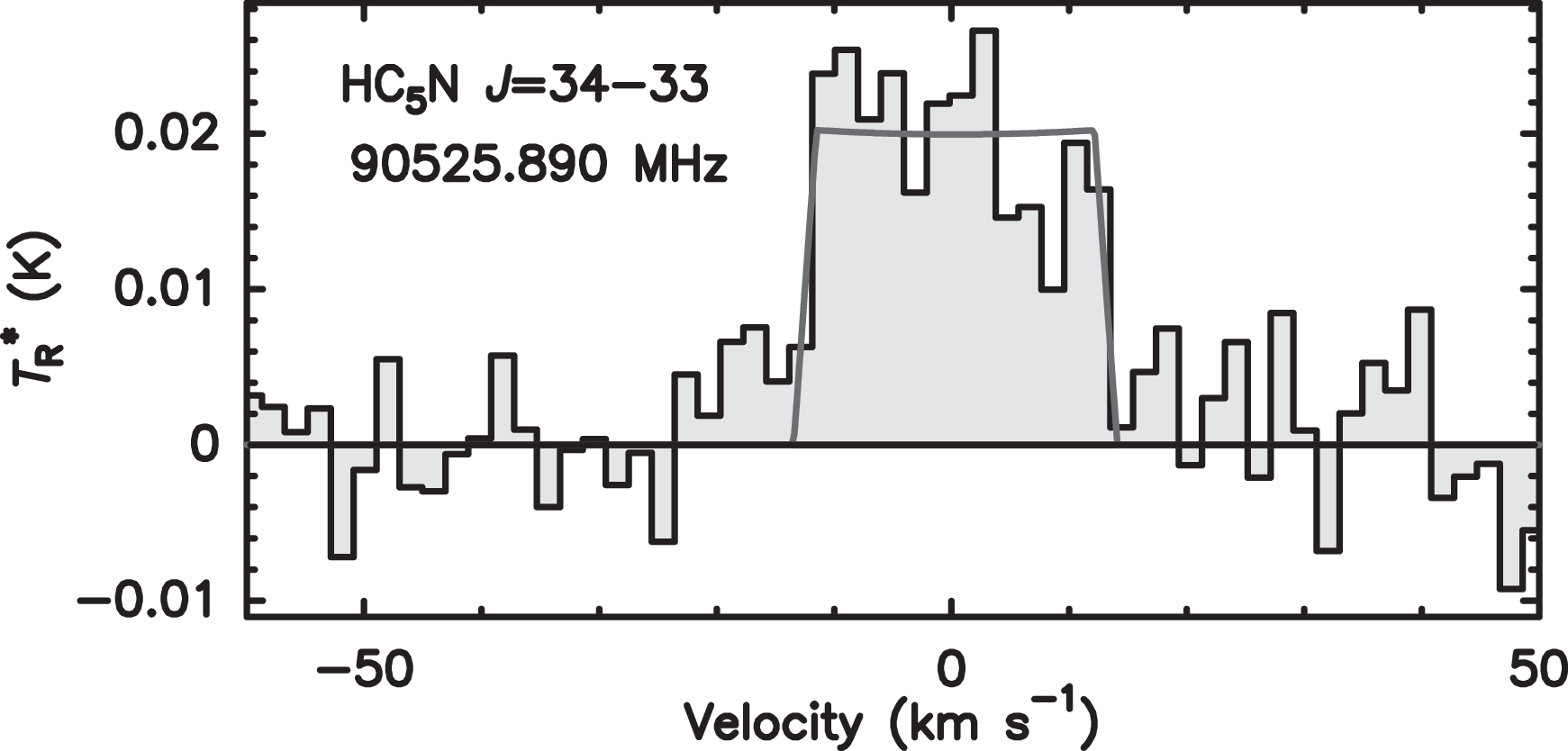}
    \includegraphics[width=0.45\textwidth]{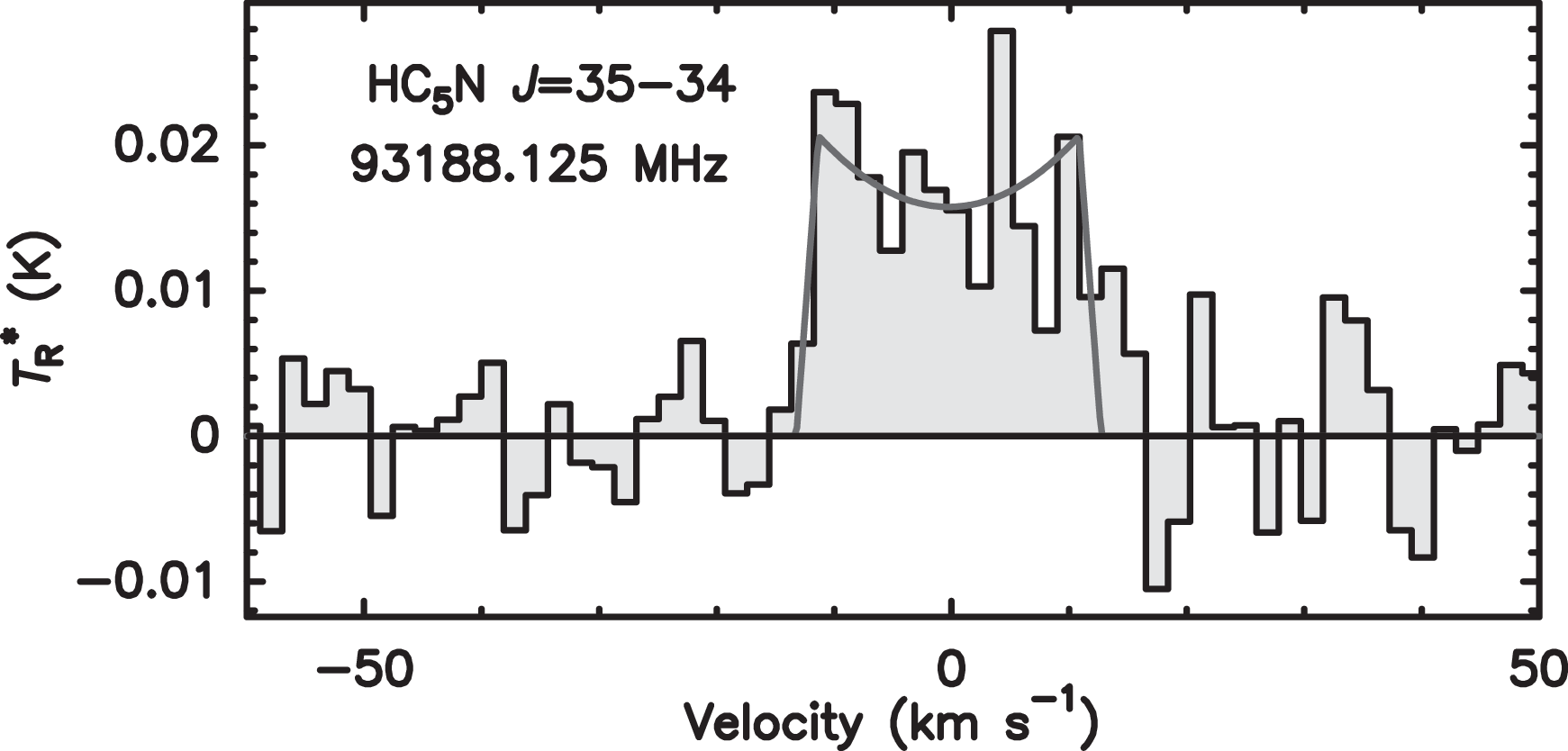}
    \includegraphics[width=0.45\textwidth]{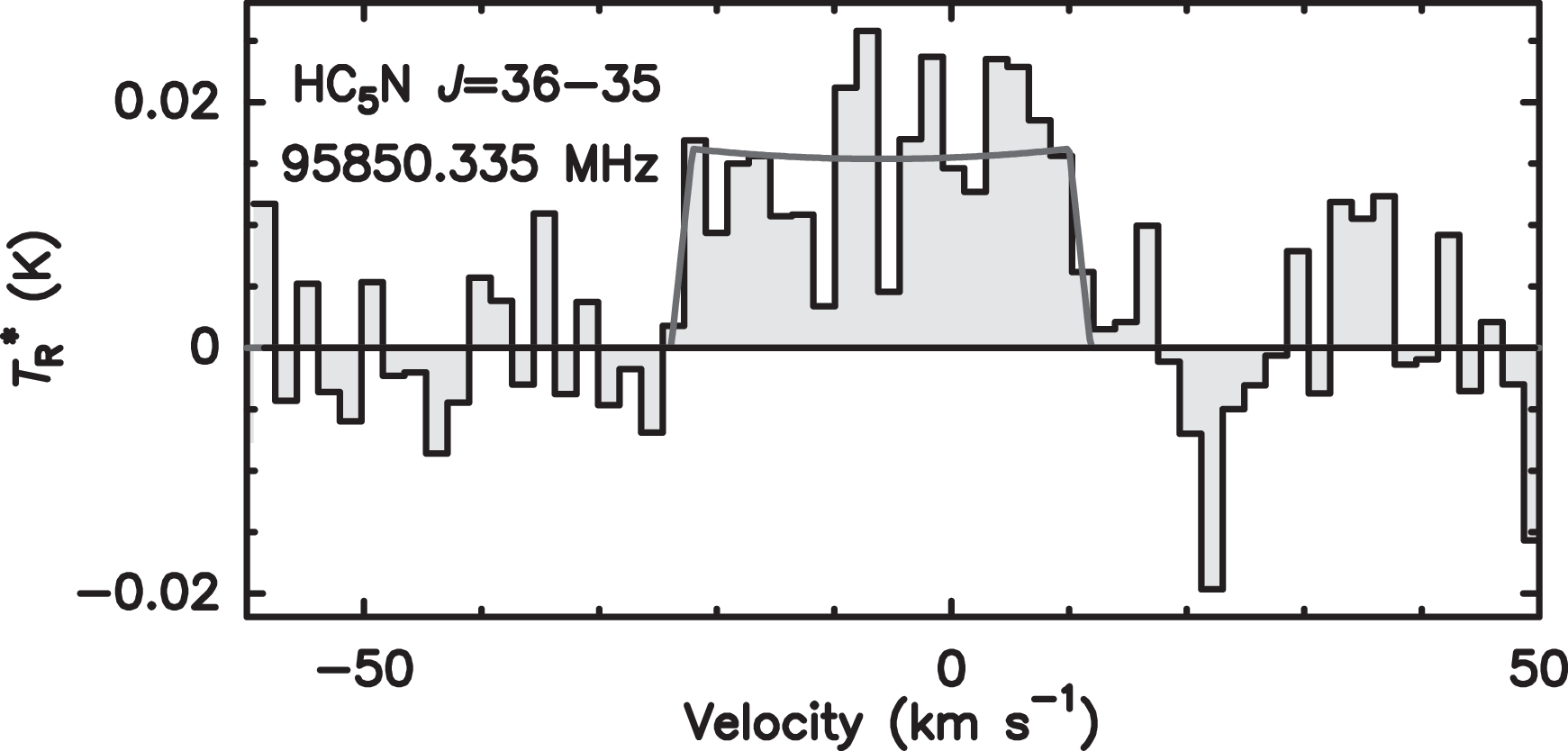}
    \includegraphics[width=0.45\textwidth]{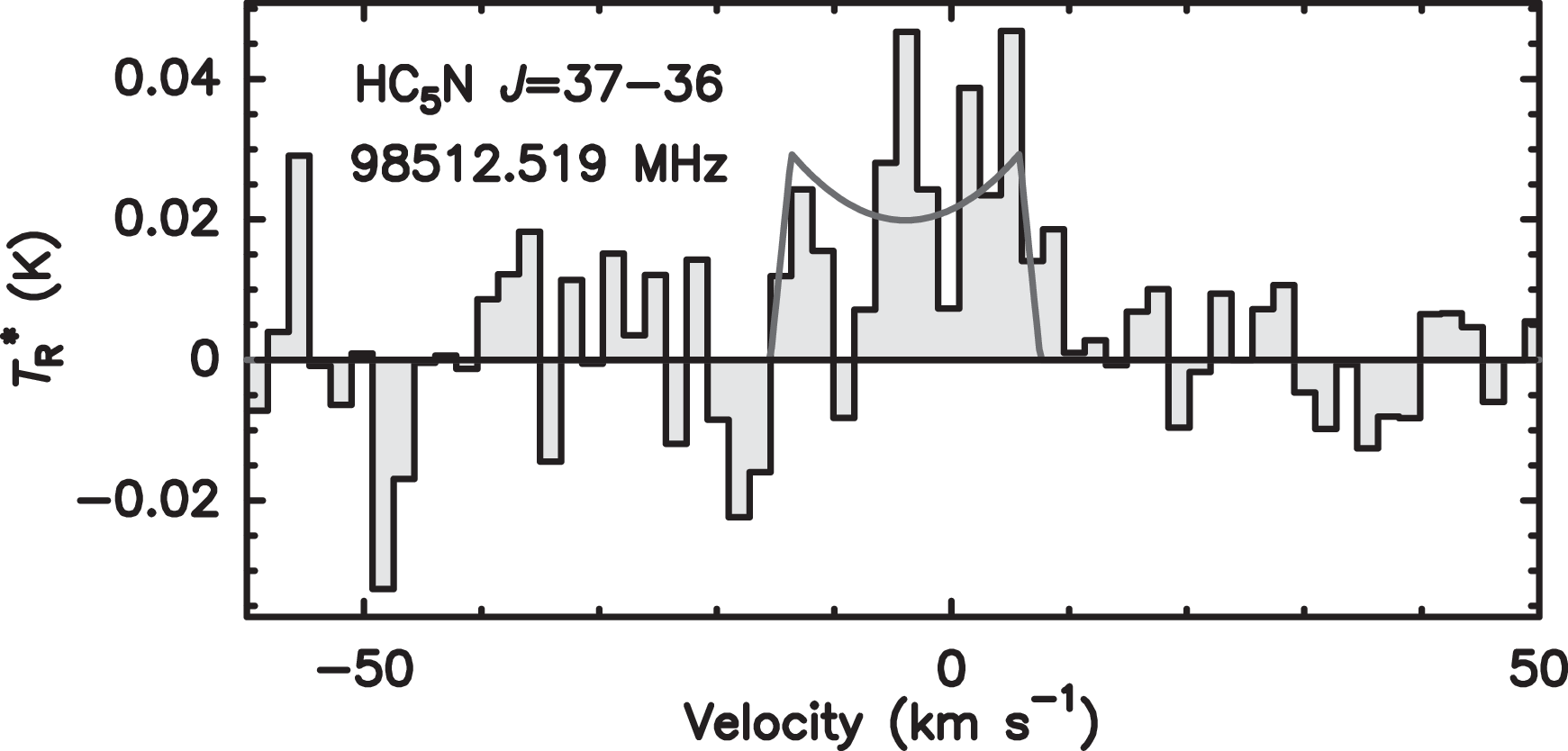}
    \includegraphics[width=0.45\textwidth]{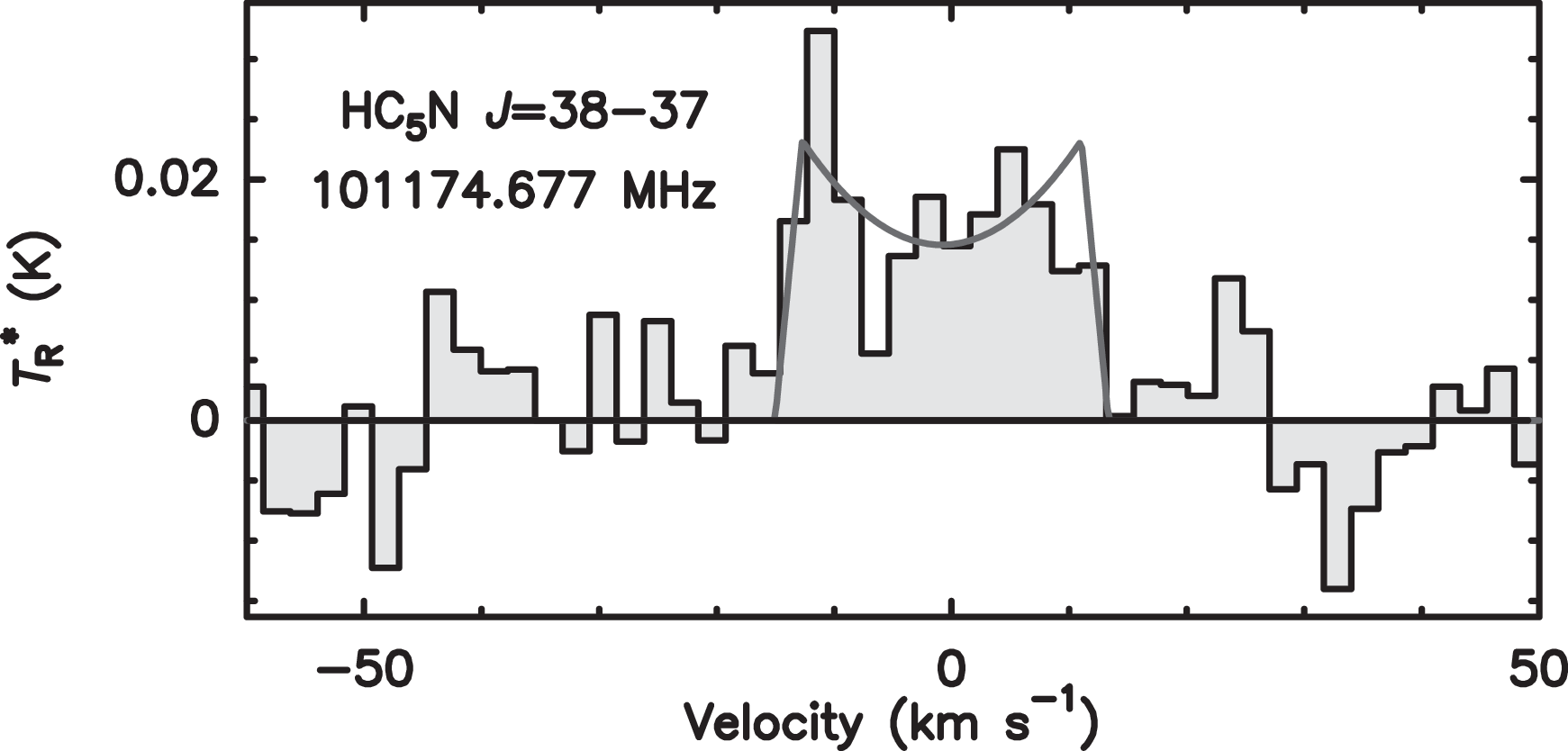}
    \includegraphics[width=0.45\textwidth]{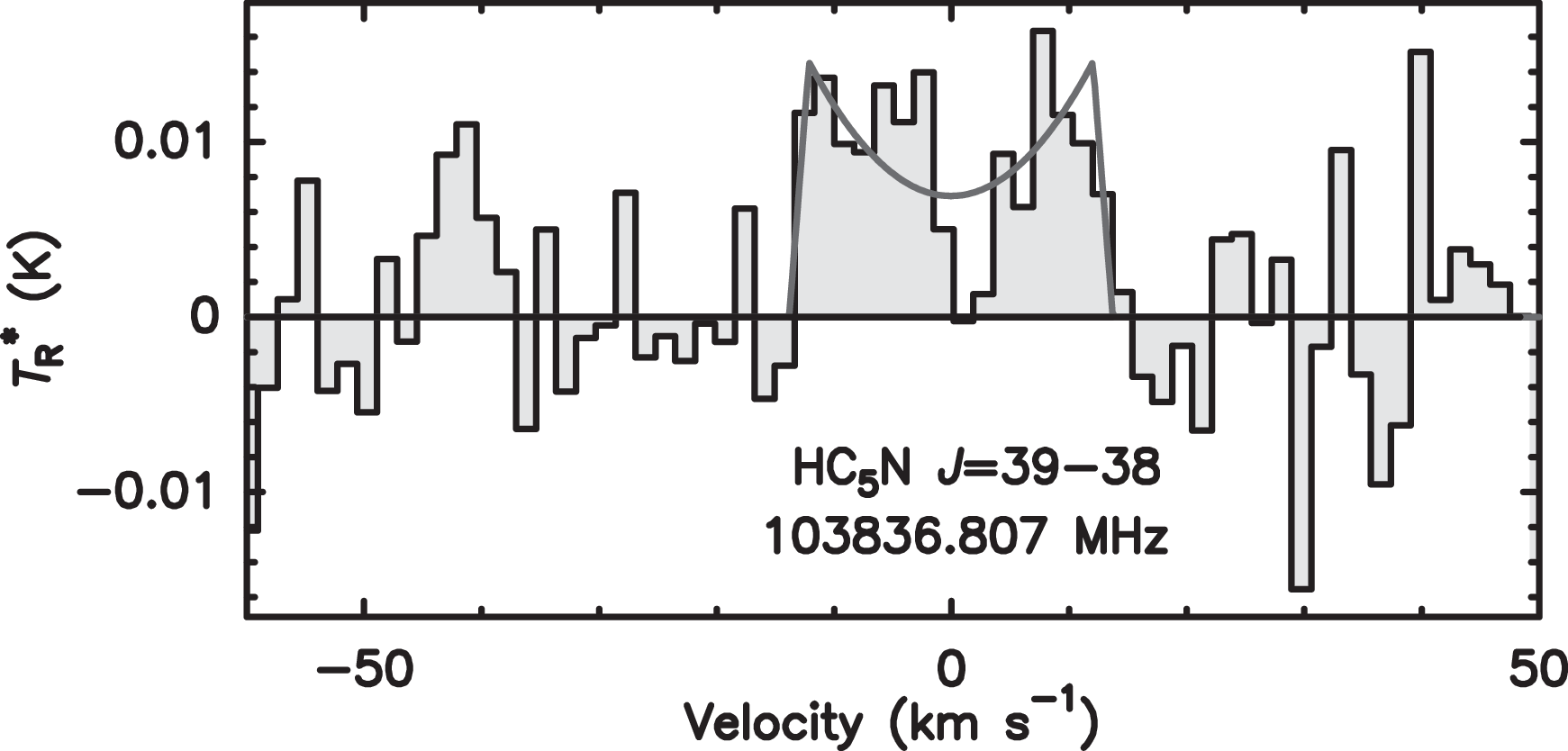}
    \includegraphics[width=0.45\textwidth]{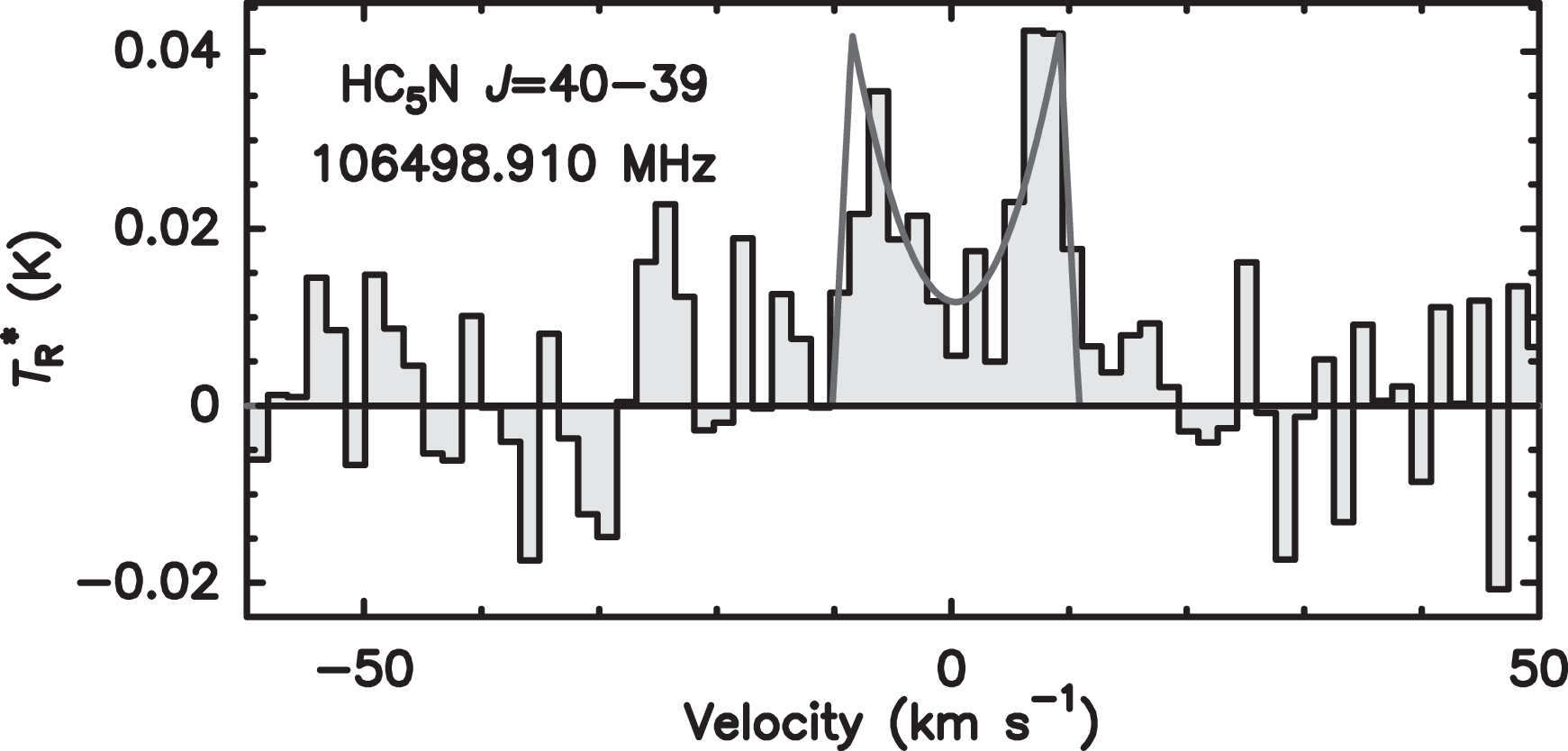}
    \includegraphics[width=0.45\textwidth]{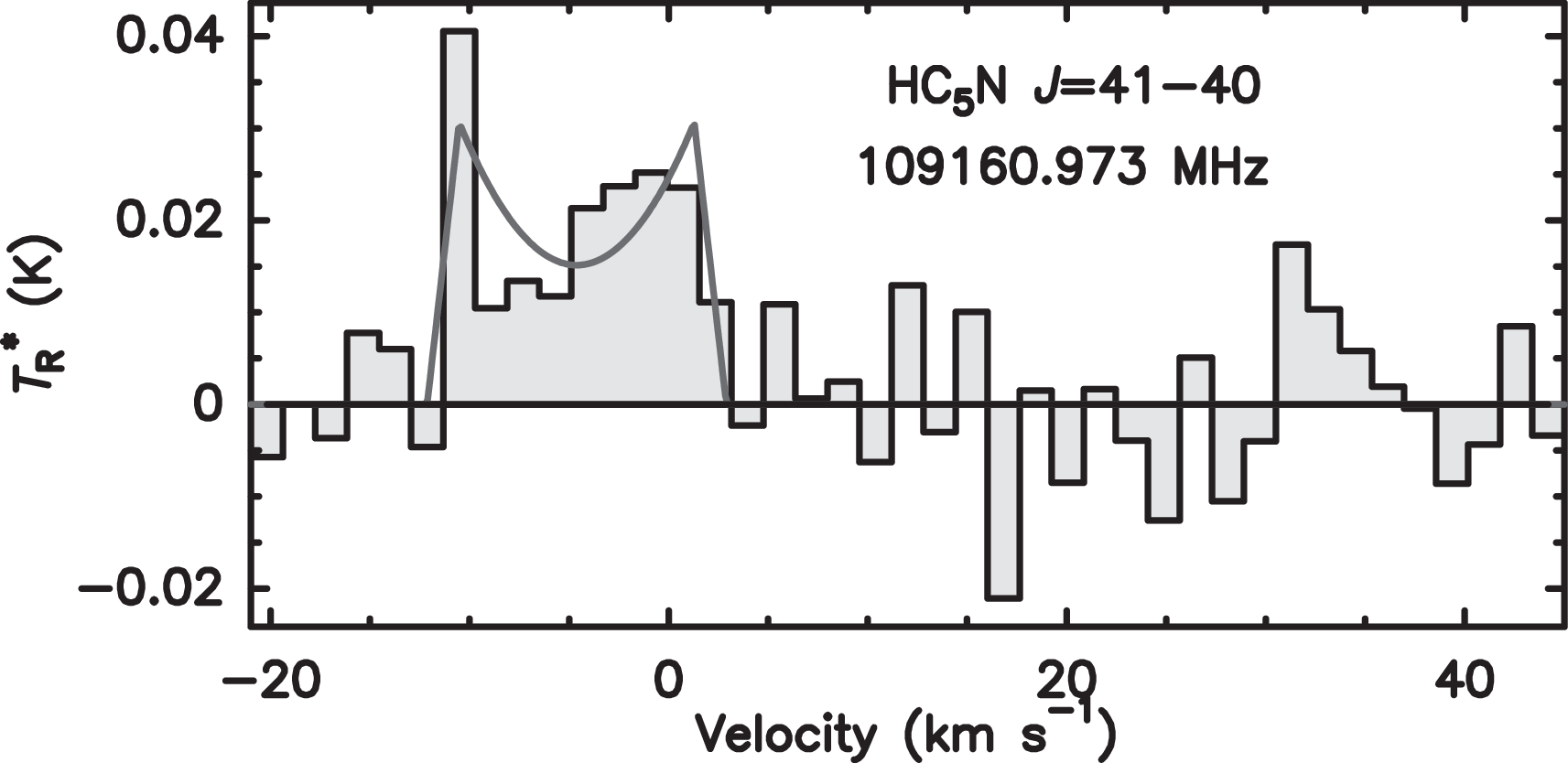}
    \caption{Same as figure~\ref{figure:3}, but for HC$_{5}$N.}
    \label{figure:11}
\end{figure*}

\clearpage

\begin{figure*}
    \centering
    \includegraphics[width=0.45\textwidth]{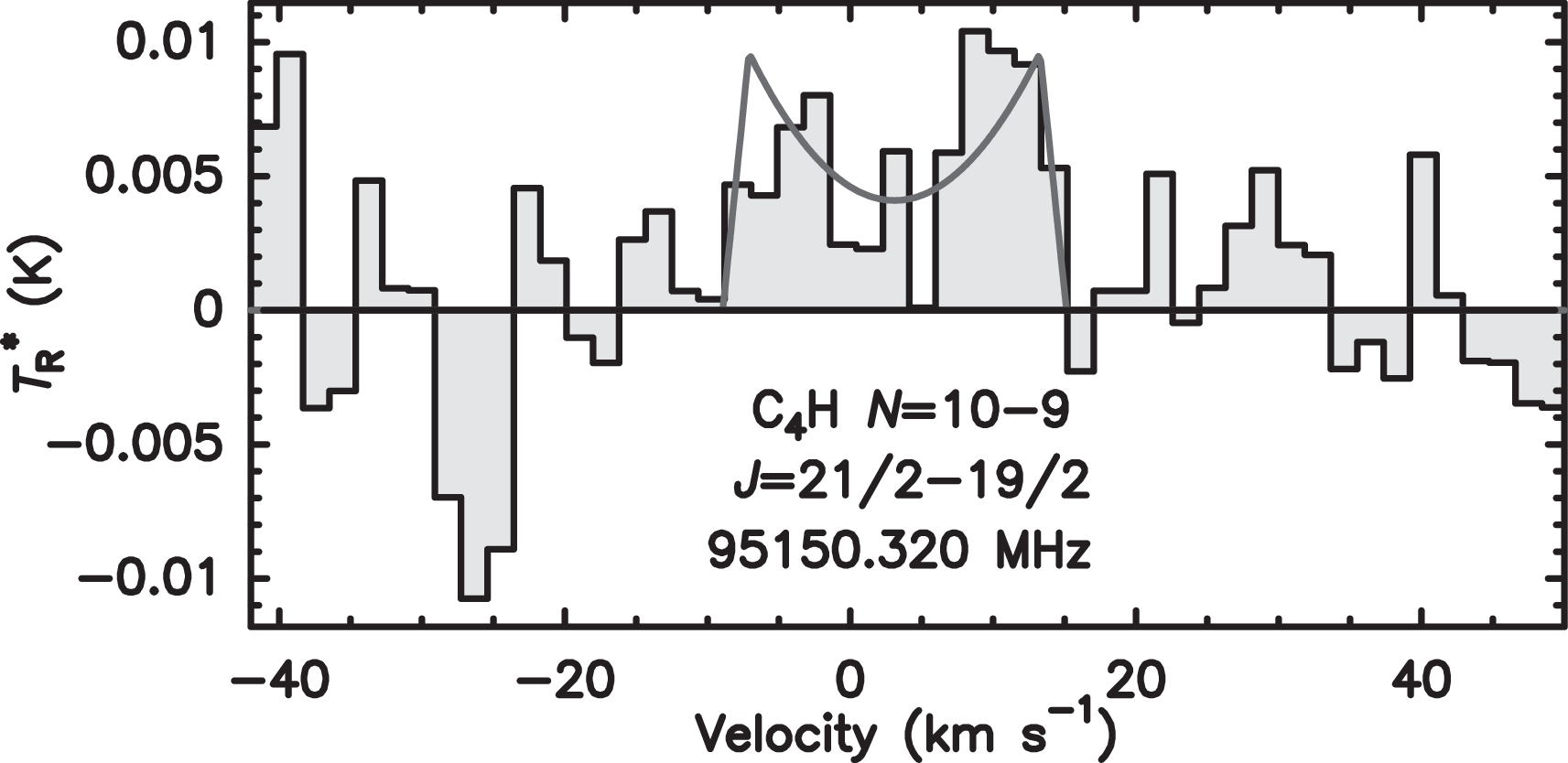}
    \includegraphics[width=0.45\textwidth]{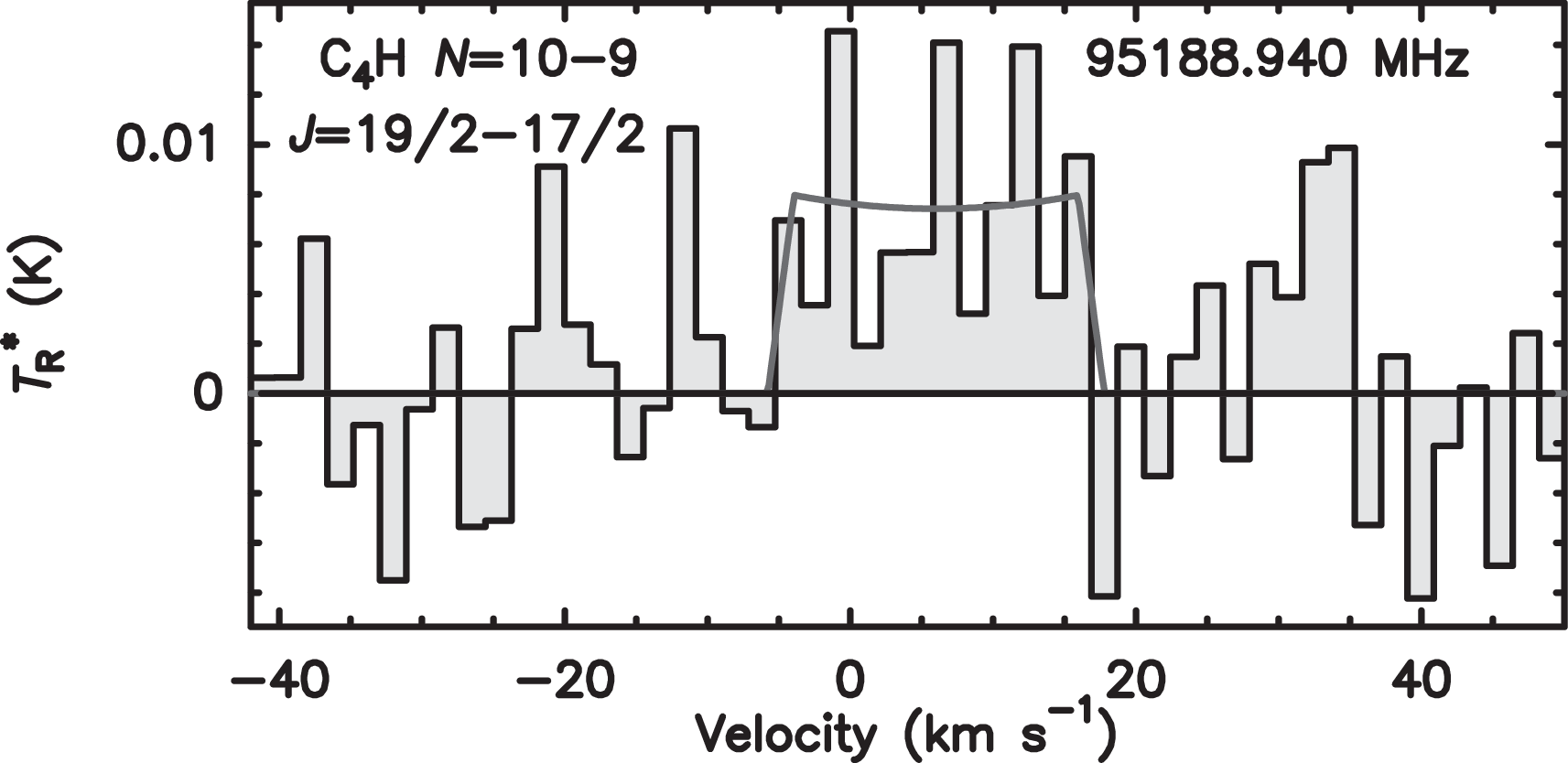}
    \includegraphics[width=0.45\textwidth]{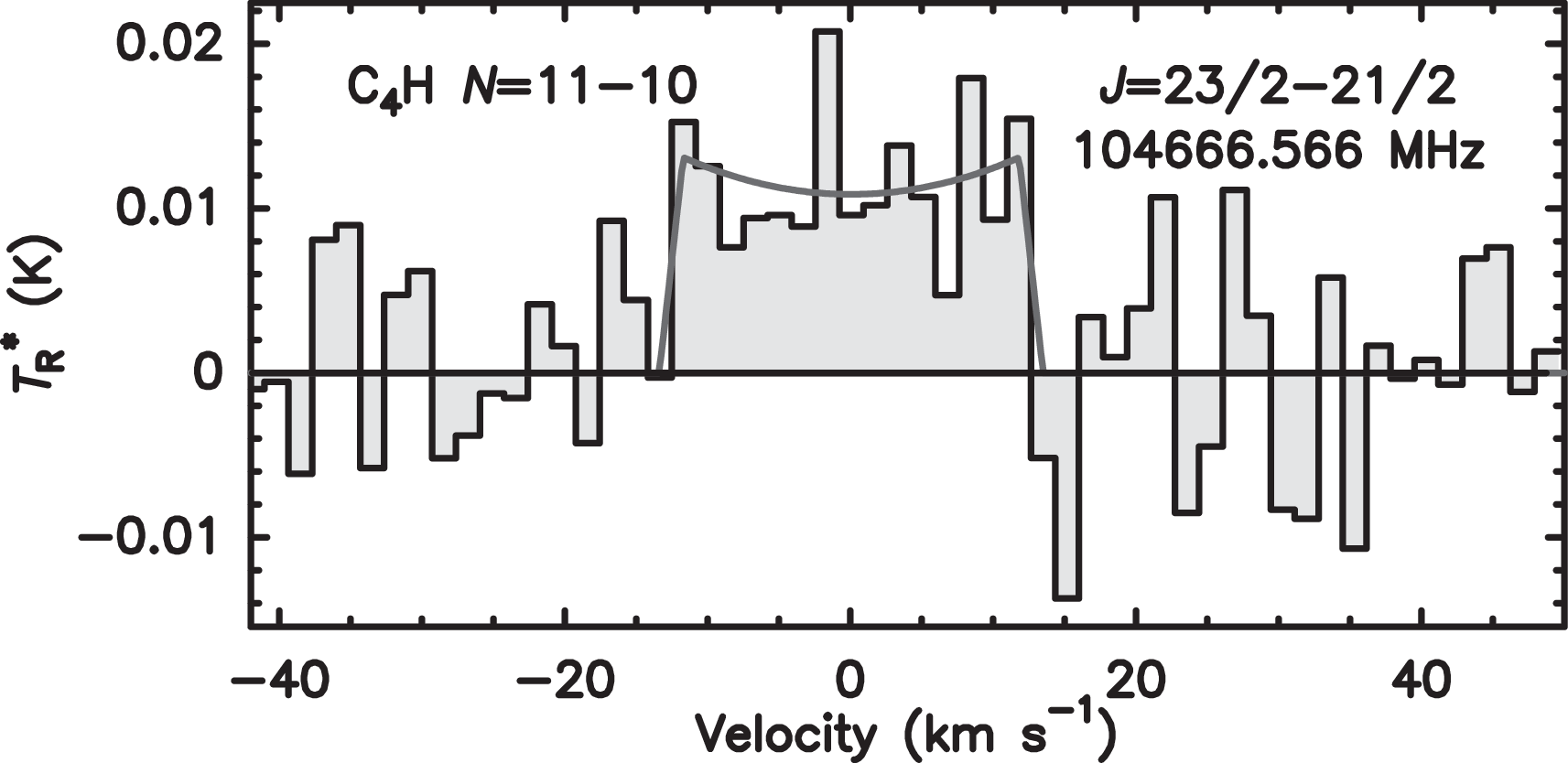}
    \includegraphics[width=0.45\textwidth]{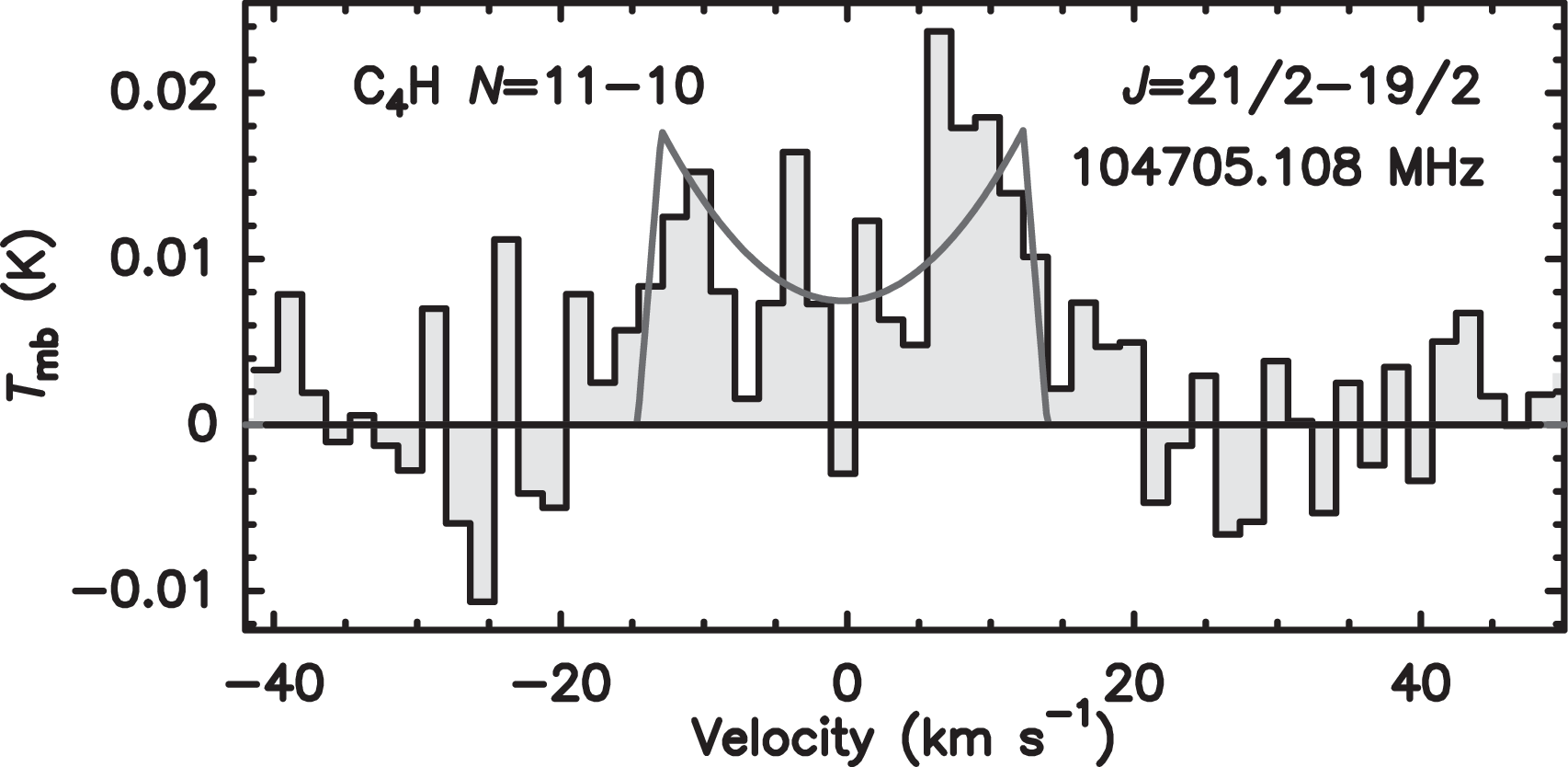}
    \includegraphics[width=0.45\textwidth]{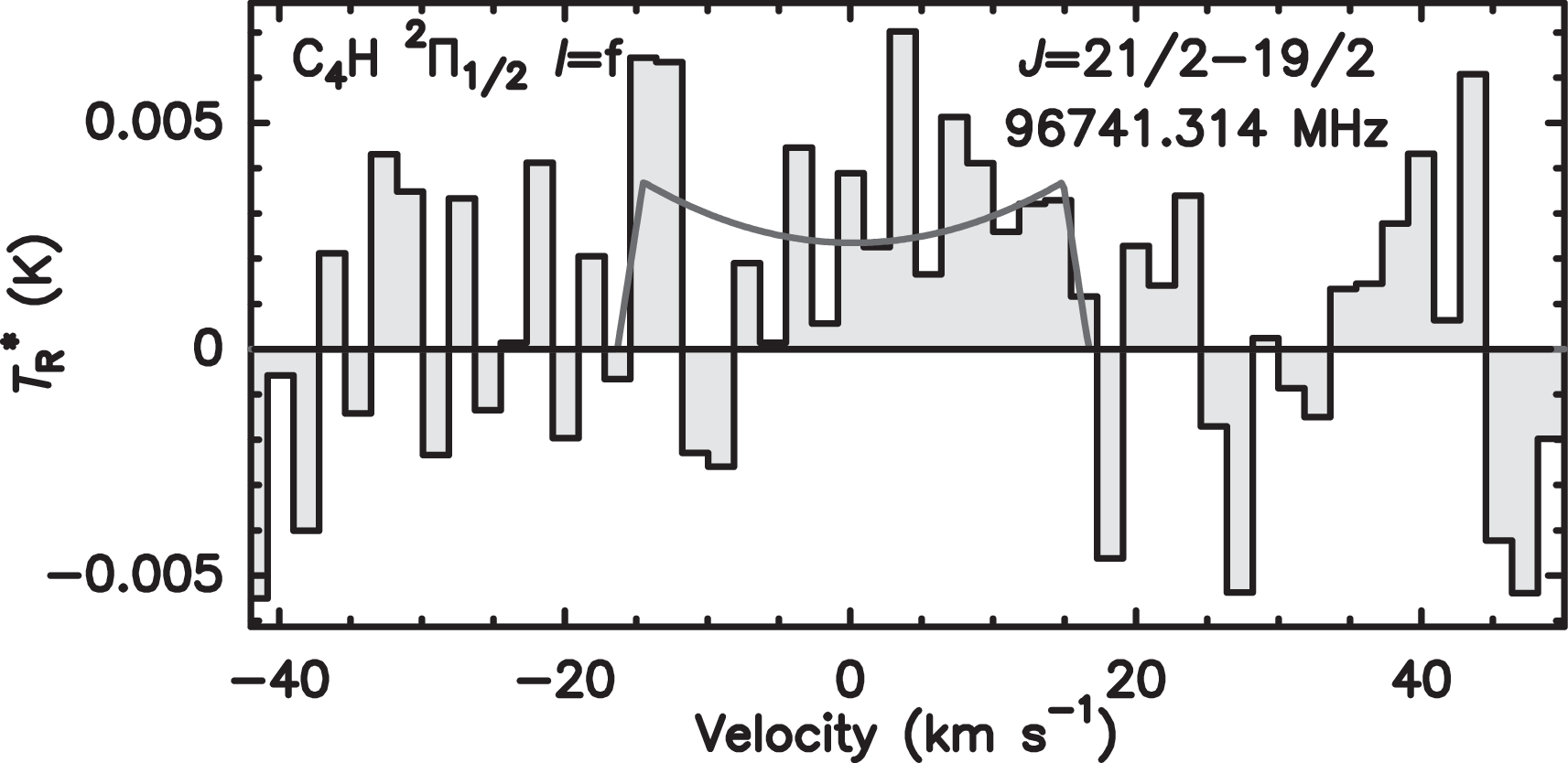}
    \caption{Same as figure~\ref{figure:3}, but for C$_{4}$H.}
    \label{figure:12}
\end{figure*}

\clearpage

\begin{figure*}
    \centering
    
    \includegraphics[width=0.45\textwidth]{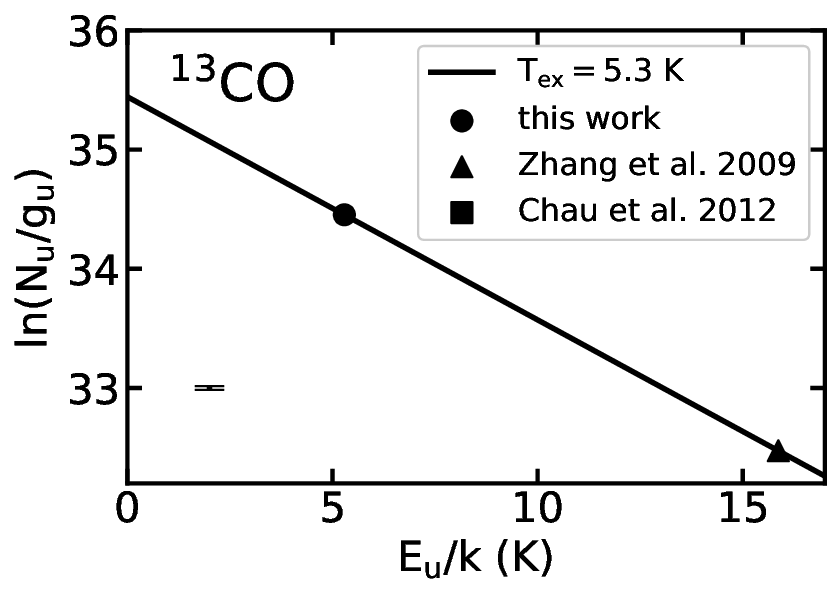}
    \includegraphics[width=0.45\textwidth]{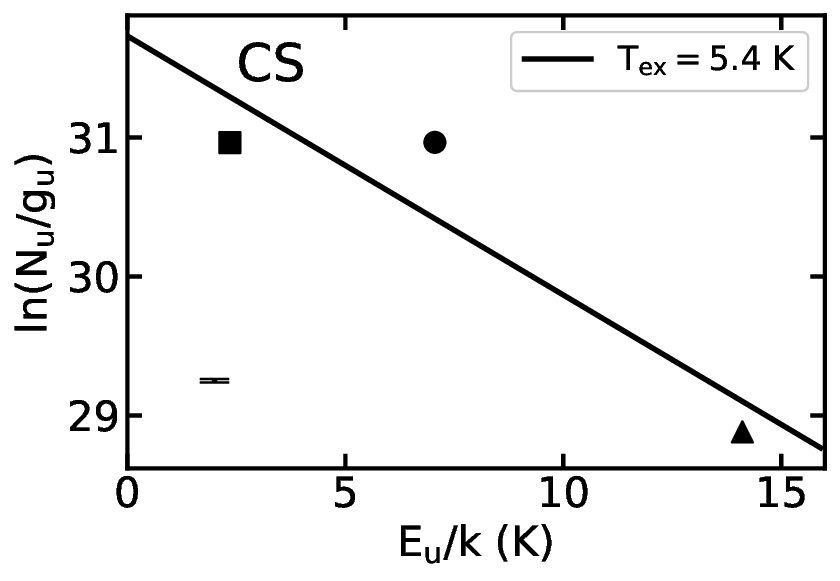}
    \includegraphics[width=0.45\textwidth]{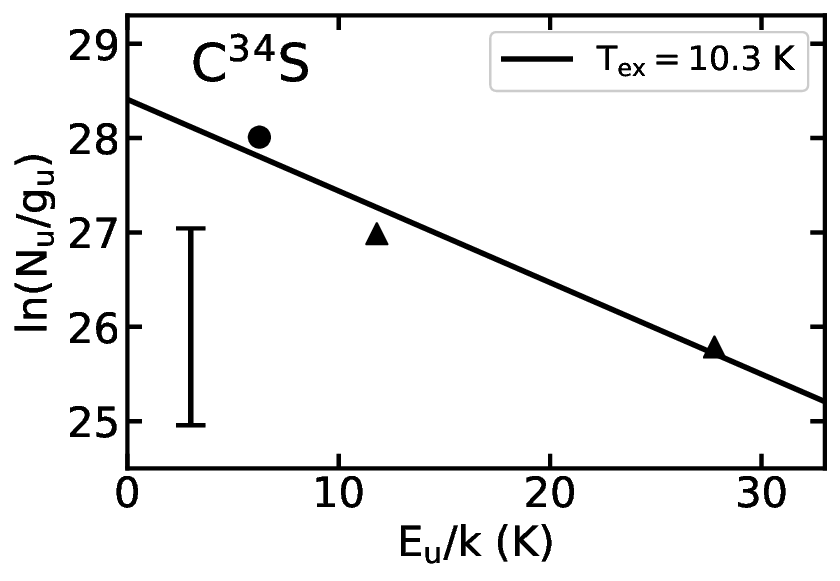}
    \includegraphics[width=0.45\textwidth]{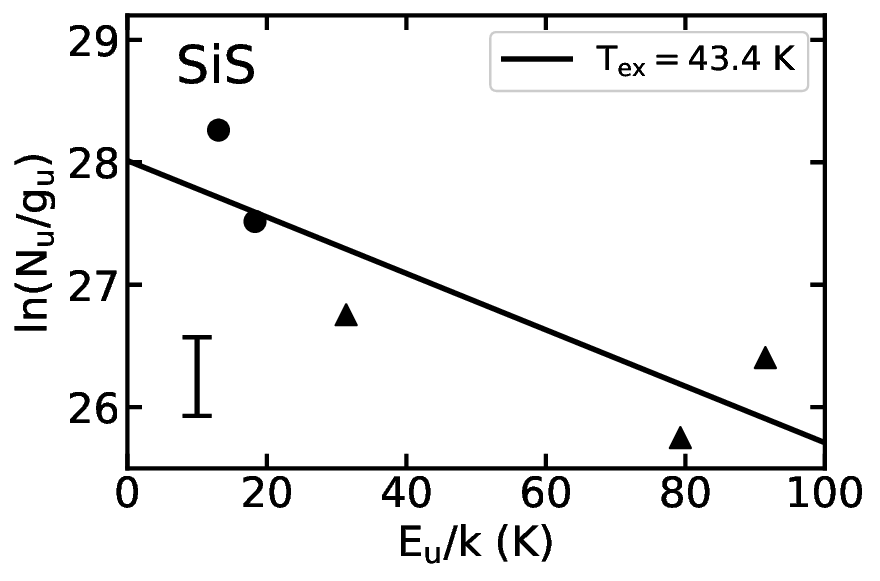}
    \includegraphics[width=0.45\textwidth]{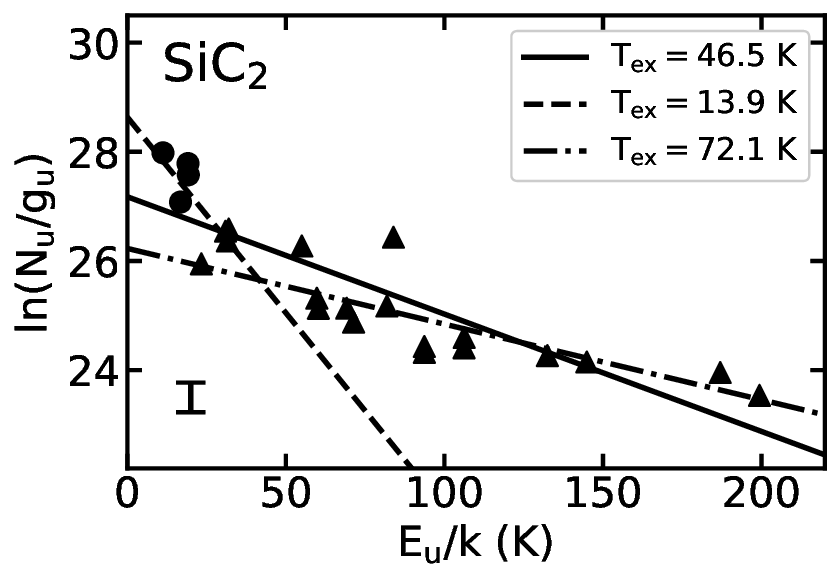}
    \includegraphics[width=0.45\textwidth]{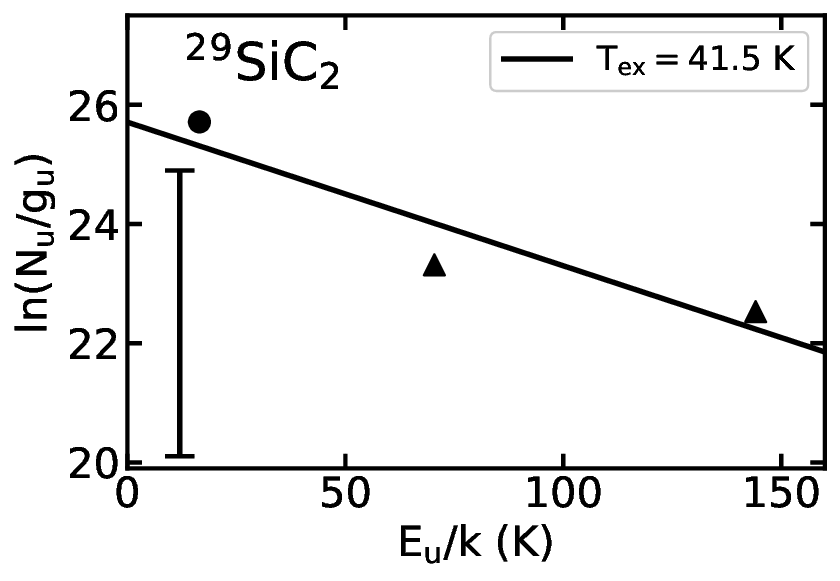}
    \includegraphics[width=0.45\textwidth]{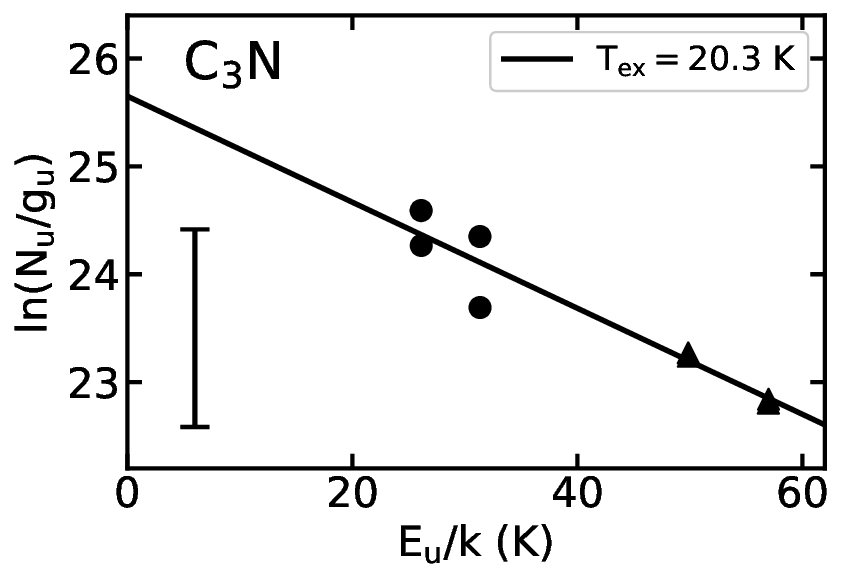}
    \includegraphics[width=0.45\textwidth]{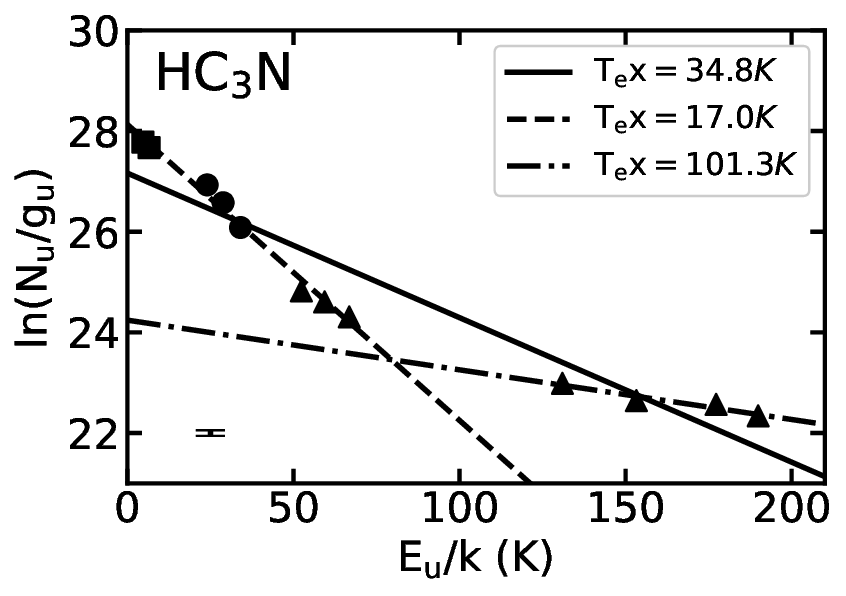}

    \caption{Rotation diagrams of the molecules detected in this work. Error bars are shown on the lower left.}
    \label{figure:13}
\end{figure*}

\clearpage

\addtocounter{figure}{-1}
\begin{figure*}
    \addtocounter{figure}{0}
    \centering

    \includegraphics[width=0.45\textwidth]{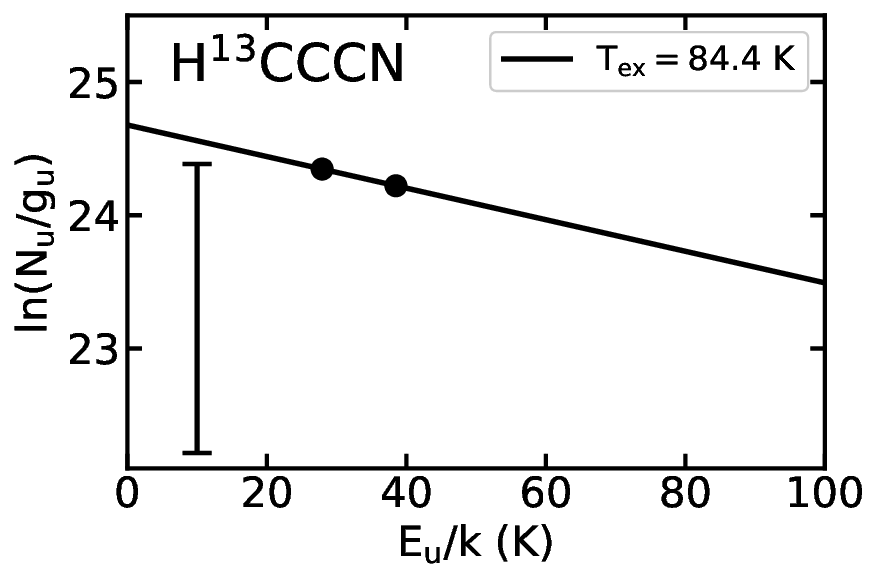}
    \includegraphics[width=0.45\textwidth]{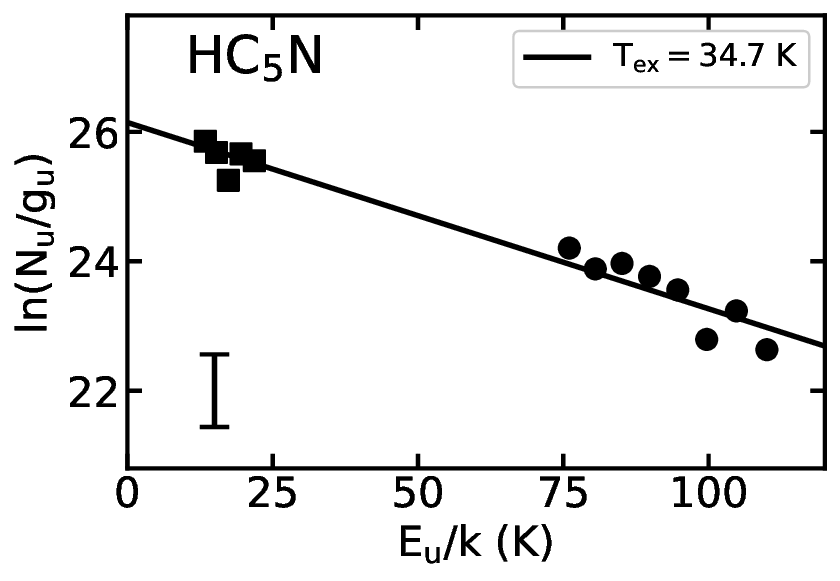}
    \includegraphics[width=0.45\textwidth]{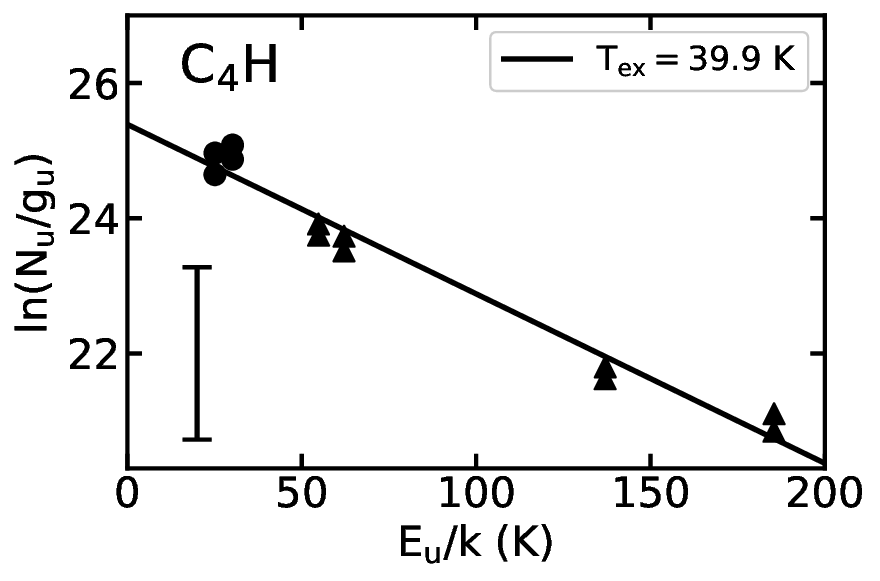}

    \caption{(Continued)}
   
\end{figure*}

\clearpage

\begin{figure*}
    \centering
    \includegraphics[width=1.0\textwidth]{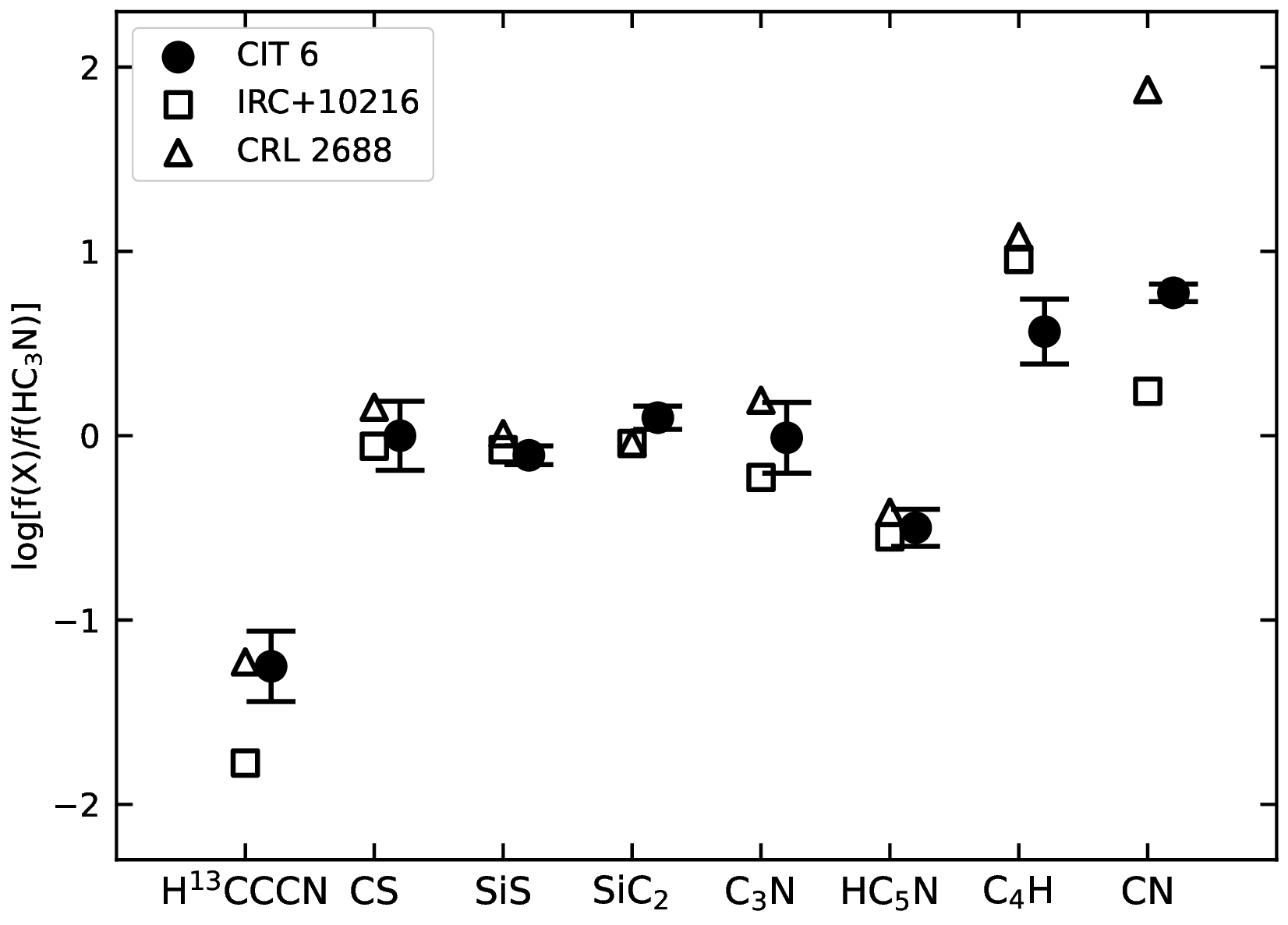}
    \caption{Comparison between the fractional abundance of the molecule X relative to HC$_{3}$N in IRC+10216, CRL\,2688, and CIT\,6.    
    The data of IRC+10216 are taken from \cite{2003A&A...402..617W}, \cite{2008ApJS..177..275H}, \cite{2015A&A...574A..56G}, and \cite{2017A&A...606A..74Z}.
    The data of CRL\,2688 are taken from \cite{2022ApJS..259...56Q}.}
    \label{figure:14}
\end{figure*}

\clearpage
\begin{table*}
\tbl{Summary of line surveys of CIT\,6.}{%
\begin{tabular}{cccc}\\
    \hline
    Frequencies  & Telescope & Reference & Line/Fre.     \\
    (GHz)        &           &           & (GHz$^{-1}$)  \\
    \hline 
    36--49        & NRO-45 m       & \cite{2012ApJ...760...66C} & 1.75 \\
    90--116       & ARO-12 m       & this work                 & 1.73  \\
    131--160      & ARO-12 m       & \cite{2009ApJ...691.1660Z} & 0.72  \\
    219--244      & SMT-10m       & \cite{2009ApJ...691.1660Z} & 1.60  \\
    252--268      & SMT-10m       & \cite{2009ApJ...691.1660Z} & 1.75  \\
    \hline
\end{tabular}}
\label{table:1}
\begin{tabnote}
\end{tabnote}
\end{table*}

\newgeometry{left=1.3cm,right=1.3cm,top=2cm,bottom=3cm}
\setlength{\tabcolsep}{0.1pt}
\renewcommand{\arraystretch}{1}
\begin{longtable}{*{8}{c}}
\caption{Molecular transitions and measurements.}
\label{table:2}\\
\hline
 Frequency & Species &  Transition & rms & $T^{\rm *}_{\rm R}$  & $\int T^{\rm *}_{\rm R}{\rm d}v$ & $V_{\rm exp}$ & Reference \\
(MHz) &   & (upper-lower) & (mK) & (K) & (K\,km s$^{-1}$) & (km\,s$^{-1}$) & \\
\hline  
\endfirsthead

\hline
Frequency & Species &  Transition & rms & $T^{\rm *}_{\rm R}$  & $\int T^{\rm *}_{\rm R}$d$v$ & $V_{\rm exp}$ & Mark \\
(MHz) &   & (upper-lower) & (mK) & (K) & (K\,km s$^{-1}$) & (km\,s$^{-1}$)\\
\hline  
\endhead

\hline
\endfoot

\hline
\multicolumn{8}{l}{\textbf{Notes} }\\
\multicolumn{8}{l}{(N) Transitions that are newly detected toward this source.}\\
\multicolumn{8}{l}{(B) Blended fine-structure lines, where the total intensity is given.}\\
\multicolumn{8}{l}{Transitions reported in the references that  are denoted by the numbers:}\\
\multicolumn{8}{l}{
(1) Zuckerman (1981); (2) Olofsson \& Rydbeck (1984); (3) Knapp \& Morris (1985);
}\\
\multicolumn{8}{l}{
(4) Bachiller et al. (1988); (5) Zuckerman \& Dyck (1989); (6) Sahai (1990); 
}\\
\multicolumn{8}{l}{
(7) Margulis et al. (1990); (8) Nyman et al.(1992); (9) Pei \& Chen (1997);
}\\
\multicolumn{8}{l}{
(10) Meixner et al. (1998); (11) Neri et al. (1998); (12) Olofsson et al. (1998);
}\\
\multicolumn{8}{l}{
 (13) Sch{\"o}ier \& Olofsson (2001); (14) Sch{\"o}ier et al. (2002); (15) Ramstedt et al. (2008);
}\\
\multicolumn{8}{l}{
(16) Milam et al. (2009); (17) Thronson \& Mozurkewich (1983); (18) Sopka et al. (1989);
}\\
\multicolumn{8}{l}{
(19) Kahane et al. (1992); (20) Groenewegen et al. 1996; (21) Woods et al.( 2003b);
}\\
\multicolumn{8}{l}{
(22) Knapp \& Chang (1985); (23) Ramstedt \& Olofsson (2014); (24)Jewell \& Snyder (1984);
}\\
\multicolumn{8}{l}{
(25) Sahai et al. (1984); (26) Henkel et al. (1985); (27) Sch{\"o}ier et al. (2007);
}\\
\multicolumn{8}{l}{
(28) Bujarrabal et al. (1994); (29) Bujarrabal et al. (1994); (30) Lindqvist et al. (2000);
}\\
\multicolumn{8}{l}{
(31) Jewell \& Snyder (1982); (32) Bachiller et al. (1997); (33) Fukasaku et al. (1994).
}
\endlastfoot
115271.2  & CO               &                   $J=1-0$           &  2.5  & 1.731  & 47.801$\pm$ 0.200  & 16.0 $\pm$ 0.1  &  (1)-(21) \\
110201.4  & $^{13}$CO        &                   $J=1-0$           &  1.1  & 0.074  & 2.681 $\pm$ 0.082  & 16.2 $\pm$ 0.1  &  (17)-(23)  \\
97981.0   & CS               &                   $J=2-1$           &  2.5  & 0.291  & 7.229 $\pm$ 0.138  & 15.2 $\pm$ 0.2  &  (12), (21), (24) \\
96413.0   & C$^{34}$S        &                   $J=2-1$           &  9.7  & 0.012  & 0.352 $\pm$ 0.062  & 9.9 $ \pm$ 0.5  & N \\
90771.6   & SiS              &                   $J=5-4$           &  5.2  & 0.026  & 0.592 $\pm$ 0.036  & 13.5 $\pm$ 0.9  &  (12), (21), (26)-(30)\\
108924.3   & SiS              &                   $J=6-5$           &  7.7  & 0.038  & 0.766 $\pm$ 0.061  & 14.9 $\pm$ 1.4  &  (21), (25), (26) \\
93063.6   & SiC$_{2}$        &         $J_{K_a,K_c}=4_{0,4}-3_{0,3}$   &  3.9  & 0.035  & 0.886 $\pm$ 0.032  & 13.8 $\pm$ 0.2  & N \\
94245.4   & SiC$_{2}$        &         $J_{K_a,K_c}=4_{2,3}-3_{2,2}$   &  5.9  & 0.019  & 0.576 $\pm$ 0.041  & 13.2 $\pm$ 0.3  & N \\
95579.4   & SiC$_{2}$        &         $J_{K_a,K_c}=4_{2,2}-3_{2,1}$   &  6.7  & 0.020  & 0.496 $\pm$ 0.047  & 11.8 $\pm$ 0.4  & N \\
115382.4   & SiC$_{2}$        &         $J_{K_a,K_c}=5_{0,5}-4_{0,4}$   &  2.5  & 0.045  & 1.279 $\pm$ 0.159  & 12.8 $\pm$ 0.3  & (21) \\
113820.1   & $^{29}$SiC$_{2}$ &         $J_{K_a,K_c}=5_{0,5}-4_{0,4}$   &  7.9  & 0.013  & 0.300 $\pm$ 0.050  & 14.5 $\pm$ 1.6  & N \\
113123.3   & CN               & $N_{J,F}=1_{1/2,1/2}-0_{1/2,1/2}$   &  1.1  & 0.030  & 0.819 $\pm$ 0.074  & 13.1 $\pm$ 0.8  & (32)  \\
113144.2   & CN               & $N_{J,F}=1_{1/2,1/2}-0_{1/2,3/2}$   &  1.1  & 0.107  & 2.598 $\pm$ 0.080  & 16.4 $\pm$ 0.6  & (32)  \\
113170.5   & CN               & $N_{J,F}=1_{1/2,3/2}-0_{1/2,1/2}$   &  10.0 & 0.124  & 2.940 $\pm$ 0.085  & 13.7 $\pm$ 0.2  & (32)  \\
113191.3   & CN               & $N_{J,F}=1_{1/2,3/2}-0_{1/2,3/2}$   &  8.4  & 0.121  & 2.857 $\pm$ 0.058  & 15.0 $\pm$ 0.3  & (32)  \\
113488.1   & CN               & $N_{J,F}=1_{3/2,3/2}-0_{1/2,1/2}$   &  1.1  & 0.415  & 16.041             & ...             & B, (16), (21), (32) \\
113491.0   & CN               & $N_{J,F}= 1_{3/2,5/2}-0_{1/2,3/2}$  & ...   & ...    & ...                & ...             &  \\
113499.6   & CN               & $N_{J,F}= 1_{3/2,1/2}-0_{1/2,1/2}$  & ...   & ...    & ...                & ...             &  \\
113508.9   & CN               & $N_{J,F}= 1_{3/2,3/2}-0_{1/2,3/2}$  & ...   & ...    & ...                & ...             &  \\
90663.5   & HNC              &                     $J=1-0$         &  4.4  & 0.057  & 1.654 $\pm$ 0.043  & 13.8  $\pm$ 0.2 & (18), (21), (28)-(30) \\
98940.0  & C$_{3}$N          &       $N_{J}=10_{21/2}-9_{19/2}$    &  7.8  & 0.013  & 0.428 $\pm$ 0.055  & 12.2 $\pm$ 0.4  & N \\
98958.8  & C$_{3}$N          &       $N_{J}=10_{19/2}-9_{17/2}$    &  8.6  & 0.016  & 0.279 $\pm$ 0.115  & 10.6 $\pm$ 1.7  & N \\
108834.3 & C$_{3}$N          &       $N_{J}=11_{23/2}-10_{21/2}$   &  7.1  & 0.017  & 0.584 $\pm$ 0.052  & 16.2 $\pm$ 0.4  & (21), (26) \\
108853.0 & C$_{3}$N          & $N_{J}=11_{21/2}-10_{19/2}$         &  6.8  & 0.015  & 0.276 $\pm$ 0.039  & 9.2  $\pm$ 0.7  & (21), (26) \\
90979.0  & HC$_{3}$N         &               $J=10-9 $             &  7.0  & 0.134  & 3.466 $\pm$ 0.076  & 14.5 $\pm$ 0.2  & (18), (21), (24), (28)-(31)\\
100076.4 & HC$_{3}$N          &               $J=11-10$             &  8.6  & 0.151  & 3.945 $\pm$ 0.084  & 15.4 $\pm$ 0.2  & N \\
109173.6 & HC$_{3}$N          &               $J=12-11$             &  1.0  & 0.154  & 4.026 $\pm$ 0.096  & 13.8 $\pm$ 0.1  & (24) \\
96983.0  & H$^{13}$CCCN      &               $J=11-10$             &  8.5  & 0.009  & 0.371 $\pm$ 0.062  & 16.1 $\pm$ 0.7  & N \\
114615.0 & H$^{13}$CCCN       &               $J=13-12$             &  1.5  & 0.011  & 0.448 $\pm$ 0.097  & 12.0 $\pm$ 0.5  & N \\
90525.9  & HC$_{5}$N         &               $J=34-33$             &  4.3  & 0.020  & 0.510 $\pm$ 0.033  & 12.8 $\pm$ 0.4  & (21), (30)  \\
93188.1  & HC$_{5}$N         &               $J=35-34$             &  4.8  & 0.016  & 0.425 $\pm$ 0.035  & 12.0 $\pm$ 0.3  & N \\
95850.3  & HC$_{5}$N         &               $J=36-35$             &  7.1  & 0.015  & 0.532 $\pm$ 0.059  & 16.9 $\pm$ 0.6  & N \\
98512.5  & HC$_{5}$N         &               $J=37-36$             &  1.1  & 0.020  & 0.502 $\pm$ 0.076  & 10.6 $\pm$ 0.7  & N \\
101174.7 & HC$_{5}$N          &               $J=38-37$             &  6.0  & 0.015  & 0.471 $\pm$ 0.047  & 13.0 $\pm$ 0.5  & N \\
103836.8 & HC$_{5}$N          &               $J=39-38$             &  5.9  & 0.007  & 0.254 $\pm$ 0.040  & 12.9 $\pm$ 0.5  & N \\
106498.9 & HC$_{5}$N          &               $J=40-39$             &  9.6  & 0.012  & 0.459 $\pm$ 0.059  & 9.7 $\pm$ 0.3   & N \\
109161.0 & HC$_{5}$N          &               $J=41-40$             &  8.1  & 0.015  & 0.294 $\pm$ 0.038  & 6.7 $\pm$ 0.0   & N \\
95150.3  & C$_{4}$H          &       $N_{J}=10_{21/2}-9_{19/2}$    &  4.0  & 0.005  & 0.139 $\pm$ 0.028  & 12.9 $\pm$ 0.9  & (21), (33) \\
95188.9  & C$_{4}$H          &       $N_{J}=10_{19/2}-9_{17/2}$    &  5.0  & 0.008  & 0.173 $\pm$ 0.032  & 10.9 $\pm$ 0.2  & (21), (33)\\
96741.3 & C$_{4}$H & $\nu_{7}$=1, $^{2}\Pi_{3/2}$, $J=21/2-19/2$, $l$=f & 3.1  & 0.002  & 0.081 $\pm$ 0.022  & 14.2 $\pm$ 0.2 & N \\
104666.6 & C$_{4}$H           &       $N_{J}=11_{23/2}-10_{21/2}$   & 6.2 & 0.011    & 0.292 $\pm$ 0.040  & 12.6 $\pm$ 0.0  & N \\
104705.1 & C$_{4}$H           &       $N_{J}=11_{21/2}-10_{19/2}$   & 5.0 & 0.007    & 0.307 $\pm$ 0.034  & 13.4 $\pm$ 0.3  & N \\
\hline

\end{longtable}
\restoregeometry

\setlength{\tabcolsep}{5pt}
\begin{table}
\tbl{Unidentified lines.}{%
\begin{tabular}{ccccccc}\\
   \hline
     Frequency &  rms & {$T^{\rm *}_{\rm R}$}  & $\int T^{\rm *}_{\rm R}$d$v$ & {$V_{\rm exp}$}\\
     (MHz)  & (mK) & (K) & (K km s$^{-1}$) & (km s$^{-1}$)\\
   \hline
    95658  & 6.5 & 0.008 & $0.310 \pm  0.056$ & $19.6 \pm 0.5$ \\
    98107  & 4.9 & 0.005 & $0.226 \pm  0.046$ & $23.3 \pm 1.4$ \\
    112822 & 7.1 & 0.046 & $1.032 \pm  0.045$ & $13.6 \pm 0.3$ \\
   \hline
\end{tabular}}\label{table:3}

\end{table}


\begin{table*}
\tbl{Excitation temperatures, column densities, and fractional abundances with respect to H$_{2}$.$^{a}$}{%
\begin{tabular}{cccccccccccc}\\
    \hline
    Species & \multicolumn{3}{c}{{$T_{\rm ex}$} (K)} & & \multicolumn{3}{c}{{$N$}(cm$^{-2}$)} & & \multicolumn{3}{c}{$f_{\rm X}$}\\
    \cline{2-4} \cline{6-8} \cline{10-12}
     & This work & Z09$^{b}$ & C12$^{c}$  & & This work & Z09 & C12 & & This Work & Z09 & C12\\
    \hline
    CO                & 5.4    & ...  & ... &  & 2.95(17) & 1.63(17) & ...     &  & 9.67(-4) & ...      & ...  \\
    $^{13}$CO         & 5.3    & ...  & ... &  & 2.47(16) & 1.35(16) & ...     &  & 7.50(-5) & ...      & ...  \\
    CS                & 5.4    & ...  & ... &  & 2.96(14) & 1.72(14) & ...     &  & 1.98(-6) & 2.0(-6)  & 7.0(-6) \\
    C$^{34}$S         & 10.3   & ...  & ... &  & 2.02(13) & 2.58(13) & ...     &  & 9.71(-8) & 3.0(-7)  & ...  \\
    SiS               & 43.4   & 29.6 & ... &  & 1.46(14) & 2.96(14) & ...     &  & 1.55(-6) & 3.4(-6)  & ...  \\
    SiC$_{2}$         & 13.9   & 57.6 & ... &  & 9.08(13) & 1.53(14) & ...     &  & 2.48(-6) & 2.4(-6)  & ...  \\
    $^{29}$SiC$_{2}$  & 41.5   & ...  & ... &  & 4.41(13) & 2.26(13) & ...     &  & 4.89(-7) & 3.6(-7)  & ...  \\
    CN                & 4.2    & ...  & ... &  & 1.56(15) & 1.22(15) & ...     &  & 1.18(-5) & 2.6(-5)  & ...  \\
    C$_{3}$N          & 20.3   & ...  & ... &  & 6.11(13) & 2.16(13) & ...     &  & 1.93(-6) & 1.9(-6)  & ...  \\
    HC$_{3}$N         & 17.0   & 41.0 & 11.8 & & 1.28(14) & 7.69(13) & 3.3(14) &  & 1.98(-6) & 1.3(-6)  & 7(-6) \\
    H$^{13}$CCCN      & 84.4   & ...  & ... &  & 2.08(13) & ...      & ...     &  & 1.11(-7) & ...      &...  \\
    HC$_{5}$N         & 34.7   & ...  & 20  &  & 1.21(14) & ...      & 1.3(14) &  & 6.27(-7) & ...      & 2.6(-6) \\
    C$_{4}$H          & 39.9   & 53.9 & ... &  & 7.43(13) & 3.04(14) & ...     &  & 7.27(-6) & 4.0 (-6) & ...  \\
    \hline
\end{tabular}}\label{table:4}
\begin{tabnote}
$^{a}$x(y) represents $x\times10^y$.\\
$^{b}$Z09 refers to \cite{2009ApJ...691.1660Z}.\\
$^{c}$C12 refers to \cite{2012ApJ...760...66C}.\\
\end{tabnote}
\end{table*}

\begin{table*}
\tbl{Isotopic abundance ratios.}{%
\begin{tabular}{ccccccc}\\
    \hline
    Isotopic Ratio & \multicolumn{3}{c}{CIT\,6} & IRC+10216 & CRL 2688$^{b}$ & Solar$^{c}$\\
    \cline{2-4} 
     & Sepcies & This work & Z09$^{a}$ &  & \\
    \hline
        $^{12}$C/$^{13}$C   & $^{12}$CO/$^{13}$CO            & 12.9 $\pm$ 0.5  & 12.1 $\pm$ 1.3 & ...        & 14.1 & 89   \\
                            & H$^{12}$CCCN/H$^{13}$CCCN      & 17.8 $\pm$ 5.5  & ...            & 59.6$^{d}$ & 17.2 & ...  \\
        $^{28}$Si/$^{29}$Si & $^{28}$SiC$_2$/$^{29}$SiC$_2$  &  5.1 $\pm$ 1.5  &  6.7 $\pm$ 3.1 & ...        & 26.8 & 19.6 \\
        $^{32}$S/$^{34}$S   & C$^{32}$S/C$^{34}$S            & 20.4 $\pm$ 4.8  &  6.7 $\pm$ 1.2 & 20.4$^{e}$ & 9.5  & 22.5  \\
    \hline
\end{tabular}}\label{table:5}
\begin{tabnote}
$^{a}$Z09 refers to \citealt{2009ApJ...691.1660Z}.\\
$^{b}$From \cite{2022ApJS..259...56Q}.\\
$^{c}$From \cite{2003ApJ...591.1220L}.\\
$^{d}$From \cite{2017A&A...606A..74Z}.\\
$^{e}$From \cite{2000A&AS..142..181C}.\\
\end{tabnote}

\end{table*}

\end{CJK*}


\begin{thebibliography}{}
%
\bibitem[Ag{\'u}ndez et al.(2014)]{2014ApJ...790L..27A} Ag{\'u}ndez, M., Cernicharo, J., Decin, L., et al.\ 2014, \apjl, 790, L27. 
\bibitem[Ag{\'u}ndez et al.(2008)]{2008Ap&SS.313..229A} Ag{\'u}ndez, M., Cernicharo, J., Pardo, J.~R., et al.\ 2008, \apss, 313, 229. 
\bibitem[Aoki et al.(2022)]{2022PASJ...74..273A} Aoki, W., Beers, T.~C., Honda, S., et al.\ 2022, \pasj, 74, 273.
\bibitem[Bachiller et al.(1988)]{1988A&A...196L...5B} Bachiller, R., Gomez-Gonzalez, J., Bujarrabal, V., et al.\ 1988, \aap, 196, L5
\bibitem[Bachiller et al.(1997)]{1997A&A...319..235B} Bachiller, R., Fuente, A., Bujarrabal, V., et al.\ 1997, \aap, 319, 235
\bibitem[Boothroyd \& Sackmann(1999)]{1999ApJ...510..232B} Boothroyd, A.~I. \& Sackmann, I.-J.\ 1999, \apj, 510, 232.
\bibitem[Bublitz et al.(2019)]{2019A&A...625A.101B} Bublitz, J., Kastner, J.~H., Santander-Garc{\'\i}a, M., et al.\ 2019, \aap, 625, A101.
\bibitem[Bujarrabal et al.(1994)]{1994A&A...285..247B} Bujarrabal, V., Fuente, A., \& Omont, A.\ 1994, \aap, 285, 247
\bibitem[Bujarrabal et al.(1994)]{1994ApJ...421L..47B} Bujarrabal, V., Fuente, A., \& Omont, A.\ 1994, \apjl, 421, L47.
\bibitem[Cernicharo et al.(2000)]{2000A&AS..142..181C} Cernicharo, J., Gu{\'e}lin, M., \& Kahane, C.\ 2000, \aaps, 142, 181. 
\bibitem[Cernicharo et al.(2011)]{2011IAUS..280..237C} Cernicharo, J., Ag{\'u}ndez, M., \& Gu{\'e}lin, M.\ 2011, in The Molecular Universe, eds. J. Cernicharo, \& R. Bachiller
(Cambridge University Press: Cambridge), 280, 237.
\bibitem[Charbonnel(1995)]{1995ApJ...453L..41C} Charbonnel, C.\ 1995, \apjl, 453, L41.
\bibitem[Charbonnel \& Do Nascimento(1998)]{1998A&A...336..915C} Charbonnel, C. \& Do Nascimento, J.~D.\ 1998, \aap, 336, 915.  
\bibitem[Chau et al.(2012)]{2012ApJ...760...66C} Chau, W., Zhang, Y., Nakashima, J.-. ichi ., et al.\ 2012, \apj, 760, 66. 
\bibitem[Cohen \& Hitchon(1996)]{1996AJ....111..962C} Cohen, M. \& Hitchon, K.\ 1996, \aj, 111, 962. 
\bibitem[Crosas \& Menten(1997)]{1997ApJ...483..913C} Crosas, M. \& Menten, K.~M.\ 1997, \apj, 483, 913. 
\bibitem[Dayal \& Bieging(1993)]{1993ApJ...407L..37D} Dayal, A. \& Bieging, J.~H.\ 1993, \apjl, 407, L37. 
\bibitem[Frost et al.(1998)]{1998A&A...332L..17F} Frost, C.~A., Cannon, R.~C., Lattanzio, J.~C., et al.\ 1998, \aap, 332, L17. 
\bibitem[Fukasaku et al.(1994)]{1994ApJ...437..410F} Fukasaku, S., Hirahara, Y., Masuda, A., et al.\ 1994, \apj, 437, 410.
\bibitem[Gong et al.(2015)]{2015A&A...574A..56G} Gong, Y., Henkel, C., Spezzano, S., et al.\ 2015, \aap, 574, A56. 
\bibitem[Groenewegen et al.(1996)]{1996A&A...306..241G} Groenewegen, M.~A.~T., Baas, F., de Jong, T., et al.\ 1996, \aap, 306, 241
\bibitem[Groenewegen et al.(1998)]{1998A&A...338..491G} Groenewegen, M.~A.~T., van der Veen, W.~E.~C.~J., \& Matthews, H.~E.\ 1998, \aap, 338, 491. 
\bibitem[He et al.(2008)]{2008ApJS..177..275H} He, J.~H., Dinh-V-Trung, Kwok, S., et al.\ 2008, \apjs, 177, 275. 
\bibitem[Henkel et al.(1985)]{1985A&A...147..143H} Henkel, C., Matthews, H.~E., Morris, M., et al.\ 1985, \aap, 147, 143
\bibitem[Herwig(2005)]{2005ARA&A..43..435H} Herwig, F.\ 2005, \araa, 43, 435. 
\bibitem[Hrivnak \& Bieging(2005)]{2005ApJ...624..331H} Hrivnak, B.~J. \& Bieging, J.~H.\ 2005, \apj, 624, 331.
\bibitem[H{\"o}fner \& Olofsson(2018)]{2018A&ARv..26....1H} H{\"o}fner, S. \& Olofsson, H.\ 2018, \aapr, 26, 1.
\bibitem[Jewell \& Snyder(1982)]{1982ApJ...255L..69J} Jewell, P.~R. \& Snyder, L.~E.\ 1982, \apjl, 255, L69.
\bibitem[Jewell \& Snyder(1984)]{1984ApJ...278..176J} Jewell, P.~R. \& Snyder, L.~E.\ 1984, \apj, 278, 176.  
\bibitem[Kahane et al.(1992)]{1992A&A...256..235K} Kahane, C., Cernicharo, J., Gomez-Gonzalez, J., et al.\ 1992, \aap, 256, 235
\bibitem[Kawaguchi et al.(2007)]{2007PASJ...59L..47K} Kawaguchi, K., Fujimori, R., Aimi, S., et al.\ 2007, \pasj, 59, L47. 
\bibitem[Knapp \& Chang(1985)]{1985ApJ...293..281K} Knapp, G.~R. \& Chang, K.~M.\ 1985, \apj, 293, 281.
\bibitem[Knapp \& Morris(1985)]{1985ApJ...292..640K} Knapp, G.~R. \& Morris, M.\ 1985, \apj, 292, 640.
\bibitem[Lindqvist et al.(2000)]{2000A&A...361.1036L} Lindqvist, M., Sch{\"o}ier, F.~L., Lucas, R., et al.\ 2000, \aap, 361, 1036
\bibitem[Lodders(2003)]{2003ApJ...591.1220L} Lodders, K.\ 2003, \apj, 591, 1220. 
\bibitem[Lovas(2004)]{2004JPCRD..33..177L} Lovas, F.~J.\ 2004, Journal of Physical and Chemical Reference Data,(AIP Publishing), 33, 177.
\bibitem[Lucas \& Gu{\'e}lin(1999)]{1999IAUS..191..305L} Lucas, R. \& Gu{\'e}lin, M.\ 1999, Asymptotic Giant Branch Stars, 191, 305
\bibitem[Lucas et al.(1986)]{1986A&A...154L..12L} Lucas, R., Omont, A., Guilloteau, S., et al.\ 1986, \aap, 154, L12
\bibitem[Margulis et al.(1990)]{1990ApJ...361..673M} Margulis, M., van Blerkom, D.~J., Snell, R.~L., et al.\ 1990, \apj, 361, 673.
\bibitem[McGuire(2018)]{2018ApJS..239...17M} McGuire, B.~A.\ 2018, \apjs, 239, 17.
\bibitem[McGuire(2022)]{2022ApJS..259...30M} McGuire, B.~A.\ 2022, \apjs, 259, 30. 
\bibitem[Meixner et al.(1998)]{1998ApJ...509..392M} Meixner, M., Campbell, M.~T., Welch, W.~J., et al.\ 1998, \apj, 509, 392.
\bibitem[Milam et al.(2009)]{2009ApJ...690..837M} Milam, S.~N., Woolf, N.~J., \& Ziurys, L.~M.\ 2009, \apj, 690, 837. 
\bibitem[M{\"u}ller et al.(2001)]{2001A&A...370L..49M} M{\"u}ller, H.~S.~P., Thorwirth, S., Roth, D.~A., et al.\ 2001, \aap, 370, L49. 
\bibitem[M{\"u}ller et al.(2005)]{2005JMoSt.742..215M} M{\"u}ller, H.~S.~P., Schl{\"o}der, F., Stutzki, J., et al.\ 2005, Journal of Molecular Structure, 742, 215. 
\bibitem[Neri et al.(1998)]{1998A&AS..130....1N} Neri, R., Kahane, C., Lucas, R., et al.\ 1998, \aaps, 130, 1. doi:10.1051/aas:1998213
\bibitem[Nyman et al.(1992)]{1992A&AS...93..121N} Nyman, L.-A., Booth, R.~S., Carlstrom, U., et al.\ 1992, \aaps, 93, 121
\bibitem[Olofsson \& Rydbeck(1984)]{1984A&A...136...17O} Olofsson, H. \& Rydbeck, G.\ 1984, \aap, 136, 17
\bibitem[Olofsson(1997)]{1997IAUS..178..457O} Olofsson, H.\ 1997, IAU Symposium, 178, 457
\bibitem[Olofsson et al.(1998)]{1998A&A...329.1059O} Olofsson, H., Lindqvist, M., Nyman, L.-A., et al.\ 1998, \aap, 329, 1059
\bibitem[Pardo et al.(2005)]{2005ApJ...628..275P} Pardo, J.~R., Cernicharo, J., \& Goicoechea, J.~R.\ 2005, \apj, 628, 275. 
\bibitem[Pardo et al.(2022)]{2022A&A...658A..39P} Pardo, J.~R., Cernicharo, J., Tercero, B., et al.\ 2022, \aap, 658, A39. 
\bibitem[Pei \& Chen(1997)]{1997PPMtO..16..226P} Pei, C.~C. \& Chen, Y.~F.\ 1997, Publications of the Purple Mountain Observatory, 16, 226
\bibitem[Pickett et al.(1998)]{1998JQSRT..60..883P} Pickett, H.~M., Poynter, R.~L., Cohen, E.~A., et al.\ 1998, \jqsrt, 60, 883. 
\bibitem[Qiu et al.(2022)]{2022ApJS..259...56Q} Qiu, J.-J., Zhang, Y., Zhang, J.-S., et al.\ 2022, \apjs, 259, 56. 
\bibitem[Qiu et al.(2023)]{2023A&A...669A.121Q} Qiu, J.-J., Zhang, Y., Nakashima, J.-. ichi ., et al.\ 2023, \aap, 669, A121. 
\bibitem[Ramstedt \& Olofsson(2014)]{2014A&A...566A.145R} Ramstedt, S. \& Olofsson, H.\ 2014, \aap, 566, A145.
\bibitem[Ramstedt et al.(2008)]{2008A&A...487..645R} Ramstedt, S., Sch{\"o}ier, F.~L., Olofsson, H., et al.\ 2008, \aap, 487, 645.
\bibitem[Sackmann \& Boothroyd(1999)]{1999ApJ...510..217S} Sackmann, I.-J. \& Boothroyd, A.~I.\ 1999, \apj, 510, 217.
\bibitem[Sahai et al.(1984)]{1984ApJ...284..144S} Sahai, R., Wootten, A., \& Clegg, R.~E.~S.\ 1984, \apj, 284, 144.
\bibitem[Sahai(1990)]{1990ApJ...362..652S} Sahai, R.\ 1990, \apj, 362, 652. doi:10.1086/169303
\bibitem[Sch{\"o}ier et al.(2002)]{2002A&A...391..577S} Sch{\"o}ier, F.~L., Ryde, N., \& Olofsson, H.\ 2002, \aap, 391, 577.
\bibitem[Sch{\"o}ier et al.(2007)]{2007A&A...473..871S} Sch{\"o}ier, F.~L., Bast, J., Olofsson, H., et al.\ 2007, \aap, 473, 871.
\bibitem[Sch{\"o}ier \& Olofsson(2001)]{2001A&A...368..969S} Sch{\"o}ier, F.~L. \& Olofsson, H.\ 2001, \aap, 368, 969.
\bibitem[Shinnaga et al.(2017)]{2017PASJ...69L..10S} Shinnaga, H., Claussen, M.~J., Yamamoto, S., et al.\ 2017, \pasj, 69, L10. 
\bibitem[Sopka et al.(1989)]{1989A&A...210...78S} Sopka, R.~J., Olofsson, H., Johansson, L.~E.~B., et al.\ 1989, \aap, 210, 78
\bibitem[Takano et al.(1992)]{1992PASJ...44..469T} Takano, S., Saito, S., \& Tsuji, T.\ 1992, \pasj, 44, 469
\bibitem[Thronson \& Mozurkewich(1983)]{1983ApJ...271..611T} Thronson, H.~A. \& Mozurkewich, D.\ 1983, \apj, 271, 611.
\bibitem[Ulrich et al.(1966)]{1966ApJ...146..288U} Ulrich, B.~T., Neugebauer, G., McCammon, D., et al.\ 1966, \apj, 146, 288. 
\bibitem[Woods et al.(2003a)]{2003A&A...402..189W} Woods, P.~M., Millar, T.~J., Herbst, E., et al.\ 2003a, \aap, 402, 189. 
\bibitem[Woods et al.(2003b)]{2003A&A...402..617W} Woods, P.~M., Sch{\"o}ier, F.~L., Nyman, L.-{\r{A}}., et al.\ 2003b, \aap, 402, 617. 
\bibitem[Zhang et al.(2017)]{2017A&A...606A..74Z} Zhang, X.-Y., Zhu, Q.-F., Li, J., et al.\ 2017, \aap, 606, A74. 
\bibitem[Zhang et al.(2020)]{2020PASJ...72...46Z} Zhang, Y., Chau, W., Nakashima, J.-. ichi ., et al.\ 2020, \pasj, 72, 46. 
\bibitem[Zhang et al.(2009a)]{2009ApJ...691.1660Z} Zhang, Y., Kwok, S., \& Dinh-V-Trung\ 2009a, \apj, 691, 1660. 
\bibitem[Zhang et al.(2009b)]{2009ApJ...700.1262Z} Zhang, Y., Kwok, S., \& Nakashima, J.\ 2009b, \apj, 700, 1262. 
\bibitem[Zhang et al.(2008)]{2008ApJ...678..328Z} Zhang, Y., Kwok, S., \& Dinh-V-Trung\ 2008, \apj, 678, 328.
\bibitem[Zhang et al.(2013)]{2013ApJ...773...71Z} Zhang, Y., Kwok, S., Nakashima, J. et al.\ 2013, \apj, 773, 71. 
\bibitem[Ziurys(2006)]{2006PNAS..10312274Z} Ziurys, L.~M.\ 2006, PNAS, 103, 12274.
\bibitem[Zuckerman \& Dyck(1989)]{1989A&A...209..119Z} Zuckerman, B. \& Dyck, H.~M.\ 1989, \aap, 209, 119
\bibitem[Zuckerman(1981)]{1981AJ.....86...84Z} Zuckerman, B.\ 1981, \aj, 86, 84.



\end{thebibliography}
\end{document}